\documentclass[fleqn,usenatbib,useAMS]{mnras}

\usepackage{graphicx}	
\usepackage{amsmath}	
\usepackage{multicol}        
\usepackage{bm}		
\usepackage{pdflscape}	
\usepackage[figuresleft]{rotfloat}
\usepackage{CJKutf8}
\usepackage{orcidlink}
\usepackage[T1]{fontenc}
\usepackage{booktabs}

\DeclareRobustCommand{\VAN}[3]{#2}
\let\VANthebibliography\thebibliography
\def\thebibliography{\DeclareRobustCommand{\VAN}[3]{##3}\VANthebibliography}

\newcommand{\SigSFR}{$\mathrm{\Sigma_{\mathrm{SFR}}}$}
\newcommand{\ha}{H$\mathrm{\alpha}$}
\newcommand{\dphi}{$\mathrm{\Delta\phi}$}
\newcommand{\re}{$R_e$}
\newcommand{\dSFR}{$\mathrm{\Delta \Sigma_{SFR}}$}
\newcommand{\dOH}{$\mathrm{\Delta log(O/H)}$}

\usepackage[T1]{fontenc}
\usepackage{ae,aecompl}

\usepackage{newtxtext,newtxmath}
\usepackage{pdflscape}
\usepackage{comment}

\title[Azimuthal ISM variations in spiral galaxies]{Quantifying azimuthal variations within the interstellar medium of $z\sim$ 0 spiral galaxies with the TYPHOON survey}

\author[Qian-Hui Chen et al.]{\ignorespaces
    Qian-Hui Chen (陈千惠)$^{\orcidlink{0000-0002-4382-1090}}$, $ ^{1,2}$\thanks{E-mail: Qianhui.Chen@anu.edu.au}
    Kathryn Grasha$^{\orcidlink{0000-0002-3247-5321}}$, $ ^{1,2,11}$\thanks{ARC DECRA Fellow}
    Andrew J. Battisti$^{\orcidlink{0000-0003-4569-2285}}$, $^{1,2}$
    Emily Wisnioski$^{\orcidlink{0000-0003-1657-7878}}$, $^{1,2}$
    \newauthor
    Zefeng Li (李泽峰)$^{\orcidlink{0000-0001-7373-3115}}$, $^{1,2,3}$
    Hye-Jin Park$^{\orcidlink{0000-0002-9809-6631}}$, $^{1,2}$
    Brent Groves$^{\orcidlink{0000-0002-9768-0246}}$, $^{2,4}$
    Paul Torrey$^{\orcidlink{0000-0002-5653-0786}}$, $^{5}$
    Trevor Mendel$^{\orcidlink{0000-0002-6327-9147}}$, $^{1,2}$
    \newauthor
    Barry F. Madore$^{\orcidlink{0000-0002-1576-1676}}$, $^{6,7}$ 
    Mark Seibert$^{\orcidlink{0000-0002-1143-5515}}$, $^{6}$
    Eva Sextl$^{\orcidlink{0009-0001-5618-4326}}$, $^{8}$
    Alex M. Garcia$^{\orcidlink{0000-0002-8111-9884}}$, $^{5}$
    Jeff A. Rich$^{\orcidlink{0000-0002-5807-5078}}$, $^{6}$
    \newauthor
    Rachael L. Beaton$^{\orcidlink{0000-0002-1691-8217}}$, $^{9,6}$\thanks{Hubble Fellow}
    and
    Lisa J. Kewley$^{\orcidlink{0000-0001-8152-3943}}$ $^{10,1,2}$
	\\
	$^{1}$Research School of Astronomy and Astrophysics, Australian National University, Canberra, ACT 2611, Australia\\
	$^{2}$ARC Centre of Excellence for All Sky Astrophysics in 3 Dimensions (ASTRO 3D), Australia\\
        $^{3}$Centre for Extragalactic Astronomy, Department of Physics, Durham University, South Road, Durham DH1 3LE, UK\\
        $^{4}$International Centre for Radio Astronomy Research, University of Western Australia, 35 Stirling Highway, Crawley WA 6009, Australia\\
        $^{5}$Department of Astronomy, University of Virginia, Charlottesville, VA 22904, USA\\
        $^{6}$The Observatories, Carnegie Institution for Science, 813 Santa Barbara Street, Pasadena, CA 91106, USA\\
	$^{7}$Department of Astronomy and Astrophysics, University of Chicago, Chicago, IL 60637, USA\\
        $^{8}$Universitäts-Sternwarte, Fakultät für Physik, Ludwig-Maximilians Universität München, Scheinerstraße 1, D-81679 München, Germany\\
        $^{9}$Department of Astrophysical Sciences, 4 Ivy Lane, Princeton University, Princeton, NJ 08544, USA\\
        $^{10}$Institute for Theory and Computation, Harvard-Smithsonian Center for Astrophysics, Cambridge, MA 02138, USA \\
        $^{11}$Visiting Fellow, Harvard-Smithsonian Center for Astrophysics, 60 Garden Street, Cambridge, MA 02138, USA\\
}

\date{Last updated XX; in original form XX}

\pubyear{2024}

\begin{document}
\begin{CJK}{UTF8}{gbsn}
	\label{firstpage}
	\pagerange{\pageref{firstpage}--\pageref{lastpage}}
	\maketitle
\begin{abstract}
Most star formation in the local Universe occurs in spiral galaxies, but their origin remains an unanswered question.
Various theories have been proposed to explain the development of spiral arms, each predicting different spatial distributions of the interstellar medium.
This study maps the star formation rate (SFR) and gas-phase metallicity of nine spiral galaxies with the TYPHOON survey to test two dominating theories: \textit{density wave theory} and \textit{dynamic spiral theory}.
We discuss the environmental effects on our galaxies, considering reported environments and merging events.
Taking advantage of the large field of view covering the entire optical disk, we quantify the fluctuation of SFR and metallicity relative to the azimuthal distance from the spiral arms.
We find higher SFR and metallicity in the trailing edge of NGC~1365 (by 0.117~dex and 0.068~dex, respectively) and NGC~1566 (by 0.119~dex and 0.037~dex, respectively), which is in line with density wave theory.
NGC~2442 shows a different result with higher metallicity (0.093~dex) in the leading edge, possibly attributed to an ongoing merging.
The other six spiral galaxies show no statistically significant offset in SFR or metallicity, consistent with dynamic spiral theory.
We also compare the behaviour of metallicity inside and outside the co-rotation radius (CR) of NGC~1365 and NGC~1566.
We find comparable metallicity fluctuations near and beyond the CR of NGC~1365, indicating gravitational perturbation.
NGC~1566 shows the greatest fluctuation near the CR, in line with the analytic spiral arms. 
Our work highlights that a combination of mechanisms explains the origin of spiral features in the local Universe.

\end{abstract}


\section{Introduction}\label{sec:intro}

Spiral galaxies constitute approximately two-thirds of all massive galaxies in the local Universe \citep{Lintott_2008, Willett_2013} and are the primary hosts for most star formation \citep{Brinchmann_2004}.
Despite their ubiquity, the underlying physics that drives the origin of spiral features remains a topic of ongoing debate. 
Numerous previous studies over the past decades have endeavoured to explain the formation of spiral arms through various theoretical frameworks \citep{Lin_1964, Toomre_1977, Athanassoula_1992, Binney_2008}.

\begin{figure}
    \centering
    \includegraphics[width = 0.5\textwidth]{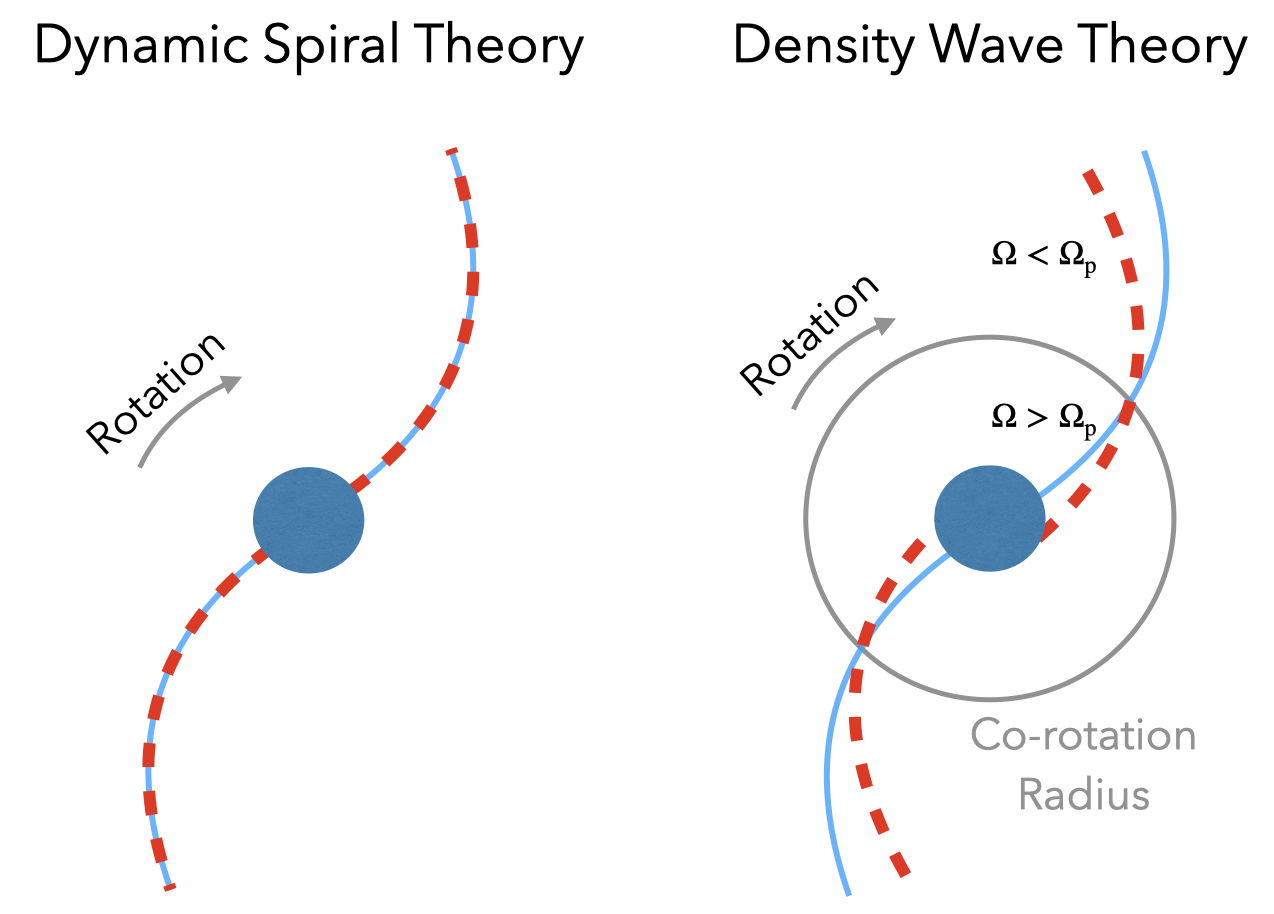}
    \caption{Theoretical expectations from the dynamic spiral theory (left) and density wave theory (right) regarding the location of young (blue) and old (red) stellar populations. The azimuthal offset between young and old stars is predicted by density wave theory but not by dynamic spiral theory, also see Fig~1 in \citet{Puerari_1997} and \citet{Martinez-Garcie_2009}. The rotation direction is based on the assumption that the observed spiral arms are trailing features, which will be applied to all spiral galaxies in this work. $\Omega$ refers to the rotational velocity of material while $\Omega_p$ denotes the rotational velocity of the spiral pattern.}
    \label{fig:theory}
\end{figure}

Among all the proposed theories, the three most widely accepted ones are the density wave theory, dynamic spiral theory, and tidal-induced spiral arms. 
While the proposed mechanisms may potentially all contribute to the formation and evolution of spiral features in a galaxy, each has specific characteristics that we detail below:

{\textit {(Quasi-stationary) density wave theory}}: Proposed by \citet{Lin_1964}, later improved and popularised by \citet{Toomre_1977}, \citet{Bertin_1996} and \citet{Shu_2016}, the density wave theory envisages long-lived spiral arms and solves the winding problem\footnote{The winding problem: in other theories, the pitch angle of the spiral features is expected to decrease to 0 over time, which is in disagreement with observations of long-lived spirals.}.
This theory proposes that spiral features are areas of greater density that rotate at a specific pattern speed across the disc, and the differential gravitational pull leads to a logarithmic spiral with a constant pitch angle \citep{Athanassoula_2010, Martinez-Garcia_2012,Davis_2015}.
As the pattern speed of the spiral arms is constant, while the rotational velocity of the stars varies radially, differential rotation occurs (right panel of Fig~\ref{fig:theory}). 
Only at the co-rotation radius (CR) are the stars and the arms expected to rotate synchronously.
Due to differential rotation, the new-borne stars rotate faster than the spiral pattern within the CR while falling behind outside the CR, which has been observed in previous works \citep{Pour-Imani_2016, Peterken_2019}.
Thus in this theory, the interstellar medium (ISM) properties and stellar populations differ on the leading edge and trailing edge of the spiral arms \citep{Gittins_2004}.
\citet{Ho_2017} observe lower gas-phase metallicity in the leading edge than the trailing edge in NGC~1365, which can be explained by their toy model.

{\textit{Dynamic spiral theory}}: \citet{Sellwood_1984} suggests that spiral arms are short-lived and recurrent features.
In numerical simulations, dynamic spiral arms arise rapidly from gravitational instability due to swing amplification \citep{Fujii_2011, Grand_2012,DOnghia_2013}.
However, this kind of spiral arms fades quickly because of the particle scattering and increased velocity dispersion of stars in the simulations.
The spiral arms heat the disc kinematically and undergo a cycle of breaking up into small segments of kpc in size.
The segments possibly will reconnect to form new, large-scale spiral patterns, which makes the spiral features ``recurrent''.
While the dynamic spiral arms may appear globally cohesive, the assemblies of segments can form at distinct times and later merge with the arms. 
This contrasts with density wave theory, where the spiral arms arise as an entity.
A kinematically cold population of stars help maintain the dynamic spiral arms while gas dissipation and accretion introduce instabilities.
There is no significant difference in rotation between the disc and the dynamic spiral arms.
Thus, the stars do not flow across the spiral features but stay still in the arm regions due to gravitational potential.
This means no coherent azimuthal variations are expected between the gas and stars in dynamical spiral theory (left panel of Fig~\ref{fig:theory}).

{\textit{Tidal-induced spiral arms}}: Tidal interactions are common and give rise to tails, bridges and spiral features \citep{Pfleiderer_1961, PikelNer1965, Toomre_1972}.
\citet{Toomre_1969} predict that the outer arms and tidal tails are material arms, rotating at a similar angular velocity with the disc, which is observed in M51 \citep{Meidt_2013}.
However, some simulations \citep{Sundelius_1987} and observations \citep{Rots_1975} argue that tidal-induced spiral arms are density waves.
Interestingly, the density wave spiral arms from tidal interactions can be quasi-stationary or kinematic --- the gas and stars do flow through the spiral arms but the spiral pattern speed decreases with radii \citep{Donner_1994,Oh_2015}.
Simulations \citep{Pettitt_2016, Pettitt_2017} report that azimuthal offsets do exist in different media in a tidal-induced spiral galaxy, in agreement with observations \citep{Schinnerer_2013, Egusa_2017}.

The fundamental physics that drives the formation of spiral features remains a topic of debate, given the various observational results among different spiral galaxies in our Universe.
One of the practical methods to discern the dominant theory is to detect the azimuthal variation across the spiral arms --- as the density wave theory predicts an observable azimuthal offset in ISM properties such as gas-phase metallicity and stellar ages while dynamic spiral theory predicts no offset.
However, existing observational evidence is conflicted with clear azimuthal variations in metallicities seen in some galaxy samples \citep{Sanchez-Menguiano_2016, Ho_2017, Vogt_2017, Ho_2018} but not in others \citep[e.g.][]{Foyle_2011, Kreckel_2019,Chen_2024}.
The discrepancy also occurs in stellar ages with some galaxies showing azimuthal variation \citep{Martinez-Garcie_2009, Sanchez-Gil_2011, Abdeen_2022} and others not \citep{Choi_2015, Shabani_2018}.
More studies using multi-wavelength imaging data report the offset between the young and old stellar populations \citep{Egusa_2004, Egusa_2009, Pour-Imani_2016, Yu_2019,Savchenko_2020, Abdeen_2020}, in line with density wave theory.

For isolated spiral galaxies, density wave and dynamic spiral theory are two widely accepted mechanisms driving the spiral features.
However, tidal forces from interactions with passing or nearby companions can also give rise to spiral arms with an offset between the gas and the stars \citep{Pettitt_2017}, making it more difficult to test density wave theory or dynamic spiral theory.
Conversely, tidal-induced spiral galaxies show distinct behaviours in their tail and bridge arms.
Gravity perturbations from the interaction can create two armed spiral galaxies \citep{Toomre_1972, Pettitt_2018}, which is likely the case in M51 \citep{Dobbs_2010}.
Spiral features created in tidal interactions are more likely to be kinematic density waves \citep[Chapter 6 in][]{Binney_1987} that will wind up faster than quasi-stationary density waves.
However, the gas moves through the spiral arms from the trailing edge to the leading edge, similar to the quasi-stationary density waves \citep{Oh_2008, Struck_2011}.
These findings indicate that the ISM and stellar distributions in spiral galaxies may be under the impact of more than one physical mechanism.

With three common theories briefly explained, we summarise the expectations of star formation surface density (\SigSFR) and gas-phase metallicity in observation from the different mechanisms.
In density wave theory, the spiral arms observed in shorter wavelengths, where the most recent star formation event has occurred, should exhibit a looser pitch angle (blue arms in the right panel of Fig~\ref{fig:theory}) than the spiral arms observed in redder wavelengths, dominated by the old stars \citep{Pour-Imani_2016}.
If star formation occurs after gas clouds pass through the minimum potential of density waves \citep[left spiral arm in Fig~1 of][]{Pour-Imani_2016}, the peaks of \SigSFR\ are predicted in the leading edge inside the CR \citep{Martinez-Garcie_2009} . 
If star formation starts as the gas clouds approach the density wave \citep[right spiral arm in Fig~1 of][]{Pour-Imani_2016}, \SigSFR\ should be higher in the trailing edge inside the CR.
In contrast, the dynamic spiral theory predicts a symmetrical \SigSFR\ distribution to the spiral arms.
Tidal induced-spiral structures will lead to kinematic density waves which result in azimuthal variations, exhibiting a decreasing pattern speed with increasing radii.
The comparison of \SigSFR\ between the leading and trailing edge, combined with the reported environmental factors, will help to disentangle the origin of spiral features.

For gas-phase metallicity, spiral arms driven by density wave theory will lead to higher metallicity in the trailing edge than the leading edge, within the CR.
This scenario is modelled in \citep{Ho_2017} as 1) self-enrichment in the trailing edge before the material reaches the spiral arms, and 2) metal mixing in the leading edge.
In dynamic spiral theory, gas-phase metallicity is expected to be symmetric on both sides of the spiral arms, as there is no differential rotation between the spiral arms and the disc.

Although many spiral galaxies in the local Universe and at higher redshift have been studied, most researches i) focus on HII regions \citep[e.g., ][]{Rozas_1996, Santoro_2022}; ii) examined azimuthal variations through visual comparison on 2D maps \citep{Kreckel_2019}, or by binning the spaxels into leading and trailing sections \citep{Ho_2018, Chen_2024}.
These previous studies offer limited quantitative analysis of the behaviour of the gas and stars as they move away from the spiral arms.
In this work, we present the azimuthal distributions of star formation rate (SFR) and gas-phase metallicity and test the toy model in \citet{Ho_2017}, with a sample of nine nearby spiral galaxies.
Our motivation is to quantitatively track the fluctuation in the ISM properties when moving in/out of the spiral arms.

We introduce our observations and sample in Sec~\ref{sec:data}. 
We present our analysis including spiral arm definition and mapping the ISM properties in Sec~\ref{sec:analysis}.
We report the fluctuation of the ISM properties when moving through the spiral arms in Sec~\ref{sec:results}.
In Sec~\ref{sec:discussion}, we will discuss the effects of CR on our results and the dominant mechanism(s) responsible for driving the spiral features in these galaxies. 
Luminosity distances are adopted from \citet{Leroy_2019}, assuming H$_0$ = 70~km~s$^{-1}$Mpc$^{-1}$ and a flat cosmology with $\Omega_m$ = 0.27.

\section{Observations and sample selection}\label{sec:data}
\subsection{Observations}\label{sec:obs}
TYPHOON\footnote{\href{https://typhoon.datacentral.org.au/}{https://typhoon.datacentral.org.au/}} is a pseudo-IFU survey of 44 galaxies observable in the southern hemisphere. 
TYPHOON uses the Wide Field CCD (WFCCD) imaging spectrograph ($18'$ $\times$ 1.65\arcsec) on the 2.5m du Pont telescope at the Las Campanas Observatory in Chile. 
The TYPHOON survey builds up a dispersed image data cube by applying the Progressive Integral Step Method (PrISM), also known as the “step-and-stare” technique. 
The large field of view (FoV) of TYPHOON, ranging from 2.3' $\times$ 18' to 6.5' $\times$ 18' in our selected spiral galaxies, allows us to observe the entire optical disc of nearby star-forming galaxies in an IFU-like manner.
The spectrograph is configured to have a resolving power of approximately R $\approx$ 850 at 7000~\AA\ and R $\approx$ 960 at 5577~\AA.
Our spectra cover the wavelength range of 3650$-$8150~\AA\ with a flux calibration accuracy of 2\% \citep{Ho_2017}.
The observations are conducted only when the seeing is smaller than the slit width of 1.65\arcsec \citep[to prevent slit loss; ][]{Grasha_2022}.
More detailed information about the TYPHOON/PrISM survey is in \citet{Ho_2017}, \citet{Chen_2023}, and Seibert et al. (In prep.). 
The raw data are reduced into 3D data cubes using a standard long-slit data reduction procedure (Seibert et al. In prep.). 
The reduced 2D spectra are later tiled together to form 3D data cubes with spectral and spatial samplings of $1.5$\AA\ and 1.65\arcsec, respectively.
The astrometric solutions are made for combined slit steps of individual nights independently and they are tied to the Gaia reference system \citep{Gaia_Collaboration_2016}.

\subsection{Sample selection}\label{sec:sample}
Our work includes bright ($M_{\mathrm{v}}$ \textless{} -20 mag) spiral galaxies that have three spiral arms or fewer.
We exclude flocculent galaxies with poorly defined arms to allow for better comparison between leading and trailing regions.
We exclude one galaxy (NGC~1068) with more than 20\% spaxels impacted by harder-component ionisation \citep[e.g., active galactic nuclei, AGN; ][]{Lamastra_2016,DAgostino_2018}, based on the Baldwin, Phillips \& Terlevich diagram \citep[BPT diagram;][]{BPT_1981}.
The remaining spiral galaxies in our sample show $<$ 2\% of their spaxels dominated by AGN (column 12 of Table~\ref{tab:info}). 
These selection criteria reduce the total sample to nine well-defined, star-forming galaxies available in this work.
Two galaxies have a strong bar (Hubble type of SB) and seven galaxies exhibit a weak bar (Hubble type of SAB).
We constrain our analysis of azimuthal variation starting from/beyond the end of bars (Sec~\ref{sec:arm_define}) to focus on the study of spiral arms.
The detailed physical parameters are listed in Table~\ref{tab:info}.

The large FoV allows us to cover the disc regions extending beyond $R_{25}$ for our entire sample.
Our sample spans a resolution of $52-170$ parsec per pixel (1.65'' per pixel), enabling us to 1) identify the central ridge line of each spiral arm, and 2)  measure changes in the ISM when moving azimuthally from the spiral arms on a spaxel level.
These azimuthal distributions are key observables to distinguish between competing spiral theories.

\begin{table*}
\scriptsize
    \newcommand{\tabincell}[2]{\begin{tabular}{@{}#1@{}}#2\end{tabular}}
    \centering
    \begin{tabular}{cccccccccccc}
        \hline
        \hline
        Galaxy & T-type & Morphology & R.A & Dec. & \tabincell{c}{Inclination\\ (degrees)} & \tabincell{c}{P.A.\\ (degrees)} & \tabincell{c}{Distance\\ (Mpc)} & \tabincell{c}{logM$_{*}$ \\ (M$_{\odot}$)} & \tabincell{c}{$R_{25}$\\ (arcmin)} & \tabincell{c}{Number of\\ spiral arms} & \tabincell{c}{Fraction of\\ excluded spaxels}\\
        (1) & (2) & (3) & (4) & (5) & (6) & (7) & (8) & (9) & (10) & (11) & (12) \\
	\hline
        NGC~1365 & 3.2 $\pm$ 0.7 & SB(s)b & 03h33m36.371s & -36d08m25.45s & 35.7 & 49.5 & 18.1 $\pm$ 0.04 &  10.75 $\pm$ 0.10 & 5.61 & 2 & 1.11\%\\
        NGC~1566 & 4.0 $\pm$ 0.2 & SAB(s)bc & 04h20m00.42s & -54d56m16.1s & 49.1 & 44.2 & 18.0 $\pm$ 0.12 & 10.67 $\pm$ 0.10 & 4.16 & 2 & 1.14\%\\
        NGC~2442 & 3.7 $\pm$ 0.6 & SAB(s)bc & 07h36m23.84s & -69d31m51.0s & 50.3 & 12.3 & 21.2 $\pm$ 2.0 & 10.56 $\pm$ 0.12$\mathrm{^a}$ & 2.75 & 2 & 0.81\%\\
        NGC~2835 & 5.0 $\pm$ 0.4 & SB(rs)c & 09h17m52.91s & -22d21m16.8s & 56.2 & 1.3 & 10.1 $\pm$ 0.12 & 9.67 $\pm$ 0.10 & 3.30 & 3 & 0.05\%\\
        NGC~2997 & 5.1 $\pm$ 0.5 & SAB(rs)c & 09h45m38.79s & -31d11m27.9s & 53.7 & 98.9 & 11.3 $\pm$ 0.12 & 10.46 $\pm$ 0.10 & 4.46 & 3 & 0.19\%\\
        NGC~4536 & 4.3 $\pm$ 0.7 & SAB(rs)bc & 12h34m27.050s & +02d11m17.29s & 73.1 & 120.7 & 15.2 $\pm$ 0.06 & 10.19 $\pm$ 0.10 & 3.80 & 3 & 0.60\%\\
        NGC~5236 & 5.0 $\pm$ 0.3 & SAB(s)c & 13h37m00.950s & -29d51m55.50s & 15.3 & 44.9 & 4.9 $\pm$ 0.04 & 10.41 $\pm$ 0.10 & 6.44 & 2 & 0.18\%\\
        NGC~5643 & 5.0 $\pm$ 0.3 & SAB(rs)c & 14h32m40.743s & -44d10m27.86s & 29.6 & 98.1 & 11.8 $\pm$ 0.12 & 10.06 $\pm$ 0.10 & 2.29 & 2 & 3.58\%\\
        NGC~6744 & 4.0 $\pm$ 0.2 & SAB(r)bc & 19h09m46.10s & -63d51m27.1s & 53.5 & 13.7 & 11.6 $\pm$ 0.12 & 10.87 $\pm$ 0.10 & 9.98 & 3 & 0.66\%\\
        \hline
        \hline
    \end{tabular}
    \caption{Physical parameters of the spiral galaxies in this study. 
     {\bf{Column 1:}} Galaxy name. 
     {\bf{Column 2:}} RC3 morphological T-types from Hyperleda (\href{http://atlas.obs-hp.fr/hyperleda/}{http://atlas.obs-hp.fr/hyperleda/}).
     {\bf{Column 3 $-$ 5:}} Morphology and J2000 Coordinates from NASA extragalactic database (NED).
     {\bf{Column 6:}} Inclination between the line of sight and polar axis from Hyperleda.
     {\bf{Column 7:}} Position angle of the major axis in the B-band, northeastward $\mathrm{^b}$.
     {\bf{Column 8 \& 9:}} Distance and stellar mass from \citet{Leroy_2019}.
     {\bf{Column 10:}} $R_{25}$, defined as the 25 mag arcsec$^2$ $B$-band isophote from NED.
     {\bf{Column 11:}} Number of spiral arms in each galaxy (Sec~\ref{sec:arm_define}).
     {\bf{Column 12:}} Fraction of spaxels excluded from BPT constraints. (Sec~\ref{sec:spaxel_select}).
     \\
     \textbf{Note.}\\
     $\mathrm{^a}$ The stellar mass and the uncertainty of NGC~2442 are from \citet{Pan_2020}.\\
     $\mathrm{^b}$ The position angle of NGC~1365 comes from the 2MASS survey \citep{Jarrett_2003}, for consistency with \citet{Ho_2017}.}
    \label{tab:info}
\end{table*}

\section{Data analysis}\label{sec:analysis}
In this section, we introduce our analyses on the TYPHOON data, starting with the emission line fitting process using {\sc{lzifu}} (Sec~\ref{sec:lzifu}). 
We select reliable spaxels with signal-to-noise ratio (SNR) limits and subsequently identify star-forming spaxels with the BPT diagram, described in Sec~\ref{sec:spaxel_select}. 
Sec~\ref{sec:arm_define} introduces the methodology to define the ridge lines of the spiral arms. 
Sec~\ref{sec:ISM} outlines how we measure the ISM properties (\SigSFR\ and gas-phase metallicity). 

\subsection{Emission line fitting}\label{sec:lzifu}
We measure the 2D emission line maps using the tool {\sc{lzifu}} \citep{Ho_2016, Ho+2016}.
The reduced emission line maps of all galaxies in the TYPHOON survey are described in Battisti et al. (in prep.). We briefly introduce the emission line fitting process of {\sc lzifu} below that is relevant to this work.

Firstly, {\sc{lzifu}} uses {\sc ppxf} \citep{Cappellari_Emsellem_2004, Cappelari_2017} to fit and subtract the continuum on a spaxel-to-spaxel basis. 
This continuum modelling is based on the {\sc miuscat} simple stellar population models \citep{Vazdekis_2012}.
The principle goal of {\sc{lzifu}} is to derive emission lines from continuum-free spectra, rather than constraining stellar parameters from continuum measurements \citep[Sec~2.2 in][]{Ho_2016} \footnote{A detailed analysis of stellar populations based on the TYPHOON data is published in \citet{Sextl_2024}. In this work, continuum fits are conducted at the spaxel level. H$\beta$ absorption line is the only stellar absorption feature that can impact the emission flux. However, H$\beta$ will not be used in our measurements of SFR or 12 + log(O/H).}.
For this work, we only adopt the emission line fits from {\sc{lzifu}}.
Secondly, {\sc lzifu} fits each emission line as a single Gaussian component using the Levenberg-Marquardt least-square method. 
In this study, we fit the following emission lines simultaneously: H$\beta$ (4861\AA), [O~\textsc{iii}]$\lambda$5007, [N~\textsc{ii}]$\lambda$6583, H$\alpha$ (6563\AA), and [S~\textsc{ii}]$\lambda\lambda$6716,31.
We tie together the velocity and velocity dispersion of all the lines.
The flux ratios of [O~\textsc{iii}]$\lambda$5007/[O~\textsc{iii}]$\lambda$4959 and [N~\textsc{ii}]$\lambda$6583/[N~\textsc{ii}]$\lambda$6548 are constrained to their theoretical values predicted by quantum mechanics \citep[3.1;][]{Gurzadyan_1997}.
Finally, {\sc lzifu} returns the outputs of emission line fluxes with corresponding error maps.

\subsection{Spaxel selection criteria}\label{sec:spaxel_select}
We apply an SNR limit of 3 to \ha\ and H$\beta$ to obtain a reliable analysis.
The spaxels with SNR (\ha\ \& H$\beta$) below 3 are excluded from all the following analyses. 
If an SNR less than 3 is detected in a spaxel for any of the following three doublet lines: [O~\textsc{iii}]$\lambda$5007, [N~\textsc{ii}]$\lambda\lambda$6548,83 and [S~\textsc{ii}]$\lambda\lambda$6716,31, we adopt a limiting value of 3$\sigma$ for the corresponding emission line(s), where $\sigma$ is the uncertainty in the line measurement \citep[see Sec~2.6 of][]{Rosario_2016}.

We correct for dust extinction using the Milky Way extinction curve from \cite{Fitzpatrick_2019} as:
\begin{equation}
	    E(B-V) = 2.5 \times \left( \frac{\log_{10}\frac{(H\alpha/H\beta)_{\mathrm{obs}}}{(H\alpha/H\beta)_{\mathrm{int}}}} {k_{H\beta}-k_{H\alpha}}\right),
\end{equation}
where $E(B-V)$ is the colour excess, $(H\alpha/H\beta)_{\mathrm{obs}}$ is the observed flux ratio measured from the data.
We adopt the intrinsic flux ratio $(H\alpha/H\beta)_{\mathrm{int}}$ of 2.86 by assuming case~B recombination at the electron temperature of 10,000~K and electron density of 100 cm$^{-3}$ \citep{Osterbrock_1989}.
We use $R_{\nu}$ = 3.1 to determine the $k$ value at each wavelength. 
For spaxels with negative $E(B-V)$ values, we assign a value of 0 to colour excess.
The $E(B-V)$ values are then used to calculate the intrinsic emission line fluxes $F_{\mathrm{int}}$ from the observed fluxes $F_{\mathrm{obs}}$, 
following \citet{Calzetti_2001}:
\begin{equation}
   F_{\mathrm{int}} = F_{\mathrm{obs}} \times 10^{0.4k_{\lambda} E(B-V)}.
\end{equation}
The extinction-corrected fluxes are used in all the following analyses to measure the \SigSFR\ and gas-phase metallicities (Sec~\ref{sec:ISM}).

We further exclude the spaxels ionised by hard components and limit our study to the photoionised spaxels by star formation by using the BPT diagram. Based on $\mathrm{ [N~\textsc{ii}]\lambda6584/{H\alpha} }$ versus $\mathrm{ [O~\textsc{iii}]\lambda5007/H\beta }$, the spaxels dominantly ionised by star formation (\ion{H}{ii} regions) are distinguished from the low-ionisation nuclear emission-line regions (LINERs) and AGN.

There are two common demarcation lines to separate AGN/LINERs and \ion{H}{ii} regions based on optical emission lines. 
\citet{Kauffmann_2003b} presents an empirical demarcation line based on the properties of $\sim$ 120,000 nearby galaxies with the Sloan Digital Sky Survey. 
\citet{Kewley_2001} models the starburst galaxies with {\sc {pegase v2.0}} and {\sc starburst99} to derive the theoretical classification scheme for AGN and \ion{H}{ii} regions. 
All spaxels over the \citet{Kewley_2001} demarcation line are excluded in the following analysis in this work. The fraction of excluded spaxels due to BPT constraints is below 4\% in all cases, as shown in the last column in Table~\ref{tab:info}.

\subsection{Spiral arm definition and \texorpdfstring{\dphi}{Delta phi}\ definition}\label{sec:arm_define}
\begin{figure*}
    \centering
    \includegraphics[width=0.9\textwidth]{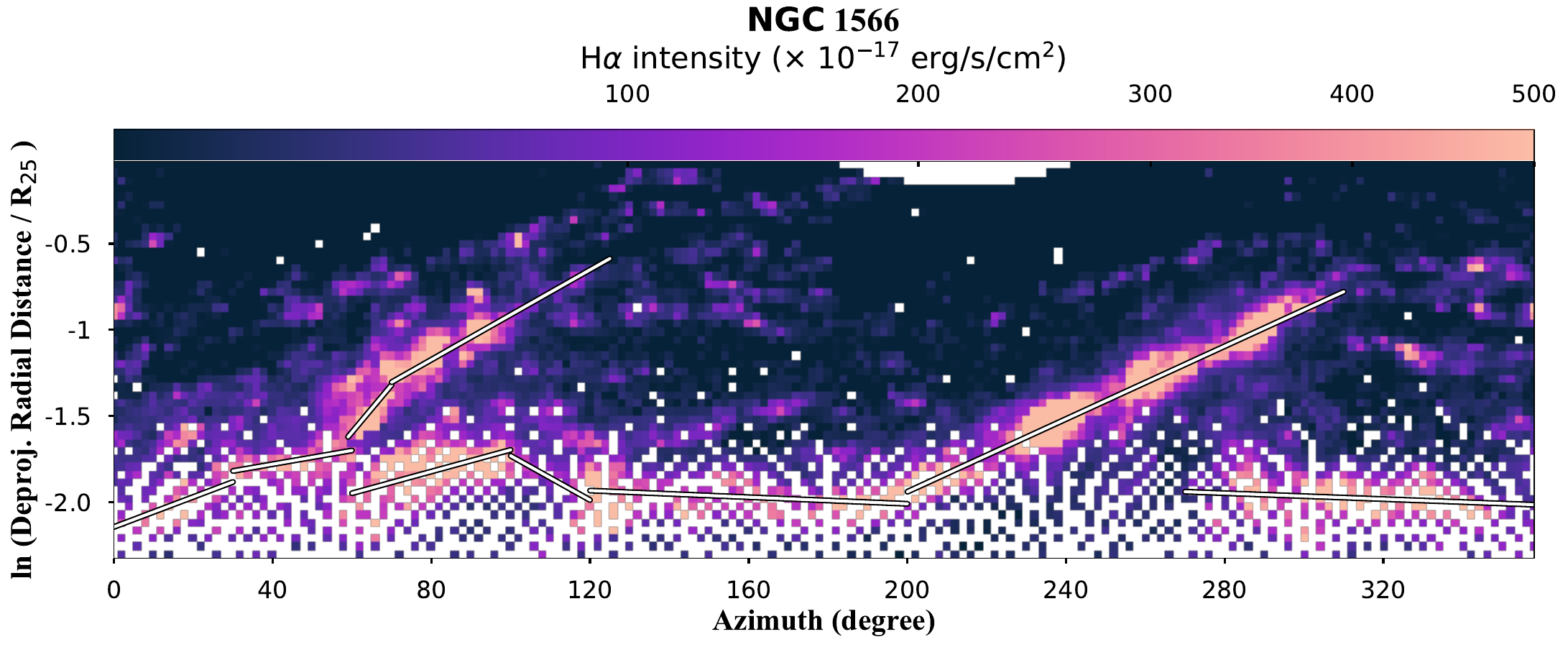}
    \caption{Deprojected radial distance in a logarithm scale versus the azimuth, colour-coded by the \ha\ intensity of NGC~1566. The best fits of the \ha-bright (intensity \textgreater 100 $\times 10^{-17}$ erg/s/cm$^{2}$) regions are shown as the black solid lines.
    The pixelation at the central region results from the reduced area as the radial distance decreases. }
    \label{fig:r_phi}
\end{figure*}

We follow the method of \citet{Ho_2018} to identify our spiral arm ridge lines:
\begin{equation}
    r(\phi) = r_0 e^{\mathrm{tan\theta_p}(\phi-\phi_0)} \label{eq:arm1}
\end{equation}
where $\mathrm{\theta_p}$ is the pitch angle, $r_0$ is the initial radius and $\phi_0$ is the initial azimuth of the spiral arm.
The spiral arms are recovered as straight lines in a plot of logarithm-scale deprojected radial distance versus azimuth (Fig~\ref{fig:r_phi}).
Eq~\ref{eq:arm1} is transformed into a linear function:
\begin{equation}
    \mathrm{ln} r(\phi) = \mathrm{ln} r_0 + \mathrm{tan\theta_p}(\phi-\phi_0) \label{eq:arm}
\end{equation}
We fit the spiral arms with Eq~\ref{eq:arm} using a non-linear least squares method\footnote{We use the curve\_fit module in python to carry out the fitting process.} using the \ha-bright regions as the input of the fitting process.  
The best-fit line is the defined spiral arm ridge line, represented as the black solid line in Fig~\ref{fig:r_phi}. 
We present the phase diagram of only NGC~1566 in this paper for demonstration.

\begin{figure}
    \centering
    \includegraphics[width=0.4\textwidth]{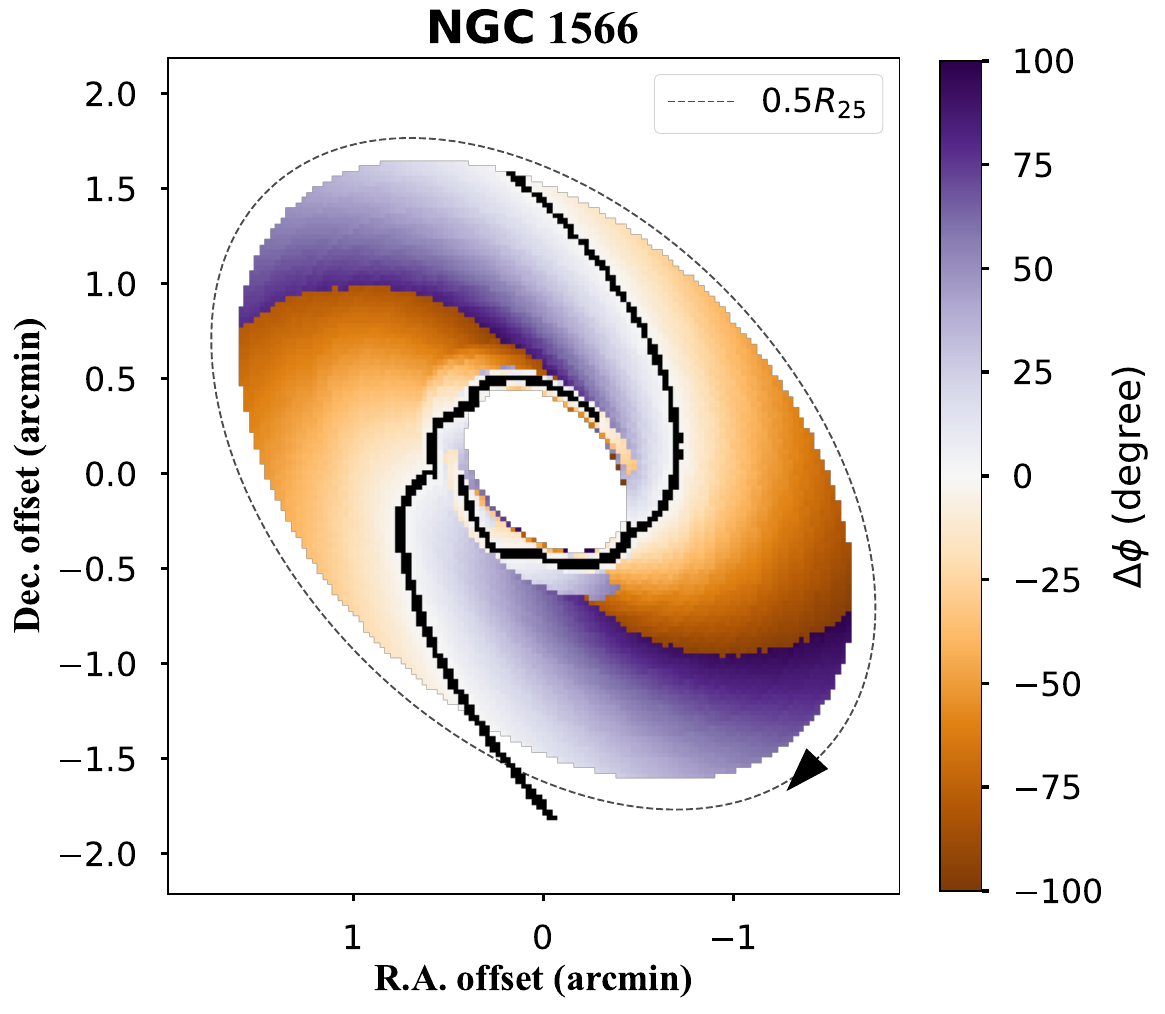}
    \caption{A measure of the angular azimuthal distance to the nearest spiral arm at a constant galactocentric distance (\dphi), taking NGC~1566 as an example, over-plotted with the spiral arm ridge lines. 
    The solid ellipse shows the location of 0.5~R$\mathrm{_{25}}$, with the circular flow indicated by the arrow.
    The orange spaxels on the leading edge are assigned negative values while the spaxels on the trailing edge are given positive values, color-coded as purple.
    The deeper colour the spaxel has, the further its azimuth is from the spiral arm.}
    \label{fig:dphi}
\end{figure}

We aim to quantitatively trace the behaviour of star formation activities as stars and gas move in and out of the spiral arms.
We adopt the parameter \dphi, which quantifies the azimuthal distance to the nearest spiral arm at a constant galactocentric distance \citep{Chen_2024}.
Given a pixel $T$ as the targeted pixel and pixel $arm1$, $arm2$, ..., $armN$ as pixels within the spiral regions at the same galactocentric distance, the \dphi\ is defined as, 
\begin{equation}
   \Delta \phi_T = -\operatorname{min}(|\phi_{arm1} - \phi_T|, |\phi_{arm2} - \phi_T|, ..., |\phi_{armN} - \phi_T|)
\end{equation}
when pixel $T$ is on the leading edge of the nearest spiral arm while 
\begin{equation}
   \Delta \phi_T = \operatorname{min}(|\phi_{arm1} - \phi_T|, |\phi_{arm2} - \phi_T|, ..., |\phi_{armN} - \phi_T|)
\end{equation}
when pixel $T$ is on the trailing edge of the nearest spiral arm. 
The \dphi\ map of NGC~1566 is presented in Fig~\ref{fig:dphi}, overplotted with the spiral arm ridge lines.

We calculate the angular distance $\theta$ that stars will travel in 10~Myr, a typical life span of O-type stars, as:
\begin{equation}\label{eq:v}
    \begin{split}
    t &= \frac{2 \pi R}{V_{\mathrm{circ}}}, \\
    \theta &= \frac{10~\mathrm{Myr}}{t} \times 360^{\circ}.
    \end{split}
\end{equation}\label{eq:V_ang}
Here, $V_{\mathrm{circ}}$ is the circular velocity and $R$ is the radial distance to the galaxy centre.
Using a typical rotational velocity of 200 km s$^{-1}$ \citep[8~kpc to the centre of Milky Way;][]{Honma_2015}, we get $\theta=$ 17$^\circ$.
For a conservative analysis, we classify spaxels with |\dphi| $< \ 20^\circ$ as spiral arm regions, which will be applied to our statistic test (Sec~\ref{sec:results_z} and Fig~\ref{fig:cdf}).

\subsection{Mapping the ISM: SFR and gas-phase metallicity}\label{sec:ISM}
\subsubsection{Mapping \texorpdfstring{\SigSFR}{Sigma SFR}\ and \texorpdfstring{\dSFR}{Delta SFR}}
The spatial distribution of star formation and gas-phase metallicities are critical components for understanding the physical evolution of spiral galaxies, including their ongoing star formation, star formation history and mixing processes in the ISM \citep{Maiolino_2019, Li_2021, Sharda_2023, Garcia_2023}.
In this section, we introduce our method to map the SFR and the gas-phase metallicity, used to investigate the effects of the spiral arms on the SFR and metallicity distribution.

\begin{figure*}
    \centering
    \includegraphics[width=0.4\textwidth]{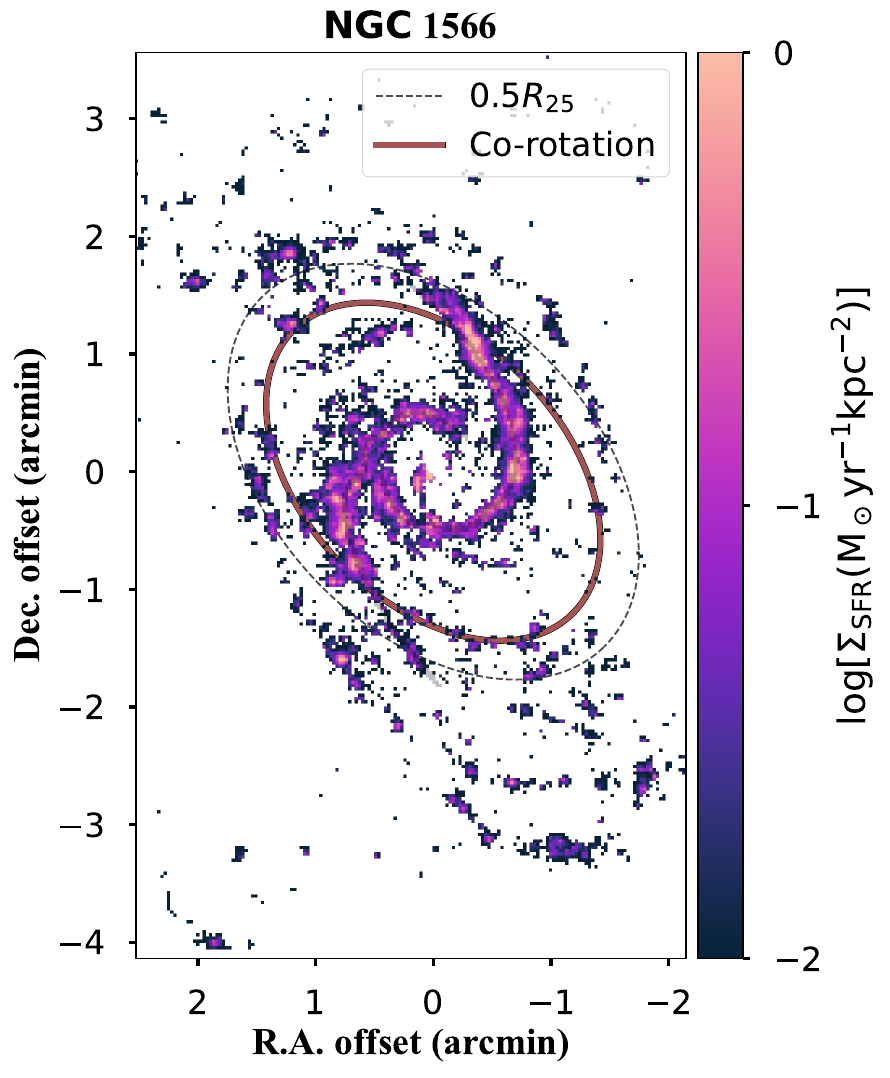}
    \includegraphics[width=0.4\textwidth]{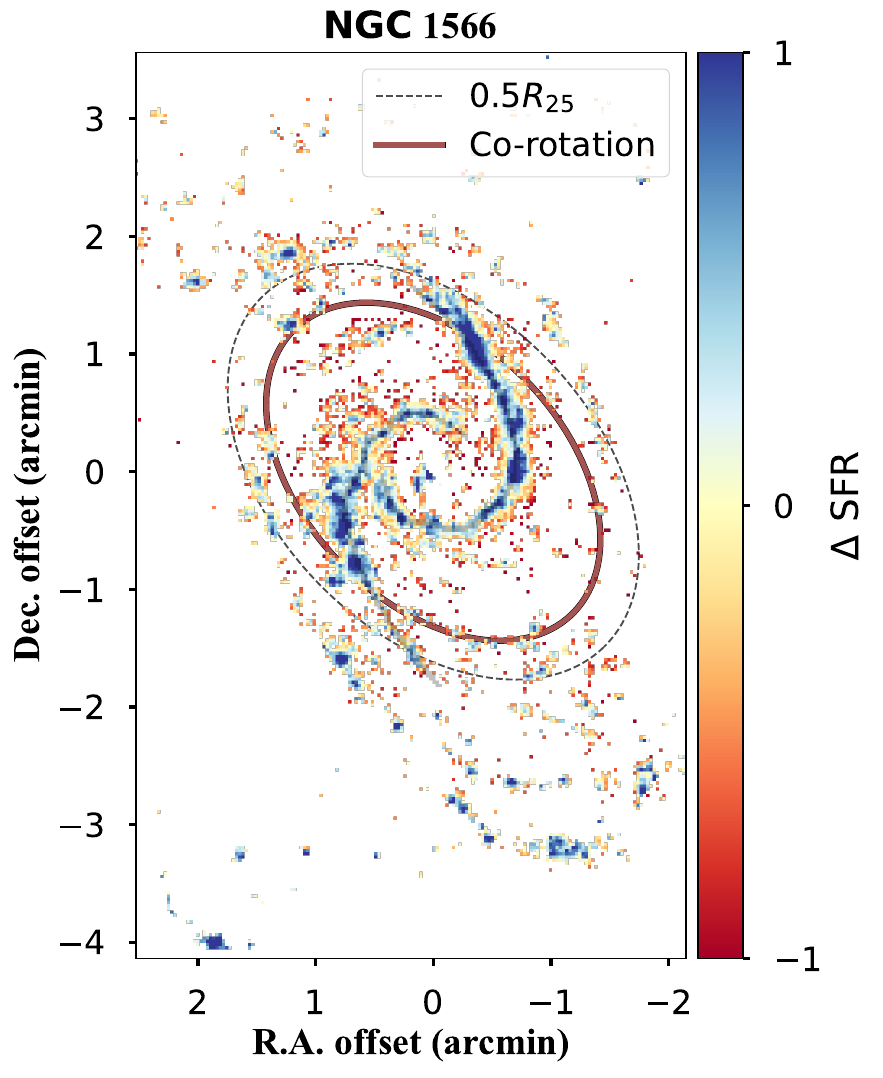}
    \caption{2D maps of derived \SigSFR\ and \dSFR\ of NGC~1566 after SNR limit and spaxel selection based on BPT diagram, overplotted with the defined spiral arms from Sec~\ref{sec:arm_define}. 
    The dashed ellipse marks the location of half $R_{25}$ and the red ellipse denotes the co-rotation radius reported in previous publications (see Sec~\ref{sec:CR})}.
    Fig~\ref{appendix_sfr} shows the SFR maps for the remaining galaxies in this work.
    \label{fig:sfr_map}
\end{figure*}

We measure the SFR by converting the extinction-corrected (i.e., intrinsic) \ha\ intensity to a SFR indicator.
Stars with masses exceeding $\sim$ 10 $\mathrm{M_\odot}$ produce a detectable flux of ionizing photons and have a short lifespan of $\lesssim$ 30 million years \citep{Calzetti_2013}. 
The \ha\ nebular emission line is thus a direct tracer of the ionizing photons powered by young, massive stars. 
Following the SFR prescription in \citet{Kennicutt_1998}, we measure the SFR as:
\begin{equation}
    \mathrm{ SFR (M_\odot yr^{-1}) = 7.9 \times 10^{-42} \times\ 4\pi} D_\mathrm{L}^2 F_{\mathrm H\alpha} \mathrm{(erg\ s^{-1}\ cm^{-2})}, \label{eq:sfr}
\end{equation}
where $F_{\mathrm{H\alpha}}$ is the flux of \ha\ per spaxel and $D_\mathrm{L}$ is the luminosity distance in Table~\ref{tab:info}.
We calculate the \SigSFR\ as follows:
\begin{equation}
    \mathrm{\Sigma_{SFR} (M_\odot yr^{-1} kpc^{-2})} = \frac{\mathrm{SFR}}{[D_\mathrm{A} (\mathrm{kpc}) \times 1.65''/180^{\circ} \times \pi] ^2},
\end{equation}
where 1.65\arcsec\ is the spaxel size of the TYPHOON data and $D_\mathrm{A}$ is the angular distance of the observed galaxy.
The measured \SigSFR\ map of NGC~1566 is shown in the left column of Fig~\ref{fig:sfr_map}, with the remaining galaxies shown in Fig~\ref{appendix_sfr}. 

We find higher \SigSFR\ (bluer spaxels) predominantly concentrated in the central regions in most of our galaxies. 
To remove the radial dependence on the \SigSFR, we subtract the radial gradient, represented by a piecewise linear function, from \SigSFR\ to get the offset \dSFR\ value, which indicates the azimuthal variation.
The residual \dSFR\ map of NGC~1365 is shown in the right column of Fig~\ref{fig:sfr_map}, with the rest of the sample presented in Fig~\ref{appendix_sfr}.

We observe positive \dSFR\ (i.e., higher \SigSFR) along the spiral arm ridge lines of NGC~1566 which is expected from our definition of spiral arms (Sec~\ref{sec:arm_define}). 
This scenario is observed in all our galaxies (Fig~\ref{appendix_sfr}).
With \dSFR\, we can better compare both sides of the spiral arms, with generally negative \dSFR\ in the leading edge (orange; \dphi\ < 0) and positive \dSFR\ in the trailing edge.
The \dSFR\ maps of other spiral galaxies can be found in Appendix~\ref{appendix_sfr} and we will further discuss the azimuthal variation in Section~\ref{sec:results_sfr}.

\subsubsection{Mapping 12 + log(O/H) and \texorpdfstring{\dOH}{Delta O/H}}\label{sec:ana_z}
Stars produce heavy metals during their lifetimes, which are subsequently released into the interstellar medium upon their death, enhancing the metal content of the ISM for subsequent generations of stars.
Consequently, gas-phase metallicity serves as a marker for preceding stellar generations, influenced by gas inflows, outflows, and depletion mechanisms. 
The spatial distribution of the gas-phase metallicity represents a snapshot in time of the production history and mixing processes. 
Deviations in the azimuthal direction of the metal distribution offer insights into the mixing process of metals with the surrounding ISM as both gas and stars orbit within the galactic potential. 
As oxygen is the most abundant metal in the gas-phase ISM, we measure the oxygen abundance through collisionally excited lines in the optical spectrum as an indicator of the gas-phase metallicity. 

We adopt the N2S2-N2\ha\ diagnostic from \citet[][hereafter D16]{Dopita_2016} to measure the gas-phase metallicity. 
The D16 diagnostic uses the \ha, [NII]$\lambda$6484 and [SII]$\lambda\lambda$6717,31 emission lines.
All four emission lines above are well-detected by the TYPHOON survey.
With the inclusion of [SII]$\lambda\lambda$6717,31 doublet lines, the D16 diagnostic is subject to contamination from diffuse ionised gas \citep[DIG;][]{ZhangK_2017,Shapley_2019, Kumari_2019}.
\citet{Poetrodjojo_2019} find that DIG has less impact on the calibrated metallicity at the resolution of the TYPHOON data, compared to MaNGA and SAMI.
Appendix~\ref{appendix_scal} further discusses the impacts of DIG with metallicity variation calculated by S-calibration \citep{Pilyugin_2016} diagnostic.
Similarly, we find higher \dOH\ in the trailing edge of NGC~1566, although the magnitude of the azimuthal variation (0.017~dex) is smaller than the one found in N2S2-N2\ha\ (0.037~dex).

\begin{figure*}
    \centering
    \includegraphics[width=0.4\textwidth, height=0.46\textwidth]{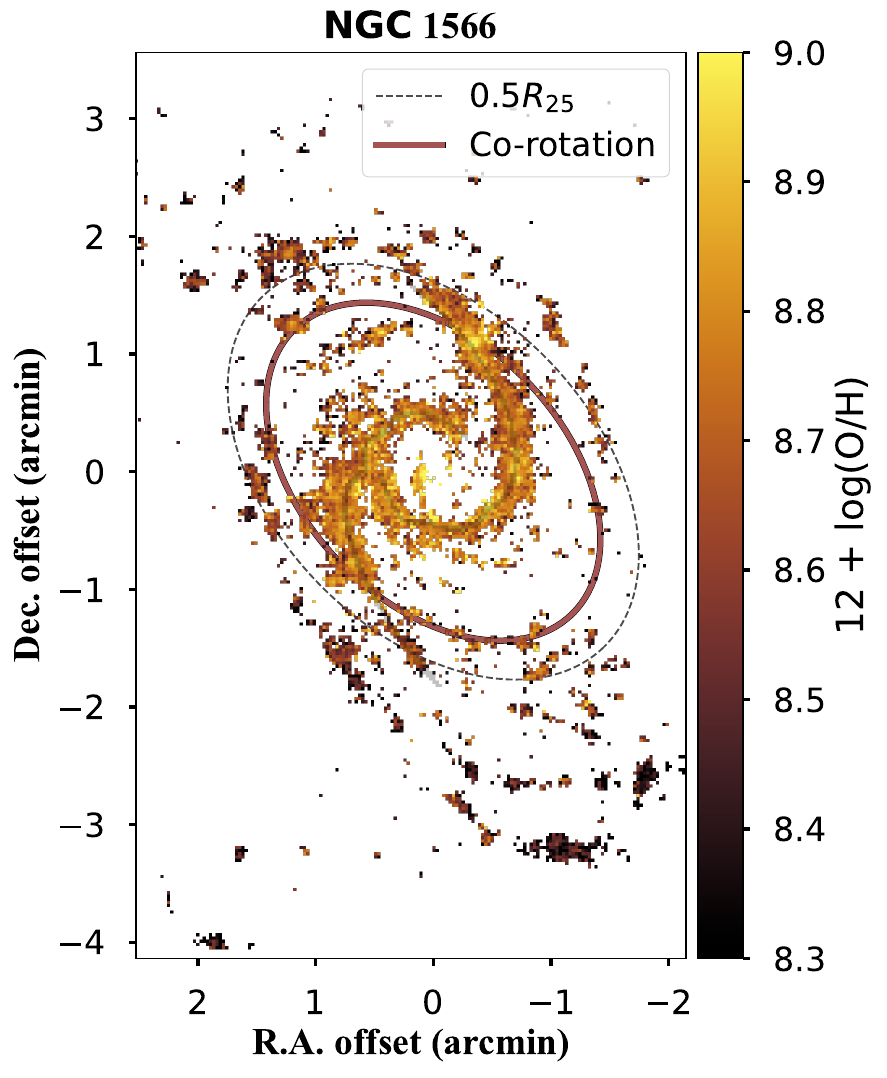}
    \includegraphics[width=0.4\textwidth, height=0.46\textwidth]{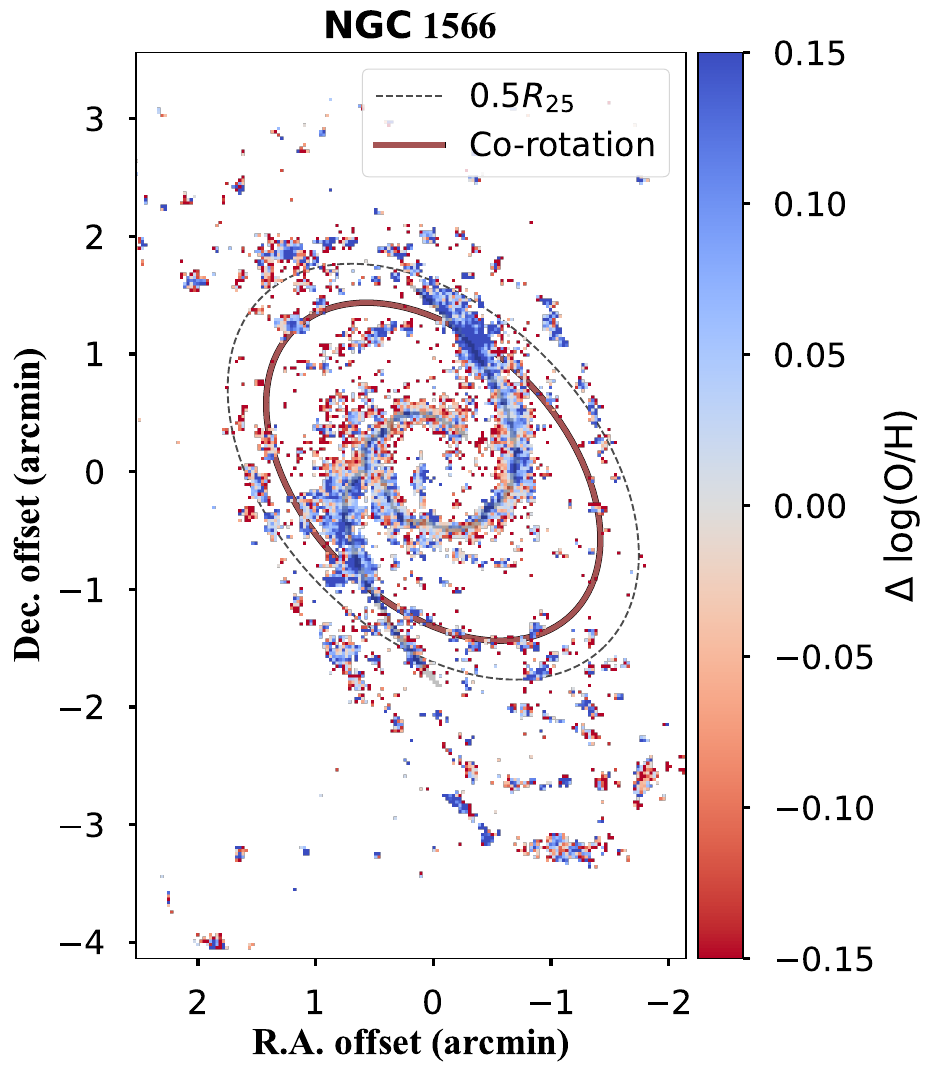}
    \caption{2D maps of derived 12 + log(O/H) and \dOH\ of NGC~1566, overplotted with the defined spiral arms from Sec~\ref{sec:arm_define}. 
    The half $R_{25}$ radius is shown as a dashed ellipse while the CR is shown as a solid ellipse.
    The 12 + log(O/H) and \dOH\ maps for the remaining sample are presented in Fig~\ref{appendix_Z}.}
    \label{fig:z_map}
\end{figure*}

We present the metallicity maps in the left panel of Fig~\ref{fig:z_map} (NGC~1566) and Fig~\ref{fig:z_map_other} for the remaining sample, overplotted with the defined spiral arms (Sec~\ref{sec:arm_define}). 
We find higher 12 + log(O/H) values in the central region (bluer spaxels) and lower 12 + log(O/H) measurements in the outskirts (redder spaxels) in all galaxies. 
The negative radial metallicity gradient is indicative of inside-out galaxy formation \citep{Tinsley_1978, Prantzos_2000}.

Similar to the methodology applied to \SigSFR, we derive the \dOH\ value for each spaxel by subtracting the radial gradient, measured through the best fit of a piecewise linear function.
The residual \dOH\ maps are shown in the right panel of Fig~\ref{fig:z_map} and the second and fourth column of Fig~\ref{appendix_Z}.
We observe generally positive \dOH\ (blue pixels) along the spiral arms in all galaxies.
To quantitatively compare the metallicity in the downstream versus upstream, we investigate the correlation between \dphi\ and \dOH\ in the spiral galaxies in Sec~\ref{sec:results_z}.

\section{Results}\label{sec:results}

\subsection{SFR}\label{sec:results_sfr}
\begin{landscape}

\begin{figure}
    \centering
    \includegraphics[width=0.42\textwidth]{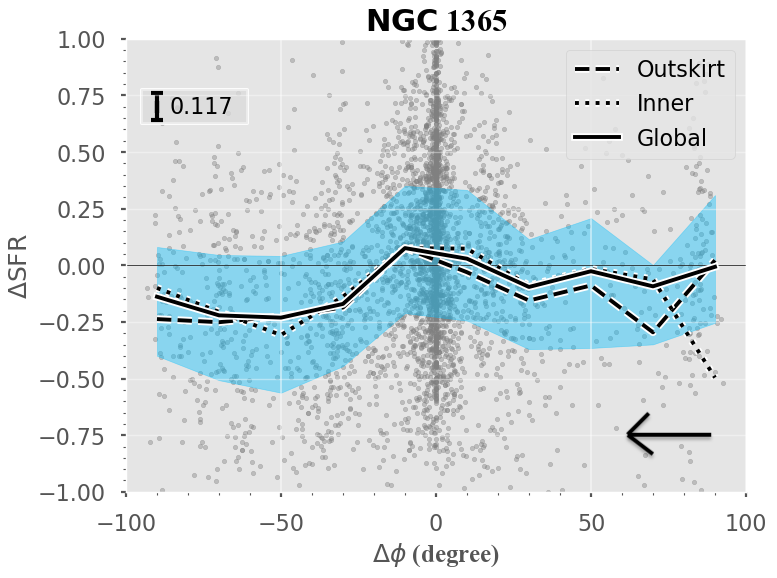}
    \includegraphics[width=0.42\textwidth]{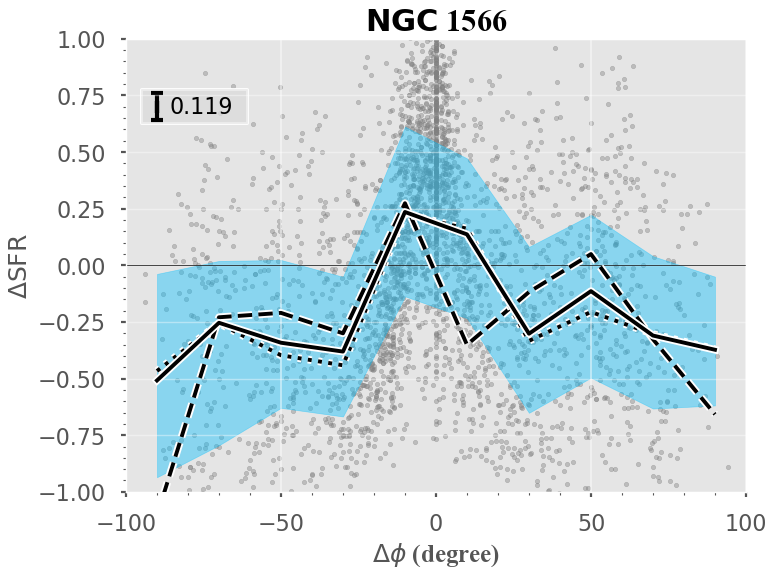}
    \includegraphics[width=0.42\textwidth]{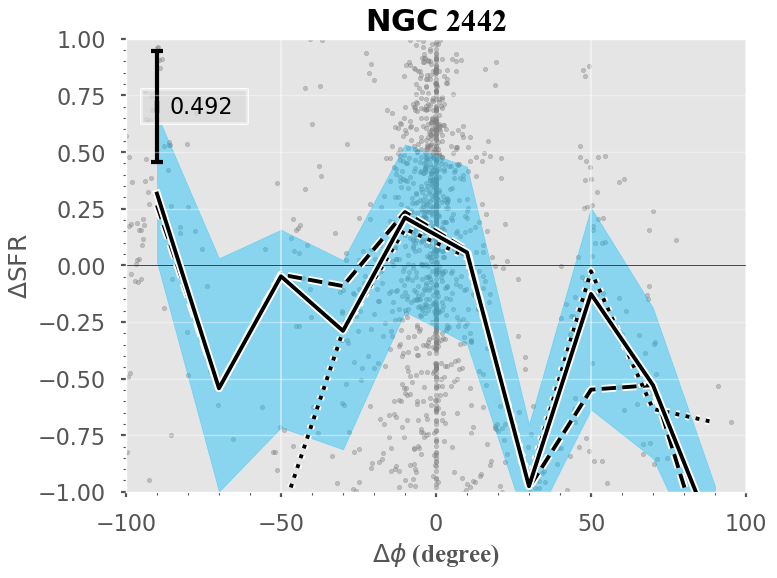}
    
    \includegraphics[width=0.42\textwidth]{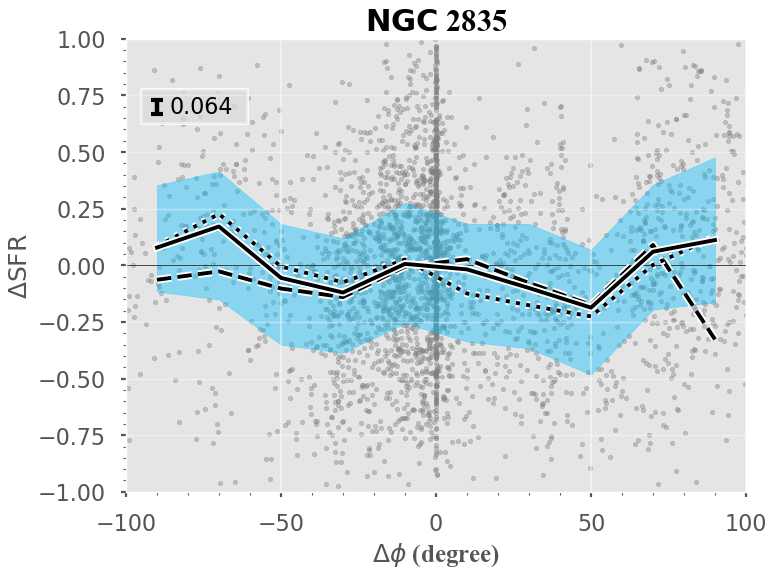}
    \includegraphics[width=0.42\textwidth]{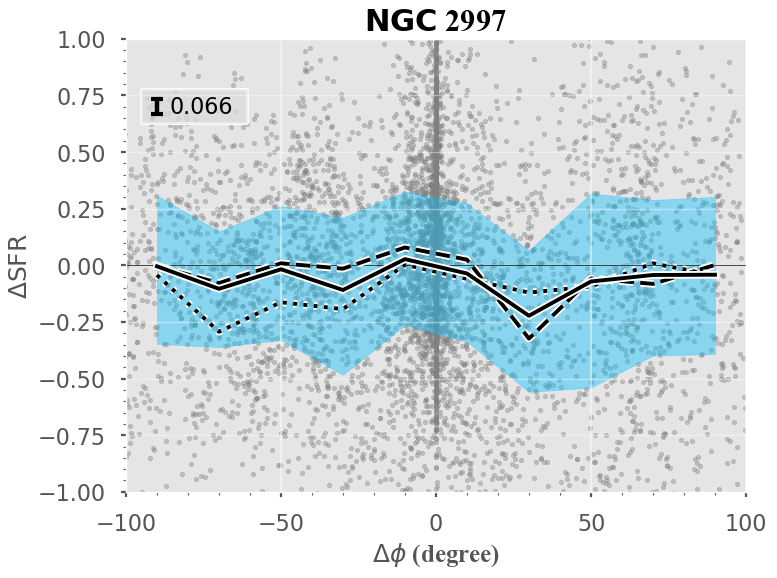}
    \includegraphics[width=0.42\textwidth]{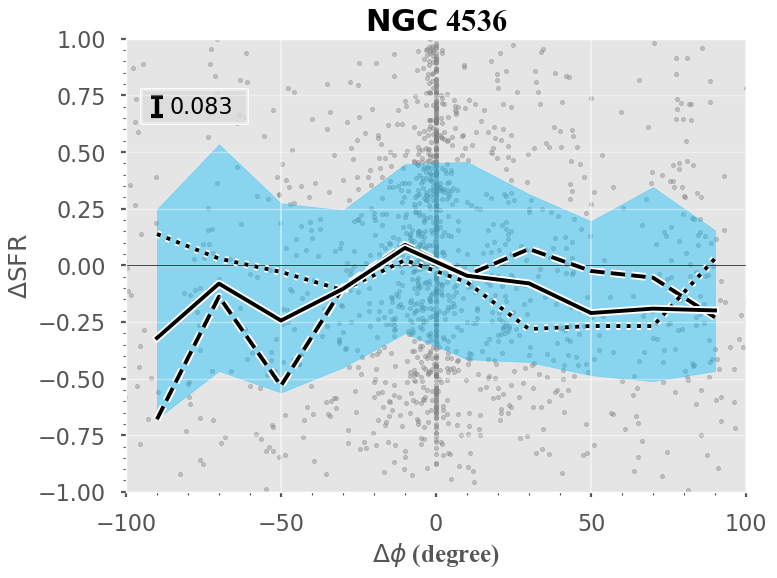}
    
    \includegraphics[width=0.42\textwidth]{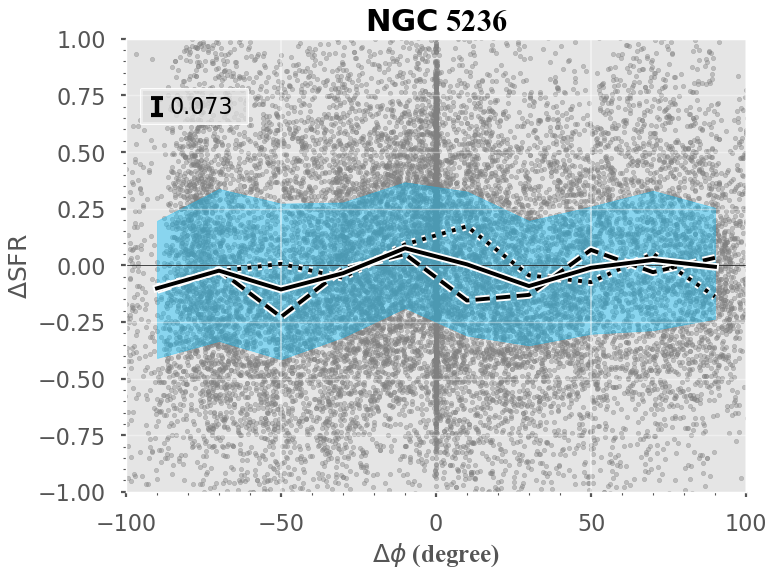}
    \includegraphics[width=0.42\textwidth]{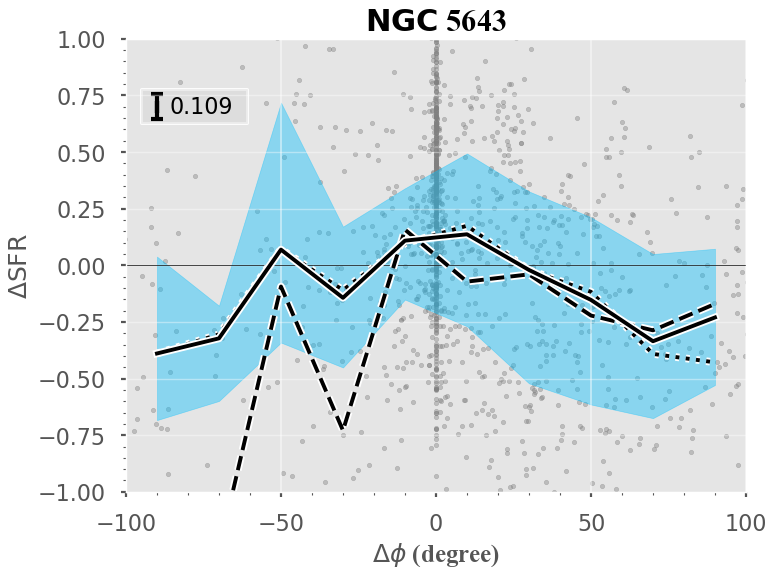}
    \includegraphics[width=0.42\textwidth]{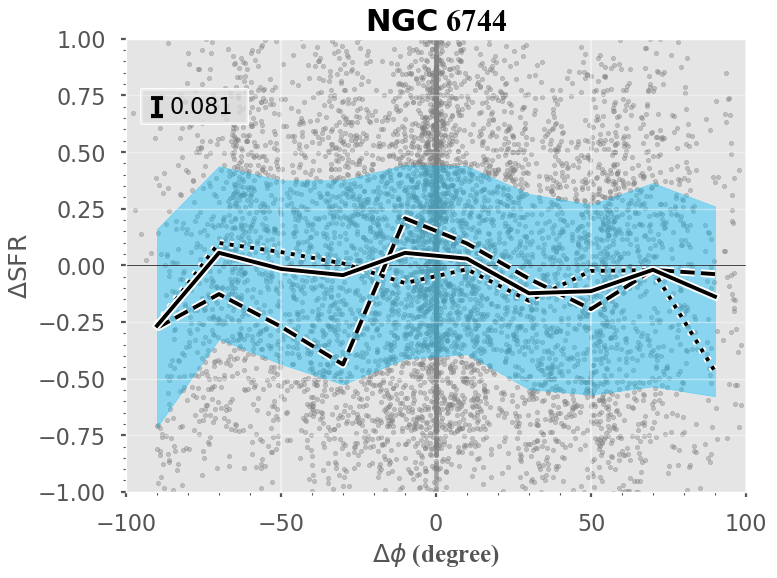}

    \caption{The fluctuation of \dSFR\ along \dphi. 
    The gas flows (shown as an arrow at the bottom right) from \dphi\ \textgreater 0 (trailing) to \dphi\ \textless 0 (leading).
    The solid black line marks the moving medians of each 20$^\circ$ \dphi\ bin, with 25\% and 75\% quartiles represented as the blue shadows.
    The average offset between \dSFR\ in the trailing and the leading edge is shown as a scale bar.
    We find subtly higher \dSFR\ in the trailing edge (\dphi $> 0$) in NGC~1365 and NGC~1566. 
    NGC~2442 shows higher \dSFR\ from -20$^{\circ}$ to -50$^{\circ}$, compared to the trailing edge.
    We do not find a significant global azimuthal offset in \dSFR\ in the other six galaxy samples.
    We find an opposite trend of \dSFR-\dphi\ in the inner region versus the outskirts only in NGC~4536, with large uncertainty and limited spaxels.
    We divide the galaxy disc into two regions: the inner region and the outskirts, with each region including half of the spaxels. 
    We trace the fluctuation of \dSFR\ in the inner region using dotted lines and in the outskirts using dashed lines and eight out of nine galaxies (except for NGC~4536) show the same \dSFR\ trend in both inner and outer regions.
    }
    \label{fig:dphi_dsfr}
\end{figure}
\end{landscape}

We examine the behaviour in SFR when moving into or out of the spiral arms by exploring the relation between \dSFR\ and \dphi\ (Fig~\ref{fig:dphi_dsfr}).
\dSFR\ denotes a higher/lower SFR within the measured spaxel compared to those at the same galactocentric distance (Sec~\ref{sec:ISM}) and \dphi\ quantifies the angular distance to the nearest spiral arm (Sec~\ref{sec:arm_define}).
Gas flows from the trailing edge (\dphi\ \textgreater 0) into the spiral arms, and then passes to the leading edge (\dphi\ \textless 0).
We measure the moving medians of each 20$^\circ$ (solid lines in Fig~\ref{fig:dphi_dsfr}), with blue shadow representing 25\% and 75\% quantiles.
The fluctuation of \dSFR\ across different \dphi\ ranges from $-$1.5~dex to 1.5~dex.

In density wave theory, star formation may occur either i) after gas clouds pass through density waves or ii) as they approach density waves, resulting in various distributions of old stars, young stars, and gas \citep{Pour-Imani_2016}. 
In scenario i/ii, it is expected to observe a gaseous blue arm with young stars on the leading/trailing edge of a stellar arm inside the CR.
Our work finds generally higher \dSFR\ on the trailing side (\dphi\ > 0), with the highest \dSFR\ near the spiral arms and lower \dSFR\ on the leading edge (\dphi\ < 0) of NGC 1365 and NGC 1566. 
This observation is consistent with density wave theory in the latter scenario, when star formation occurs as gas clouds approach the potential minimum \citep[right spiral arms in Fig~1 of][]{Pour-Imani_2016}.
In NGC~2442, we find decreasing \dSFR\ in the trailing edge and increasing \dSFR\ in the leading edge (\dphi $< 0$).
Between \dphi\ of -20$^{\circ}$ and -50$^{\circ}$, the trailing edge shows higher \dSFR\ than the corresponding leading edge, which is likely contributed by the interaction affecting this galaxy (further discussed in Sec~\ref{sec:dominant}).
In NGC~2835, NGC~2997, NGC~4536, NGC~5236, NGC~5643 and NGC~6744, we do not observe significant differences on either side of the spiral arms.
This indicates that the \SigSFR\ is comparable in these six galaxies and this finding aligns with the predicted symmetrical \SigSFR\ from dynamic spiral theory.
We do note that there is a large scatter in the \dSFR-\dphi\ trend and it is possible that the non-detected offset in \SigSFR\ could be due to noise obscuring the trend.

The varying motions of material inside and outside the CR might obscure and even eliminate the azimuthal offset.
To address the potential obscuration of detecting azimuthal offset, separate analyses of \dphi-\dSFR\ (Fig~\ref{fig:dphi_dsfr}) inside versus outside the CR or at various radial regions (if CR is unknown) are necessary.
Among our sample, only three spiral galaxies have CR reported from the literature (see Sec~\ref{sec:CR}).
To ensure consistency in our analyses, we examine the \dphi-\dSFR\ trend at various radii for all galaxies, without using the reported CR for three of our samples.
We divide each galaxy disc into two radial ranges, with an equal number of spaxels in each section\footnote{The radial cut for each galaxy is: NGC~1365$-$10.98~kpc (0.37$R_{25}$), NGC~1566$-$8.80~kpc (0.41$R_{25}$), NGC~2442$-$10.35~kpc (0.61$R_{25}$), NGC~2835$-$5.81~kpc (0.58$R_{25}$), NGC~2997$-$7.95~kpc (0.54$R_{25}$), NGC~4536$-$7.43~kpc (0.50$R_{25}$), NGC~5236$-$4.66~kpc (0.38$R_{25}$), NGC~5643$-$4.76~kpc (0.42$R_{25}$), NGC~6744$-$10.54~kpc (0.35$R_{25}$). This is also applied to the separation of inner and outer regions in Sec~\ref{sec:results_z} and Fig~\ref{fig:dphi_dz}.}.
The fluctuation of \dSFR\ along \dphi\ in the inner (outer) radial bin is presented as dotted (dashed) lines in Fig~\ref{fig:dphi_dsfr}.
We observe the same \dSFR\ - \dphi\ trend between the inner and outer regions in the eight of our spiral galaxies.
NGC~4536 is the only spiral galaxy exhibiting an opposite \dSFR\ trend between the inner and outer regions, which still fluctuates in between 25\% and 75\% quantiles (blue shadow).

\subsection{12 + log(O/H)}\label{sec:results_z}
As introduced in Sec~\ref{sec:intro}, spiral arms driven by density wave theory will lead to higher metallicity in the trailing edge than the leading edge, while dynamic spiral arms will not show azimuthal variations.
Similar to \SigSFR\ (Sec~\ref{sec:results_sfr}), we will quantify the metallicity offsets on each side of the spiral arms.

\begin{landscape}

\begin{figure}
    \centering
    \includegraphics[width=0.43\textwidth]{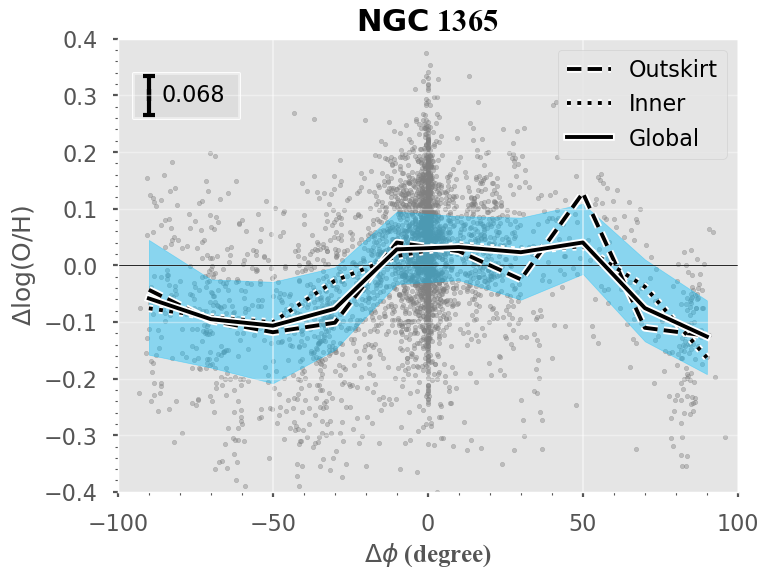}
    \includegraphics[width=0.43\textwidth]{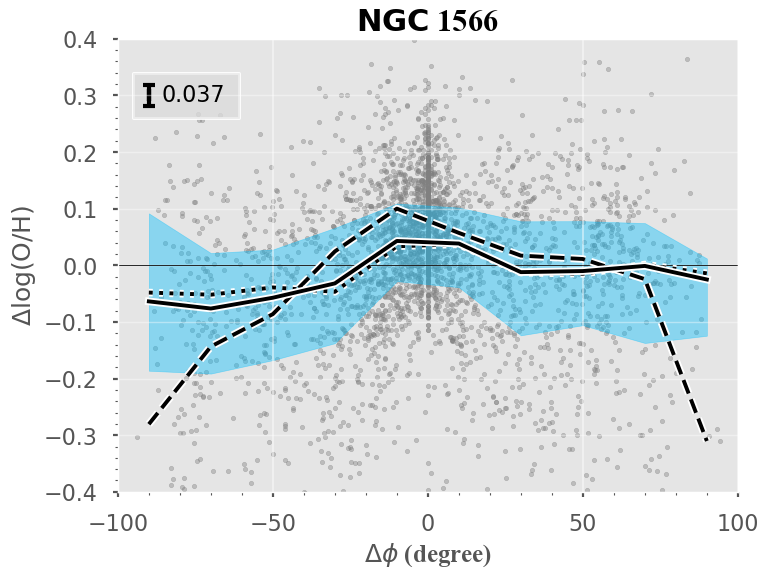}
    \includegraphics[width=0.43\textwidth]{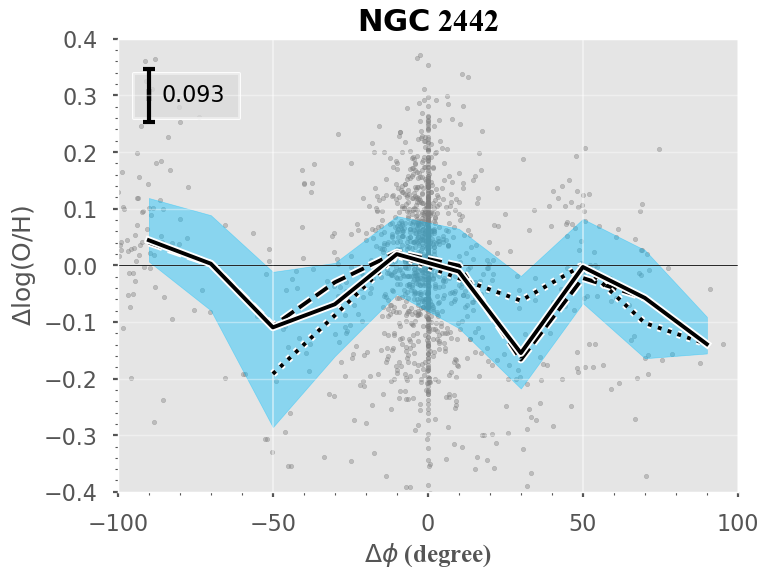}
    
    \includegraphics[width=0.43\textwidth]{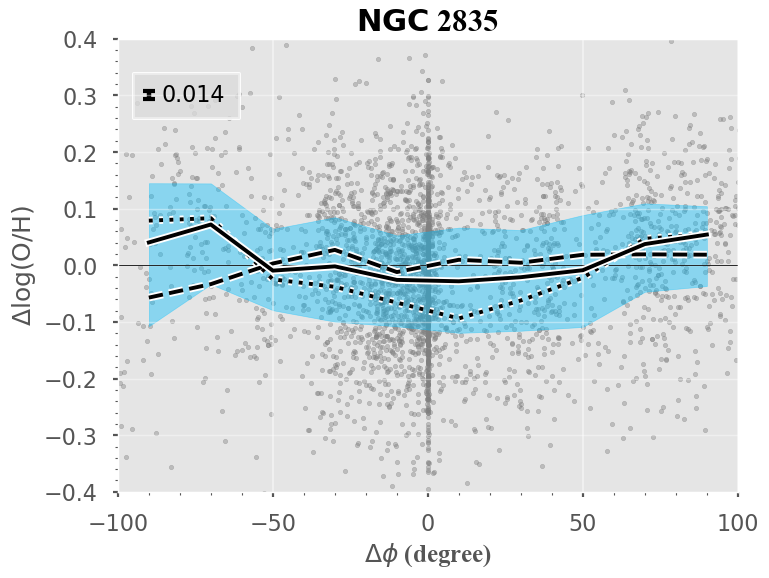}
    \includegraphics[width=0.43\textwidth]{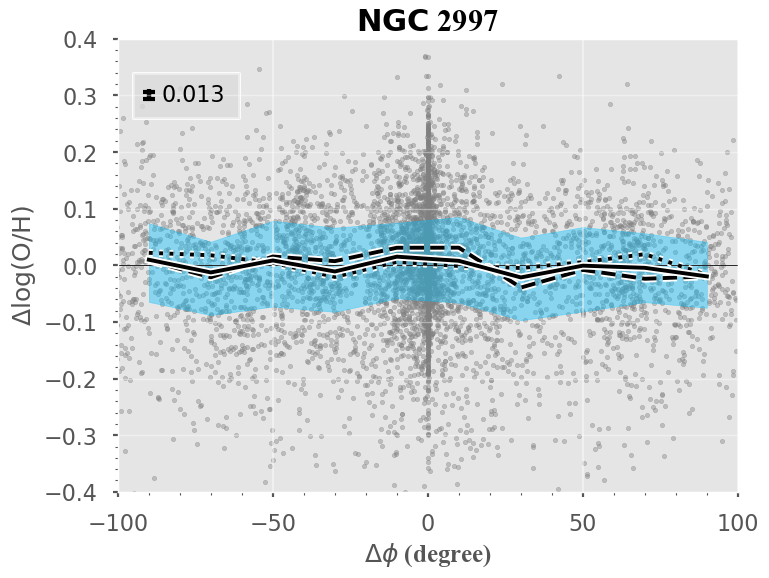}
    \includegraphics[width=0.43\textwidth]{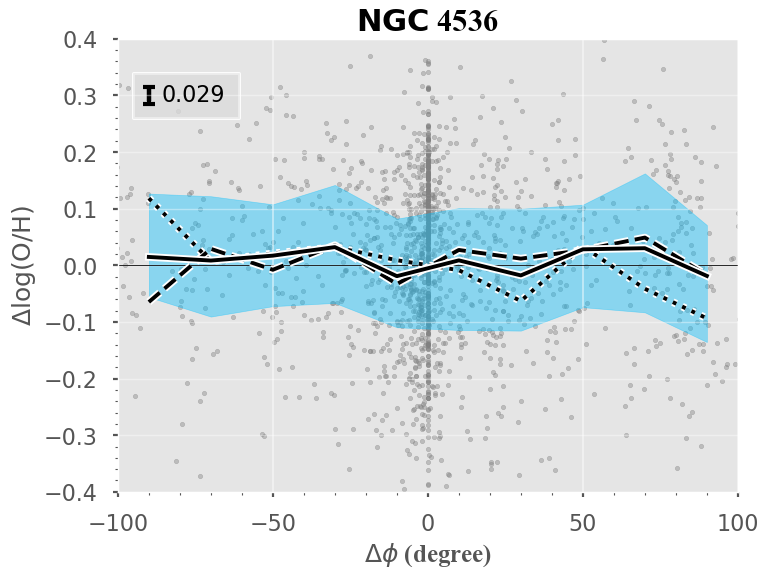}
    
    \includegraphics[width=0.43\textwidth]{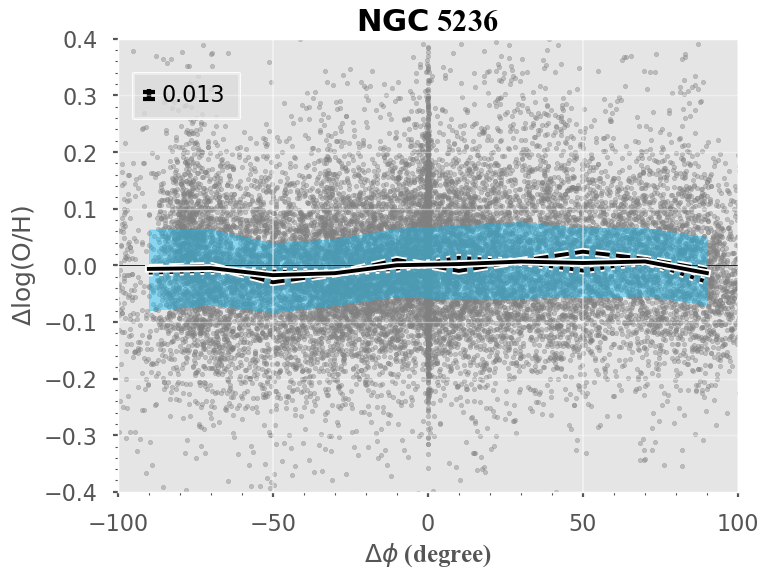}
    \includegraphics[width=0.43\textwidth]{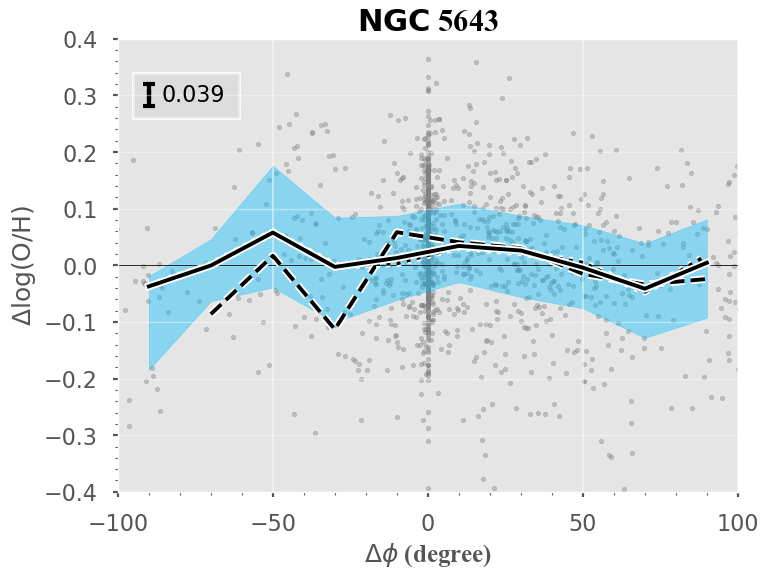}
    \includegraphics[width=0.43\textwidth]{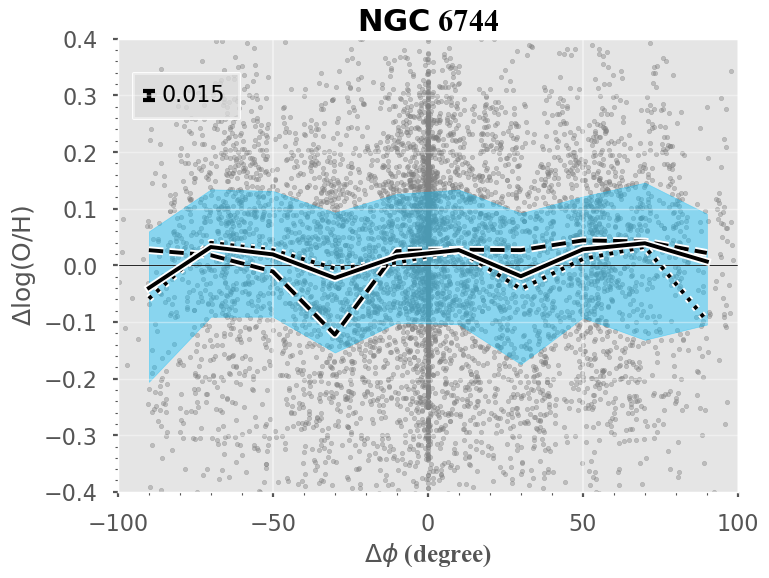}
    \caption{Similar to Fig~\ref{fig:dphi_dsfr} but for \dOH.
    The gas flows (leftward) from 100$^{\circ}$ to 0$^{\circ}$ then towards -100$^{\circ}$.
    The higher the \dOH\ is, the more metal-rich the spaxel is, compared to other spaxels at the same galactocentric distance.
    We find significant offsets ($> 0.1$ dex) in the metallicity of NGC~1365 and NGC~1566 and NGC~2442 (\dphi $> 50$).
    We find no significant azimuthal variation in the other six spiral galaxies.
    We do not find any opposite trend of \dOH-\dphi\ in the inner region versus the outskirt. 
    The \dOH\ in the inner region of NGC~2835 is significantly different from the \dOH\ in the outskirt, indicative of environmental effects in the outskirt.}
    \label{fig:dphi_dz}
\end{figure}
\end{landscape}

Fig~\ref{fig:dphi_dz} shows \dOH\ versus \dphi, showing the azimuthal fluctuation of the relative metal content once the global radial trend has been removed (Sec~\ref{sec:ana_z}).
The spaxels with positive \dOH\ represent regions of enriched gas, while spaxels with negative \dOH\ values have less enriched gas.
Similar to Sec~\ref{sec:results_sfr} with \SigSFR, we measure the moving medians of each $20^\circ$ \dphi\ (solid black lines in Fig~\ref{fig:dphi_dz}).
The 25\% and 75\% quantiles are shown as a blue-shaded region.

In NGC~1365, we observe increasing \dOH\ from $100^{\circ}$ to $20^{\circ}$ and a drop of \dOH\ from $-20^{\circ}$ to $-50^{\circ}$.
Both NGC~1365 and NGC~1566 show higher \dOH\ in the trailing edge (\dphi \textgreater 0) than in the leading edge (\dphi \textless 0).
This finding is consistent with the toy model in \citet{Ho_2017}, which predicts i) a build-up of metal-rich gas in the trailing edge when the material rotates forward to the spiral arms; ii) a decrease in metallicity due to the mixing and diluting process when the material passes the spiral arms.
We remind the readers that the model in \citet{Ho_2017} assumes that gas overtakes the spiral patterns (i.e., inside the CR).
In NGC~2442, we observe decreasing metallicity from \dphi\ $\sim$ 70$^{\circ}$ to \dphi\ $\sim$ 20$^{\circ}$ and increasing metallicity at \dphi\ \textless -20 $^{\circ}$.
This does not align with either dynamic spiral theory or density wave theory.
We attribute the azimuthal variation in NGC~2442 to the ongoing merging event (Sec~\ref{sec:dominant}).
In NGC~2835, NGC~2997, NGC~4536, NGC~5236, NGC~5643 and NGC~6744, we find no offset on both sides of the spiral arms, indicating no observed azimuthal variation in metallicity.
This lack of azimuthal variation aligns with the prediction from the dynamic spiral theory.
However, the uncertainty of gas-phase metallicity and detection limit might also be attributed to the absent azimuthal variation.
Further statistical analysis is carried out to test the reliability of the observed azimuthal variation in gas-phase metallicity (below).

The density wave theory predicts material to show opposite kinematics inside and outside the CR.
The opposite motion leaves a caveat to interpreting the azimuthal variation without analysis of different radial ranges.
Similar to Fig~\ref{fig:dphi_dsfr}, we divide the disc into two sections: inner region and outskirt, with an equal number of spaxels in each section.
We present the azimuthal trend in \dOH\ separately with dashed (outskirts) and dotted (inner) lines in Fig~\ref{fig:dphi_dz}.
We do not observe a significant difference between the inner and outer regions in eight of our samples, except for NGC~2835.
This divergence is evidence of the gas accretion from the circumgalactic medium of NGC~2835, which was detected in the flattened metallicity radial profile \citep{Chen_2023}.
The higher \dOH\ near the spiral arms in the outskirts, compared to the inner region, supports the notion that spiral arms facilitate gas radial migration (Sec~\ref{sec:radial_migra}).

\begin{figure*}
    \centering
    \includegraphics[width=0.3\textwidth]{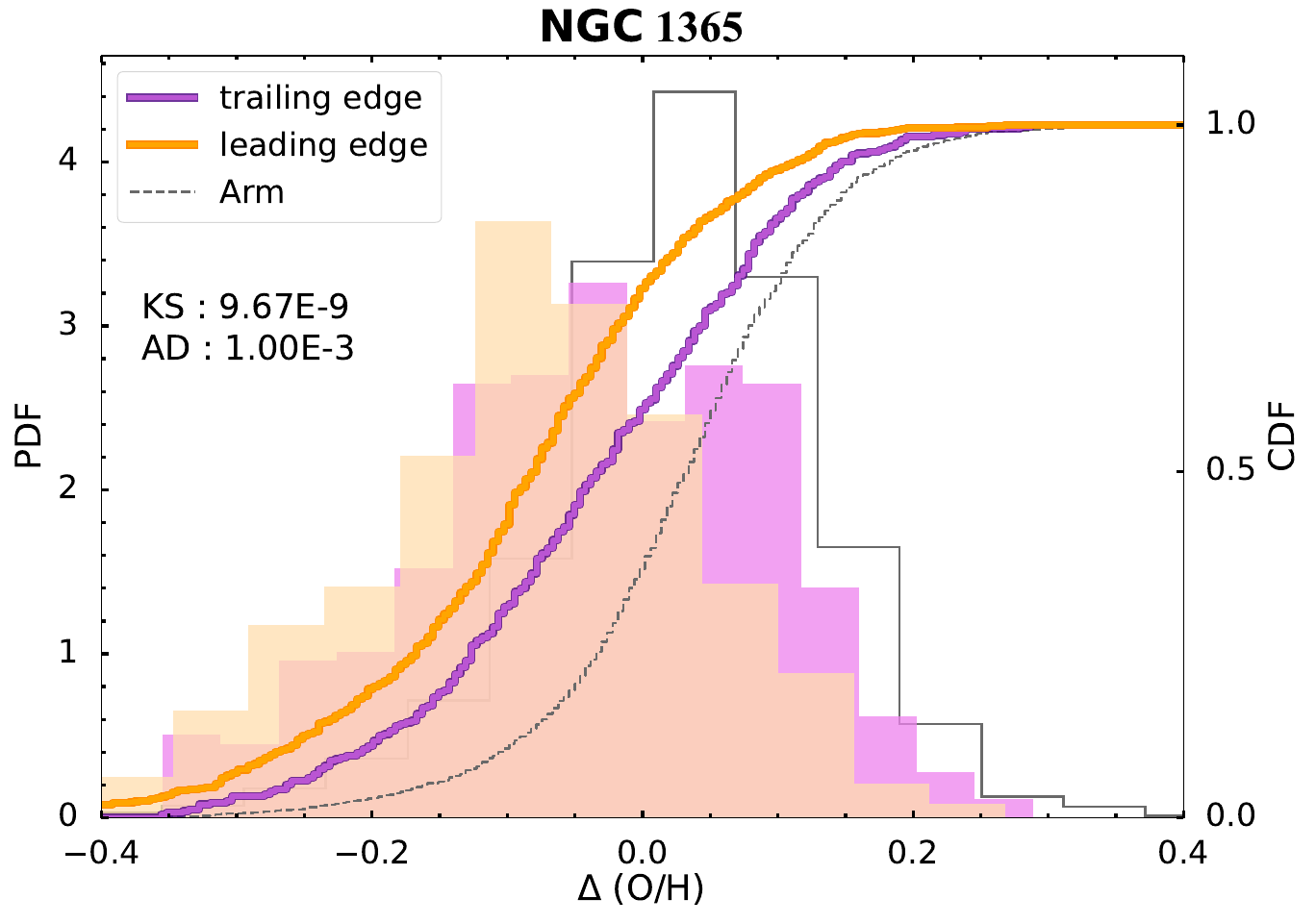}
    \includegraphics[width=0.3\textwidth]{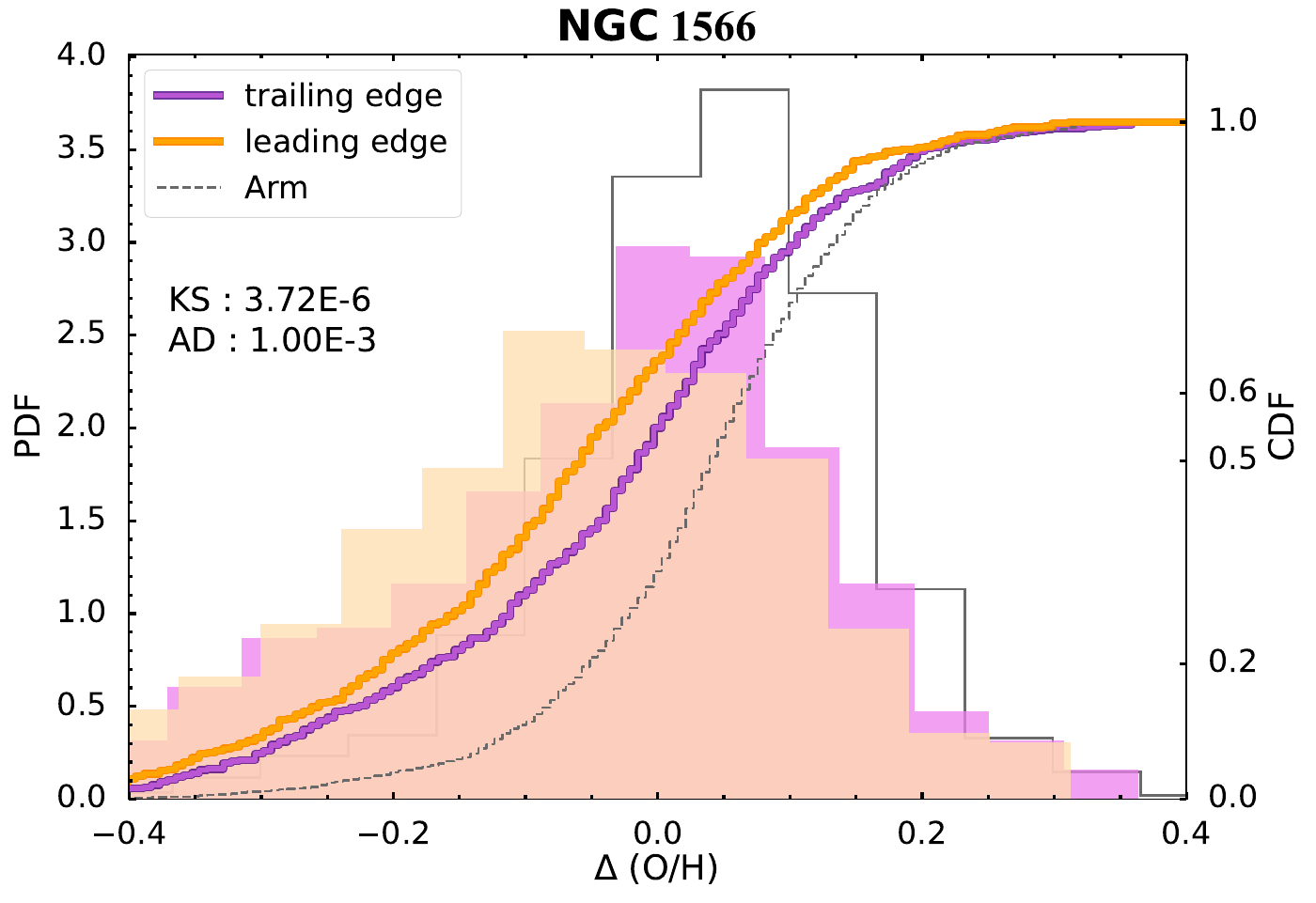}
    \includegraphics[width=0.3\textwidth]{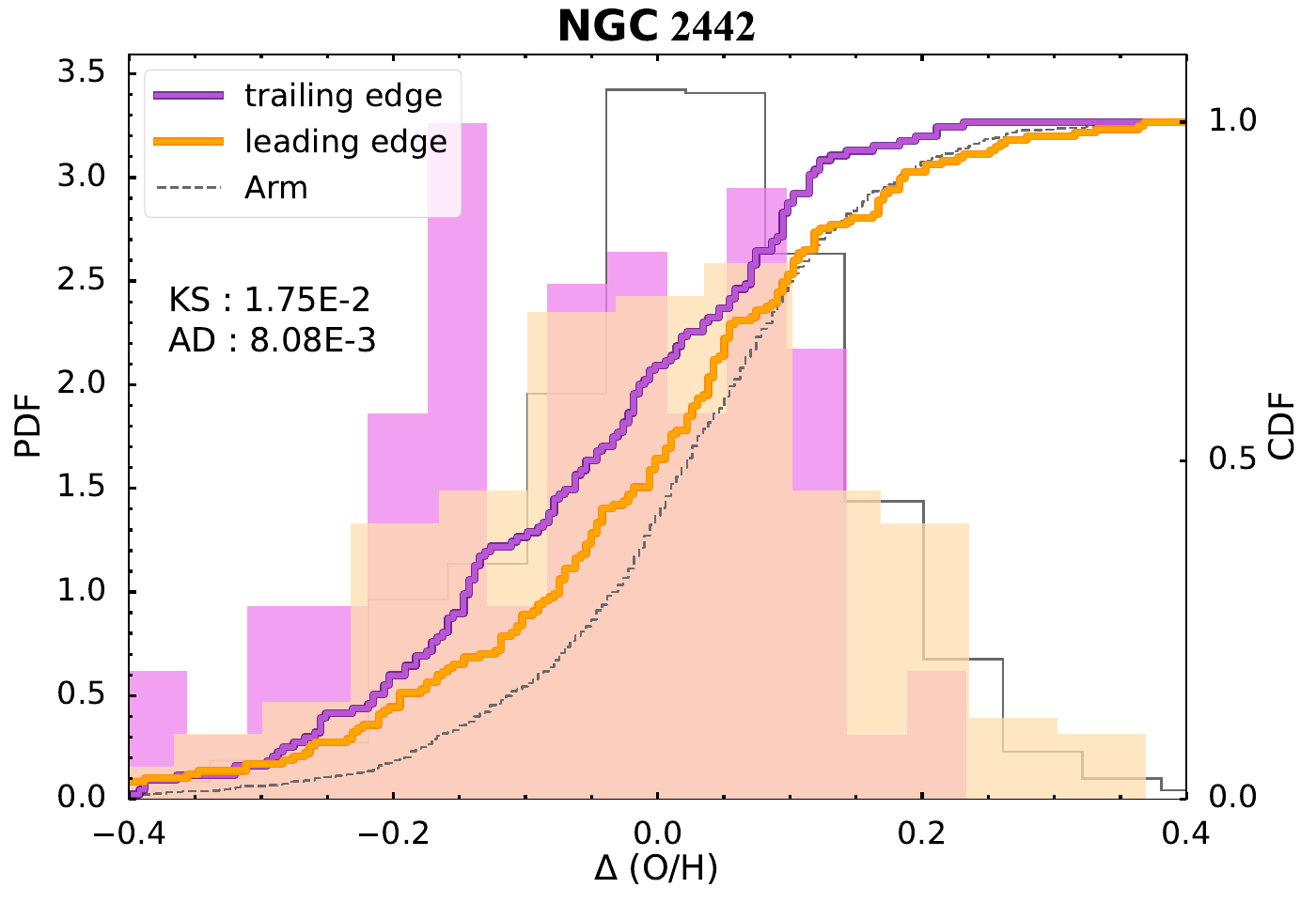}
    
    \includegraphics[width=0.3\textwidth]{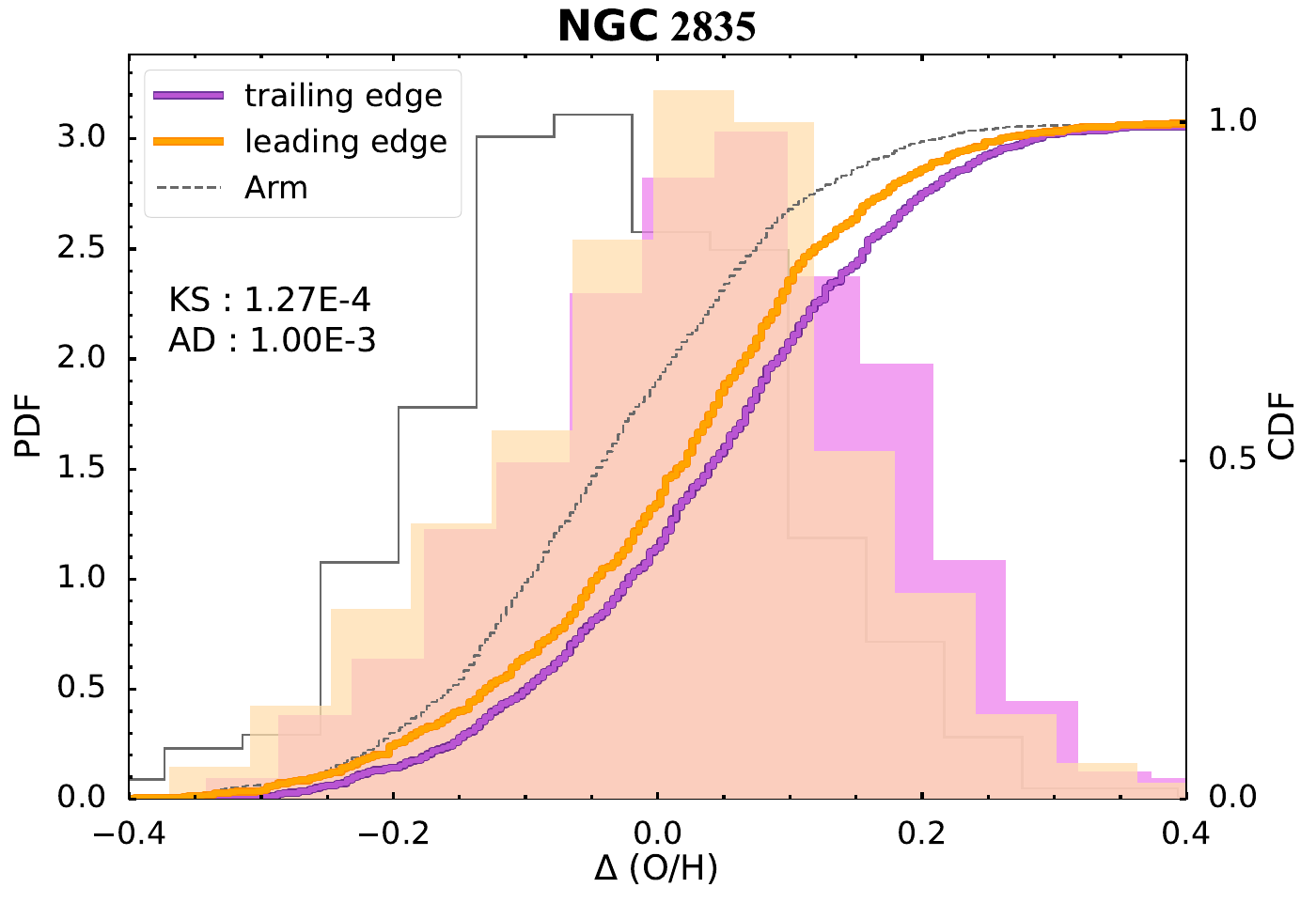}
    \includegraphics[width=0.3\textwidth]{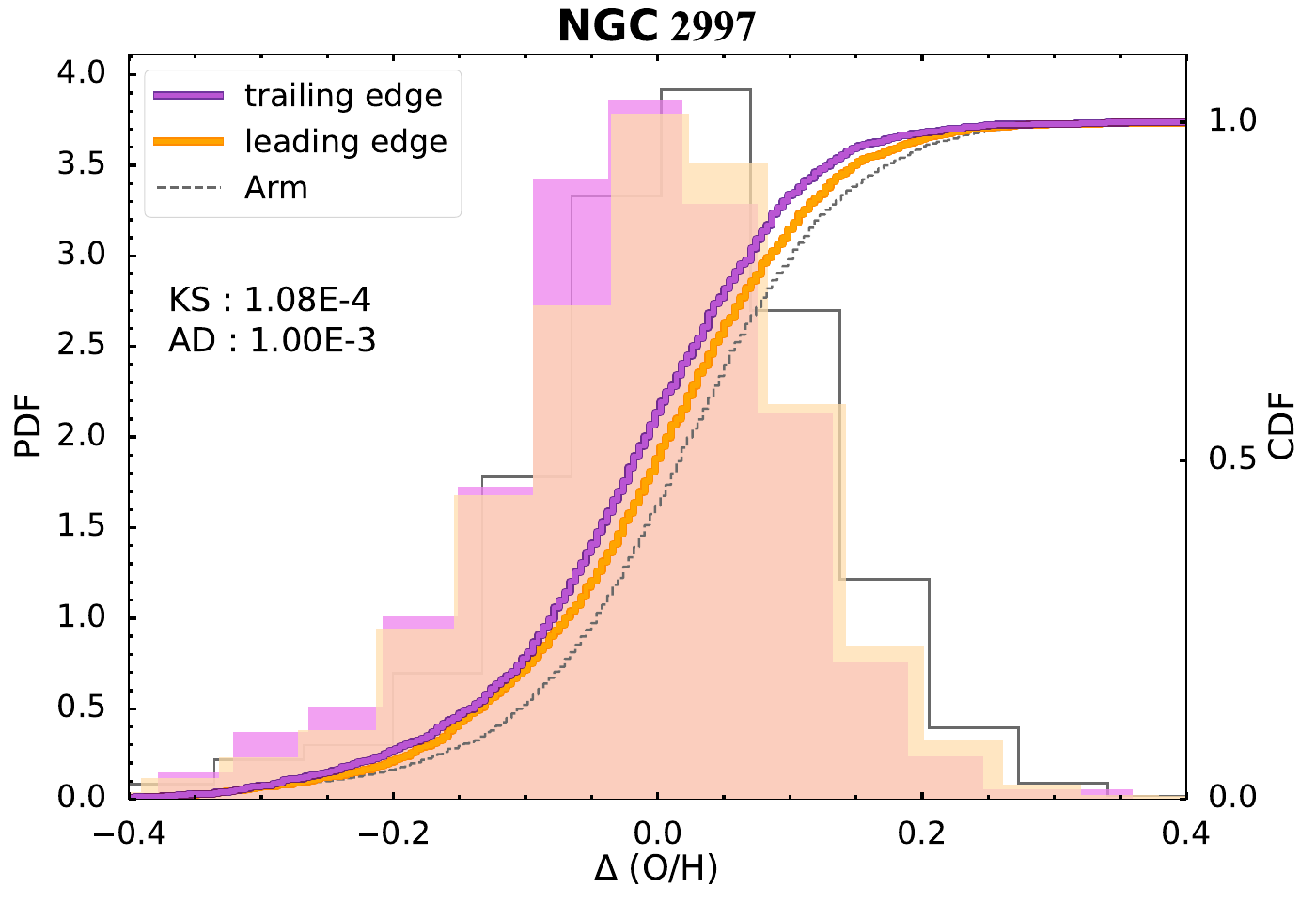}
    \includegraphics[width=0.3\textwidth]{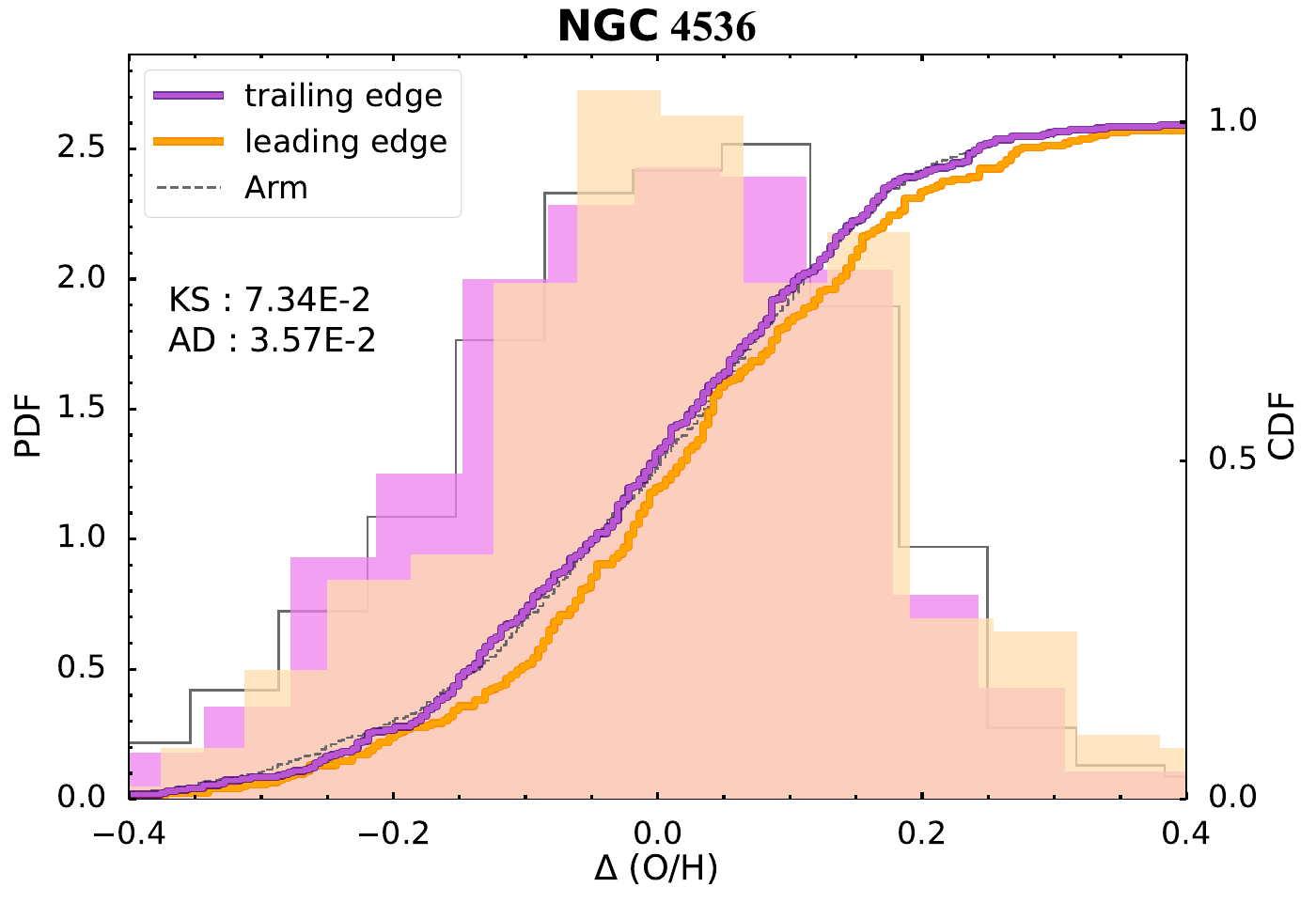}
    
    \includegraphics[width=0.3\textwidth]{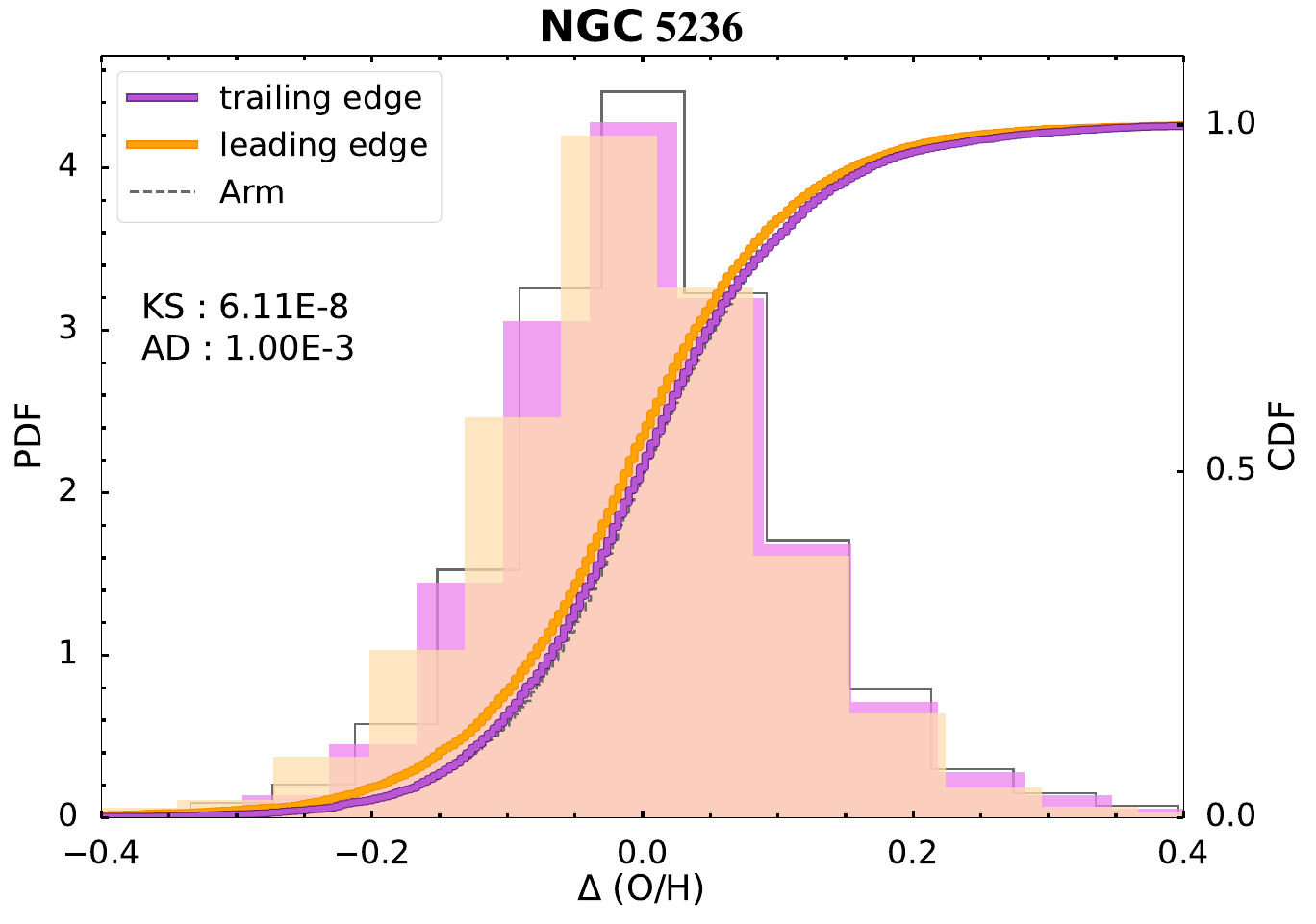}
    \includegraphics[width=0.3\textwidth]{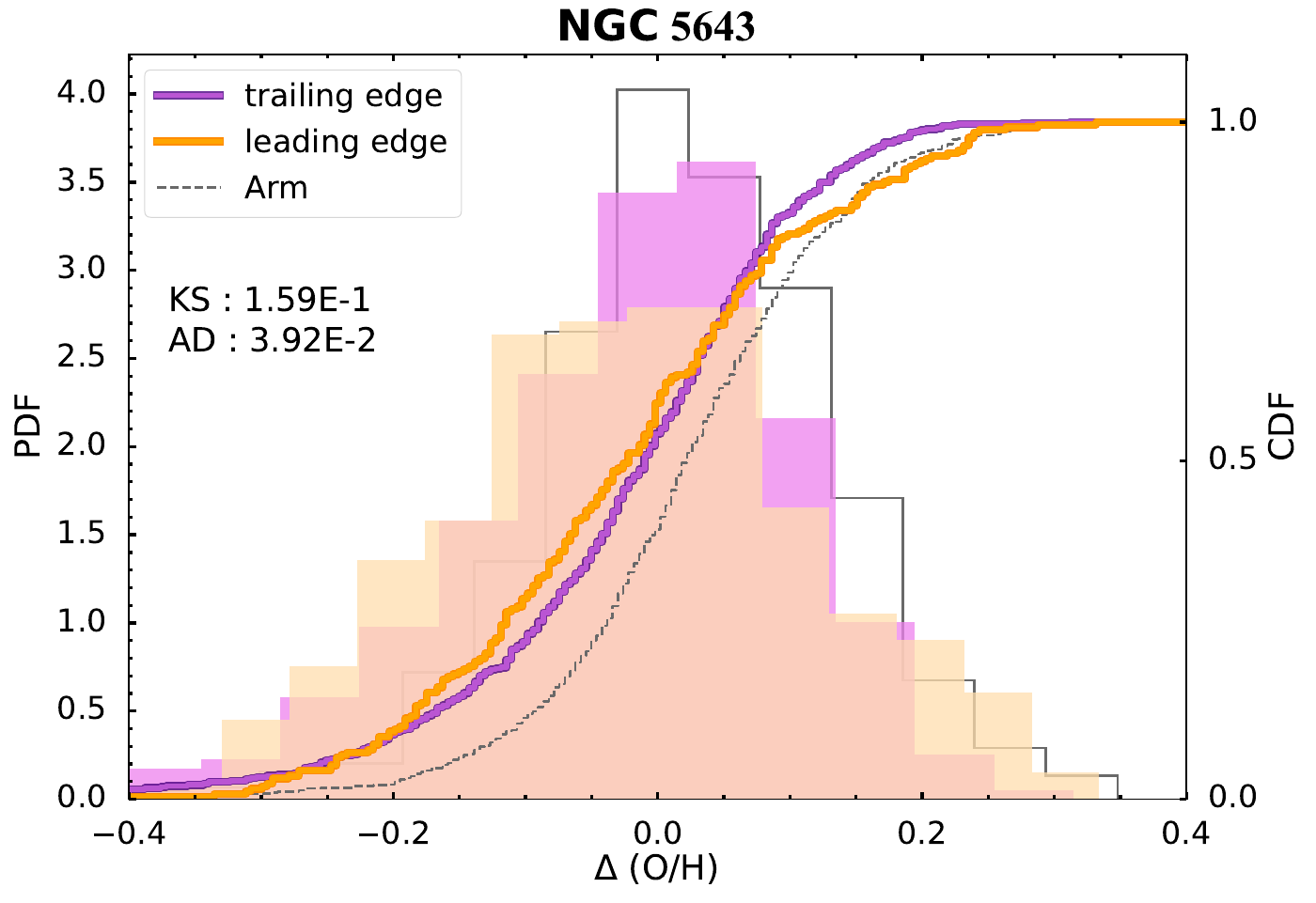}
    \includegraphics[width=0.3\textwidth]{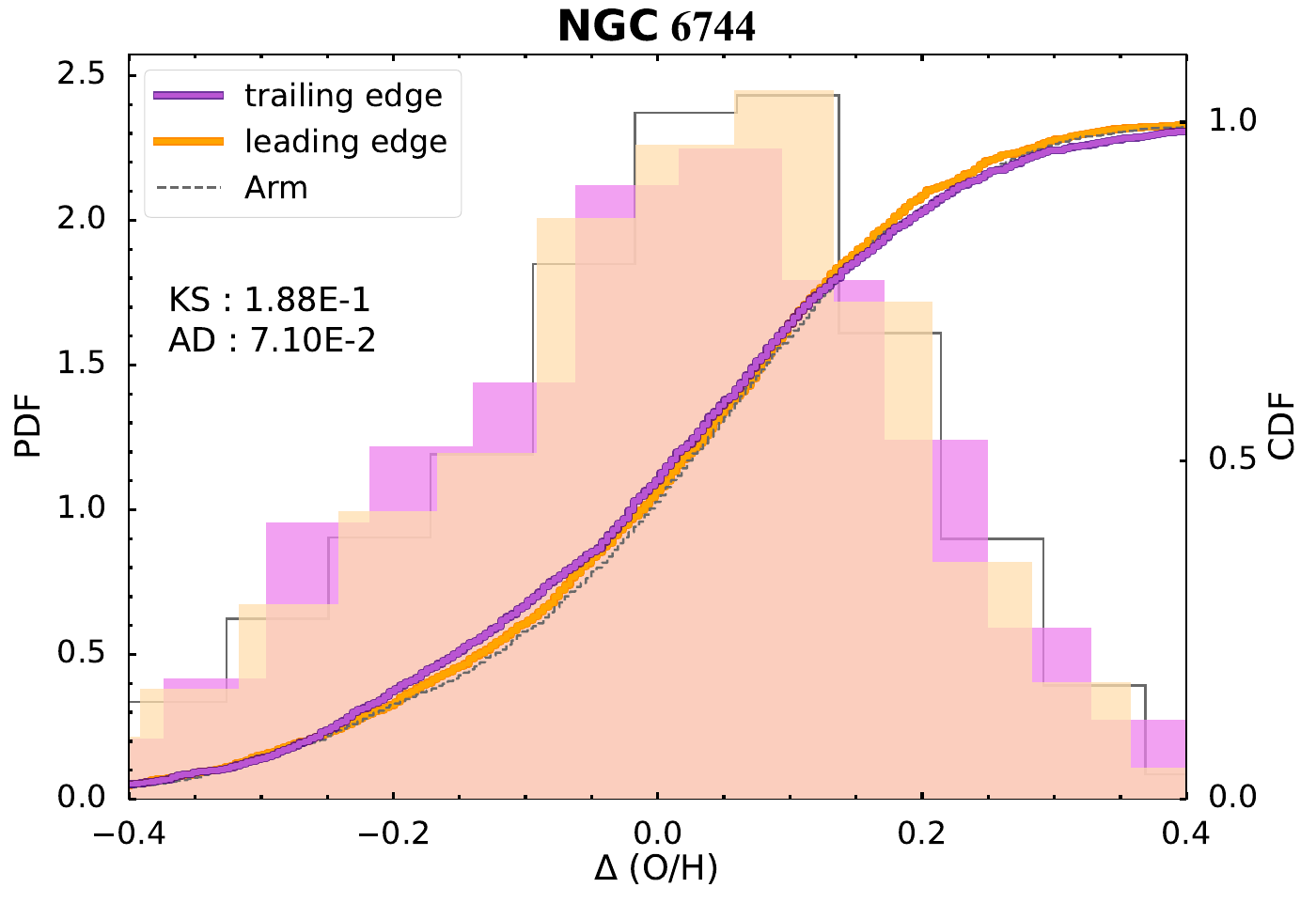}
    \caption{Histograms and CDF diagrams of \dOH, colour-coded by their location to the spiral arms: purple marks the trailing edge of the spiral arms and orange marks the leading edge.
    We leave the gap within $-20^{\circ}$ \textless \dphi\ \textless $20^{\circ}$ as the spiral arm region (black hollow histograms and black CDFs), which are not included in the leading/trailing edge regions.
    NGC~1365 and NGC~1566 show higher \dOH\ in the trailing edge (purple) than the leading edge (orange), while NGC~2442 presents higher \dOH\ in the leading edge.
    The other six galaxies show similar metallicity CDFs on both sides of the spiral arms.
     }
    \label{fig:cdf}
\end{figure*}

To assess whether the metallicity on both sides of the spiral arms is drawn from the same parent distribution, we present histograms and cumulative distribution functions (CDFs) in Fig~\ref{fig:cdf}, colour-coded by their location in their spiral arms: trailing (purple), leading (orange) and spiral arm (black) region.
We conduct KS tests and Anderson–Darling (AD) tests, with $p$-values from both tests presented in Fig~\ref{fig:cdf}.
The $D$-value, representing the maximum absolute difference between the CDFs of metallicity on both sides, assesses whether metallicity distributions deviate from each other.

We test the detection limit of the statistical framework with the bootstrap resampling method. 
For each pixel in the galaxy, we calculate the error ($\sigma_Z$) through error propagation.
In each bootstrapping trail, we generate a random metallicity map from a Gaussian distribution with a mean value of observed metallicity and standard deviation of $\sigma_Z$.
After repeating the procedure 1000 times, we compute the mean values of $D$-values, $p$-values and significance level, as well as the standard deviation. 
We summarise the $D$-values, KD/AD test results and their uncertainty in Tab~\ref{tab:$D$-value}.

Fig~\ref{fig:cdf} shows that the trailing edge (purple) of NGC~1365 and NGC~1566 exhibit systematically higher metallicity than the leading edge (orange), with a $p$-value\footnote{There is a lower limit of 1$\times 10^{-3}$ for $p$-value from the AD test.} of 1$\times 10^{-3}$ and 1.52$\pm 2.51 \times 10^{-3}$ from the AD test, respectively.
The $D$-values of 0.171 and 0.114 also suggest the azimuthal variation in the metallicity of NGC~1365 and NGC~1566.
Combined with the observed azimuthal variation seen in \SigSFR (Fig~\ref{fig:dphi_dsfr}), this suggests that density wave theory drives the spiral features in both NGC~1365 and NGC~1566.
In NGC~2442, the $D$-value of 0.142 and the $p$-value of 4.80$\pm 5.57\times 10^{-2}$ from AD test indicate that the metallicity on both sides of the spiral arms is drawn from different parent distributions.
However, NGC~2442 shows higher metallicity values in the leading edge (orange) instead of the trailing edge (purple), the opposite of what was observed in both NGC~1365 and NGC~1566.
The kinematic and star-forming properties of NGC~2442 are indicative of what is expected from a typical interacting system shortly after the initial collision \citep{Mihos_1997}. 
We thus attribute the opposite azimuthal variation trends present in metallicity to the ongoing merger.

In NGC~2835, NGC~2997, and NGC~5236, the AD test returns a $p$-value of (1.11$\pm 0.72) \times 10^{-3}$, (1.15$\pm 0.87) \times 10^{-3}$, and 1$\times 10^{-3}$, rejecting the hypothesis that the metallicities distributions are drawn from the same parent distribution.
Interestingly, the metallicity offset between \dphi\ $> 0$ and \dphi\ $< 0$ is absent in Fig~\ref{fig:dphi_dz} and absent in the CDFs with a $D$-value of 0.084 $\pm$ 0.010, 0.075 $\pm$ 0.008, and 0.043 $\pm$ 0.003, respectively.
The small $D$-value in the metallicity of NGC~2835, NGC~2997 and NGC~5236 aligns with the prediction from dynamic spiral theory.
The small $p$-value can be driven by 1) the tail of the distribution, as the mean values of the metallicity on both sides of the spiral arms are not distinguishable,
and/or 2) the environmental effects such as gas accretion (Sec~\ref{sec:dominant}) enhancing the metallicity asymmetry.
We notice that NGC~2835 has surprisingly metal-poor spiral arms, compared to the interarm regions. This may result from the gas accretion from the circumgalactic medium \citep[similar to NGC~2915;][]{Werk_2010}.

The substantial scatter observed in both \SigSFR\ and metallicity can obscure the detection of azimuthal offset, leading to a discrepancy between the absence of an azimuthal offset and the statistical tests of NGC~2835, NGC~2997, and NGC~5236. 
Previous studies on star clusters in NGC~5236 have reported: i) a small fraction of higher \SigSFR\ in the leading edge of one arm \citep{Silva-Villa_2012}, and ii) an azimuthal age gradient \citep{Bialopetravicius_2020, Abdeen_2022}. 
These observational results, different from our work, may be attributed to the various analyses performed on star clusters and on spaxel levels. Measuring metallicity with less uncertainty and conducting deeper observations in fainter inter-arm regions could provide more evidence. 
In this paper, using the TYPHOON survey, we will maintain our discussion of the non-detected azimuthal variation in these three galaxies as a preference for dynamic spiral theory. 

NGC~4536, NGC~5643 and NGC~6744 show highly similar metallicity CDF between the leading edge and trailing edge, with a $p$-value larger than 0.05.
These symmetrical distributions in metallicity show a preference for the dynamic spiral theory when explaining the formation of spiral features in these five galaxies.
We test our detection limit with 1000 times bootstrapping within the metallicity measurement uncertainty.
The lower than 0.05 $p$-values, even with the uncertainty, suggest the absent azimuthal variation in NGC~4536, NGC~5643 and NGC~6744 are not obscured by the detection limit.

\begin{table*}
\centering
    \begin{tabular}{*5{c}}
    \toprule
    Galaxy & $D$-value & $z$-score & $p$-value (KS) & Significance level (AD)\\
    \midrule
    \textbf{NGC~1365} & 0.171 $\pm$ 0.013 & 4.70$\sigma$ & (1.33 $\pm$ 5.41) $\times$ 10$^{-6}$*  & 1.00 $\times$ 10$^{-3}$*\\
    \textbf{NGC~1566} & 0.114 $\pm$ 0.015 & 2.92$\sigma$ & (1.18 $\pm$ 3.27) $\times$ 10$^{-3}$* & (1.51 $\pm$ 2.51) $\times$ 10$^{-3}$*\\
    NGC~2442 & 0.142 $\pm$ 0.025 & 1.25$\sigma$ & (1.05 $\pm$ 1.12) $\times$ 10$^{-1}$ & (4.80 $\pm$ 5.57) $\times$ 10$^{-2}$* \\
    \textbf{NGC~2835} & 0.084 $\pm$ 0.010 & 2.56$\sigma$ & (0.52 $\pm$ 1.05) $\times$ 10$^{-2}$* & (1.11 $\pm$ 0.72) $\times$ 10$^{-3}$*\\
    \textbf{NGC~2997} & 0.075 $\pm$ 0.008 & 3.43$\sigma$ & (2.99 $\pm$ 7.78) $\times$ 10$^{-4}$* & (1.15 $\pm$ 0.87) $\times$ 10$^{-3}$*\\
    NGC~4536 & 0.070 $\pm$ 0.017 & 0.31$\sigma$ & (3.78 $\pm$ 2.48) $\times$ 10$^{-1}$ & (9.82 $\pm$ 8.18) $\times$ 10$^{-2}$\\
    \textbf{NGC~5236} & 0.043 $\pm$ 0.003 & 4.30$\sigma$ & (0.85 $\pm$ 2.05) $\times$ 10$^{-5}$* & 1.00 $\times$ 10$^{-3}$*\\
    NGC~5643 & 0.073 $\pm$ 0.012 & 0.55$\sigma$ & (2.92 $\pm$ 1.60) $\times$ 10$^{-1}$ & (1.30 $\pm$ 0.67) $\times$ 10$^{-1}$\\
    NGC~6744 & 0.030 $\pm$ 0.005 & 0.60$\sigma$ & (2.74 $\pm$ 1.47) $\times$ 10$^{-1}$ & (8.49 $\pm$ 5.33) $\times$ 10$^{-2}$\\
    \bottomrule
    \end{tabular}
    \caption{$D$-values, $z$-scores, $p$-values from the KS tests, significance level from AD tests and their uncertainty. 
    The $D$-value indicates the largest vertical distance between the CDFs of the leading and trailing edge \dOH.
    The $z$-scores quantify how significant the $p$-values (KS) are. A $z$-score of 1 corresponds to being 1 standard deviation away from the mean in a Gaussian distribution.
    The 1$\sigma$ uncertainty in $D$-values, $p$-values and significance level is the standard deviation of 1000 iterations of bootstrap resampling.
    We use asterisks to highlight $p$-values and significance levels below 0.05. 
    Galaxies with both $p$-values and significance levels below 0.05 are highlighted in bold.}
    \label{tab:$D$-value}    
\end{table*}

\section{Discussion}\label{sec:discussion}

\subsection{Global galaxy properties and azimuthal variations}\label{sec:glo_properties}
In this work, three of our nine galaxies (NGC~1365, NGC~1566, NGC~2442) show statistically significant azimuthal variation in metallicity.
Similarly, \citet{Kreckel_2019} found subtle azimuthal variation in half of their galaxy samples but not always associated with the spiral features.
It is important to study the correlation between the global properties of spiral galaxies and the presence of azimuthal variation in the gas-phase metallicity.
This will bring us hints on which type of spiral galaxies tend to exhibit observable azimuthal variation in metallicity.

\begin{figure*}
    \centering
    \includegraphics[width = 0.24\textwidth, height=1.4in]{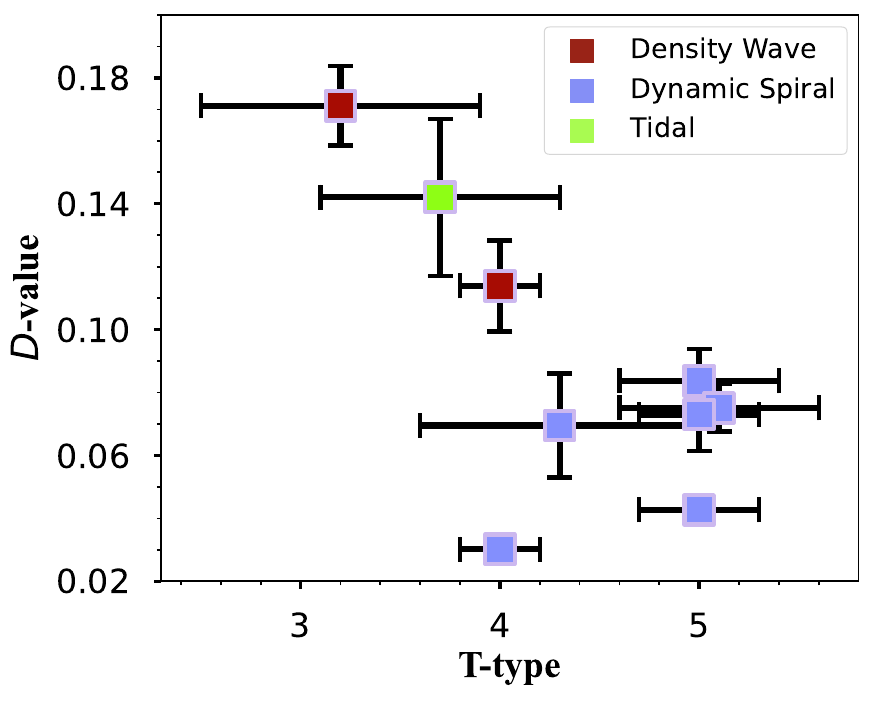}
    \includegraphics[width = 0.24\textwidth, height=1.4in]{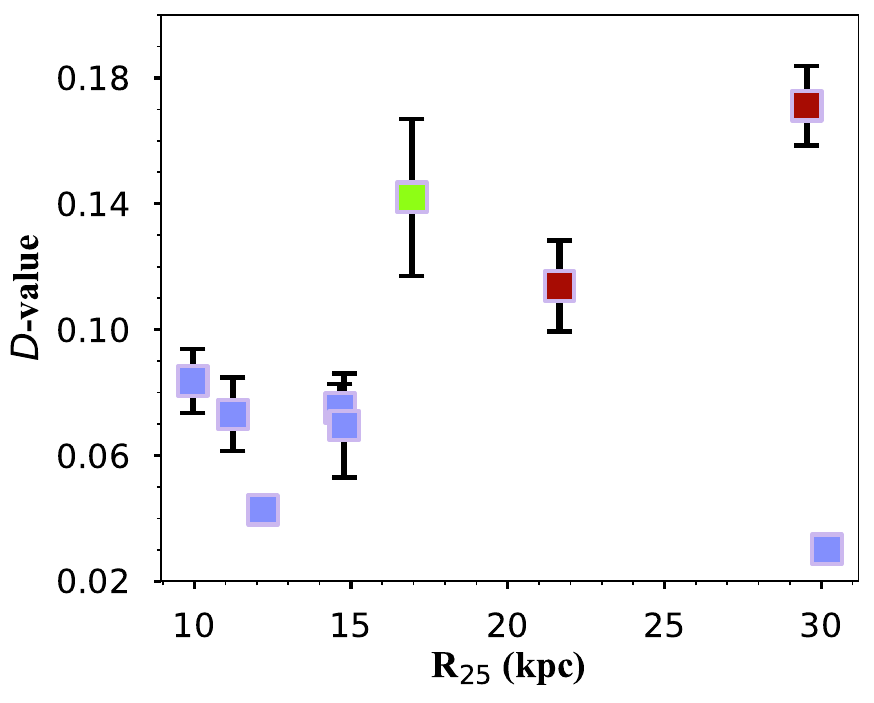}
    \includegraphics[width = 0.24\textwidth, height=1.4in]{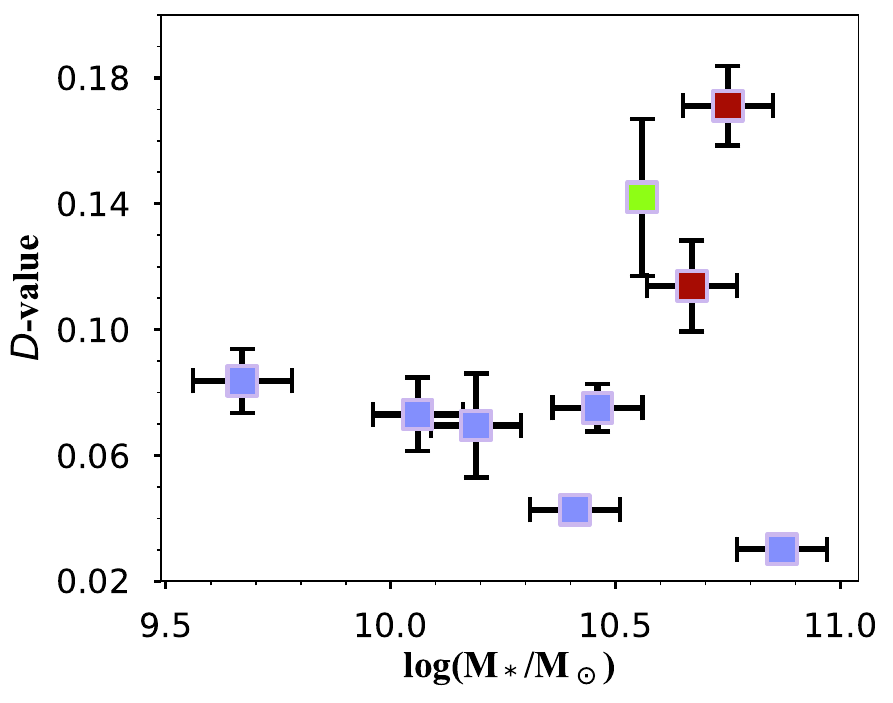}
    
    \includegraphics[width = 0.24\textwidth, height=1.4in]{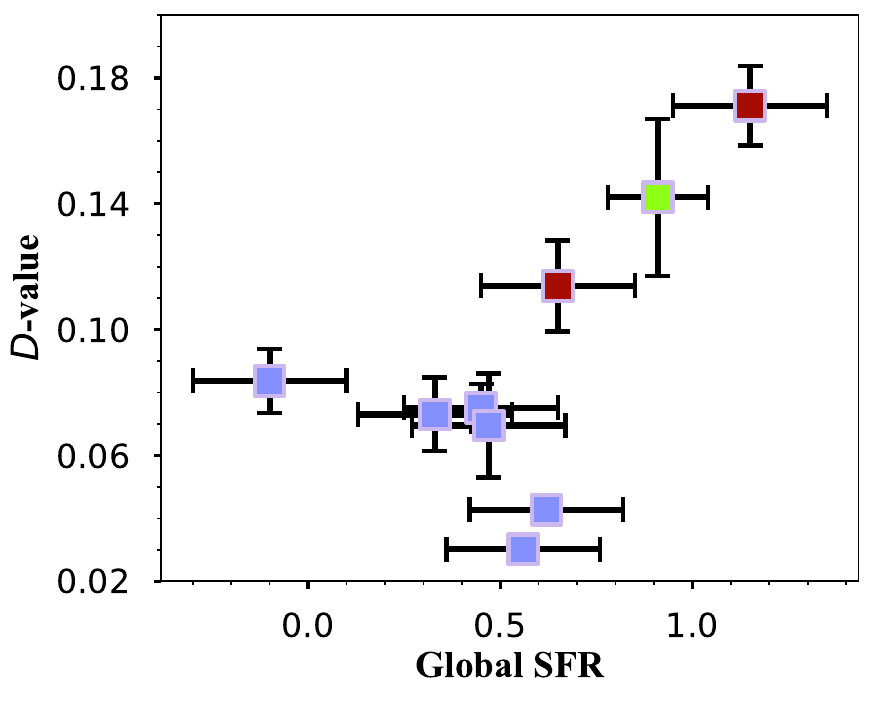}
    \includegraphics[width = 0.24\textwidth, height=1.4in]{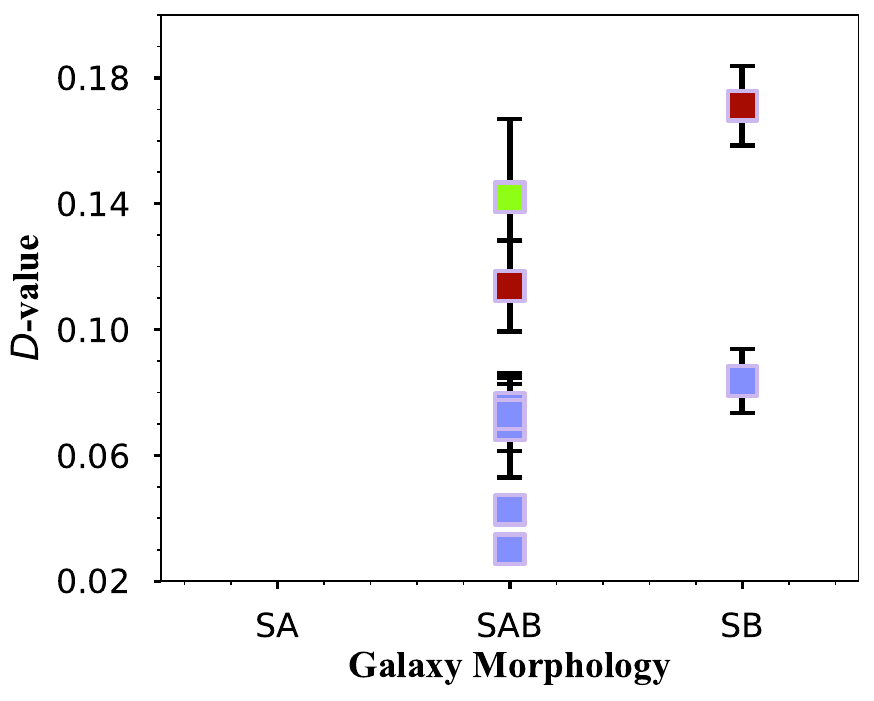}
    \includegraphics[width = 0.24\textwidth, height=1.4in]{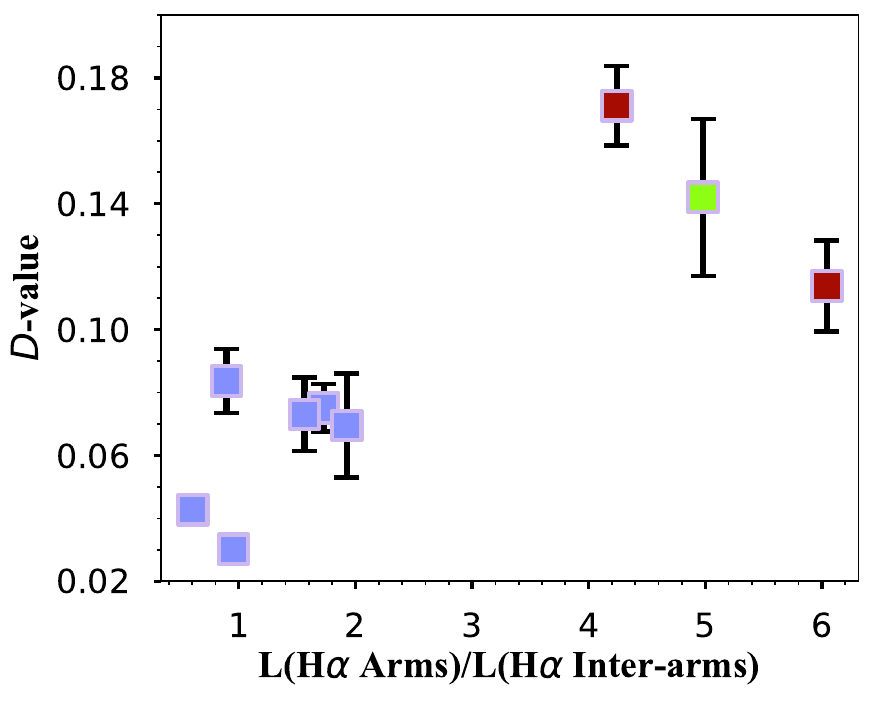}
    \includegraphics[width = 0.24\textwidth, height=1.4in]{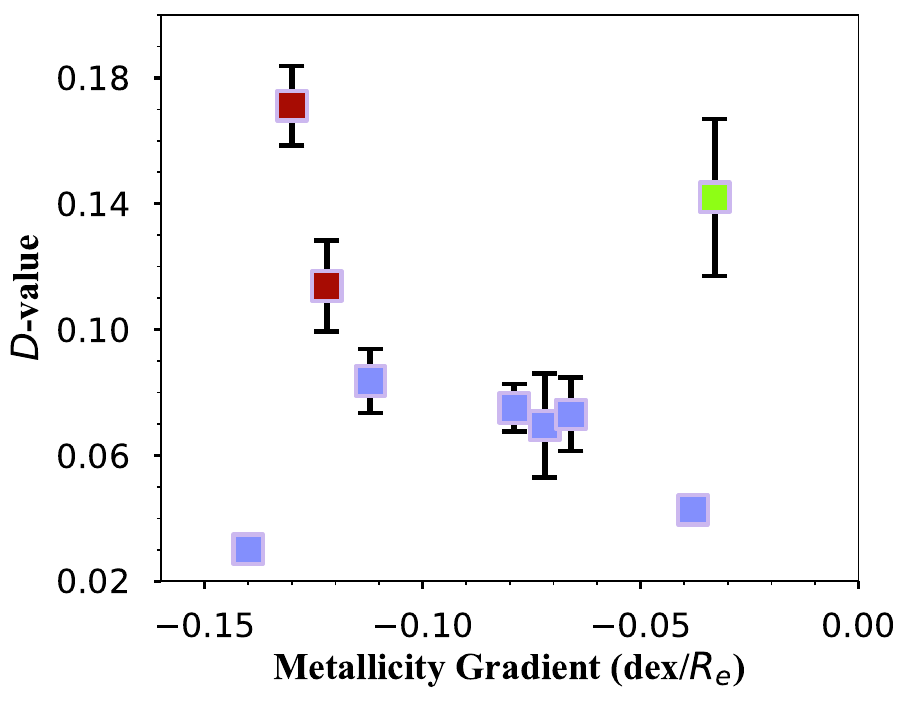}
    \caption{The relation between global properties and $D$-value drawn from metallicity CDFs (Fig~\ref{fig:cdf} and Tab~\ref{tab:$D$-value}). 
    The higher the $D$-value is, the greater the metallicity azimuthal variation is.
    All galaxies are colour-coded by their dominant mechanism that drives the origin of spiral arms.
    T-type is the numerical morphological type adopted from \href{http://atlas.obs-hp.fr/hyperleda/}{http://atlas.obs-hp.fr/hyperleda/}, with 3 referring to Sb, 4 referring to Sbc and 5 referring to Sc in the \citet{Vaucouleurs_1959} morphological classification.
    We use $R_{25}$ to represent the size of the galaxy disc, taken from Tab~\ref{tab:info}.
    The stellar mass is the same as Tab~\ref{tab:info}.
    The global SFR are taken from \citet{Leroy_2019} $\mathrm{^a}$.
    The presence of a bar is taken from the morphological information in Tab~\ref{tab:info}.
    The \ha\ luminosity ratios between the spiral arms (|\dphi| \textless 20$^{\circ}$) and the inter-arm regions describe the strength of spiral features.
    The radial gradient of metallicity is defined as the slope of the best linear fit on spaxels, in dex kpc$^{-1}$ (further discussed in Sec~\ref{sec:radial_migra}). 
    \\
    {\bf Note.}\\
    $\mathrm{^a}$: the global SFR of NGC~2442 is taken from \citet{Pancoast_2010} as it is not included in \citet{Leroy_2019}.
    }
    \label{fig:global}
\end{figure*}

In Fig~\ref{fig:global}, we show the correlation between $D$-values calculated from metallicity CDFs (Fig~\ref{fig:cdf}) and global galaxy properties, including the tightness of spiral arms (T-type), disk size, stellar mass, global SFR, the presence of a bar, amplitude of spiral arms and metallicity gradient.
The points are colour-coded by their dominant mechanism that drives the spiral features (Sec~\ref{sec:results_z} and further discussed in Sec~\ref{sec:dominant}).
In Fig~\ref{fig:global}, we find that more open-armed galaxies (large T-type) tend to have low $D$-values while less open-armed galaxies (small T-type) exhibit large $D$-values. 
There is a weak but positive correlation between $R_{25}$ and $D$-values, with one outlier galaxy, NGC~6744.
Our finding suggests that more extended galaxies tend to have stronger azimuthal variation in metallicity compared to compact galaxies.
We present the strength of spiral arms using the \ha\ luminosity ratio between the spiral arm regions (|\dphi| \textless 20$^{\circ}$) and the inter-arm regions.
Our galaxies show larger $D$-values with stronger spiral arms, which suggests that galaxies with more pronounced spiral arms tend to have greater metallicity offset on both sides of spiral arms.
Similarly, \citet{Sanchez-Menguiano_2020} find the metallicity difference of arm versus interarm is larger ($\sim$ 0.015~dex) in grand-design than in flocculent galaxies.
More studies on nearby galaxies can fill up the gap between strong spiral arms ($\frac{L_{\mathrm{arm}}}{L_{\mathrm{inter-arm}}}$ \textgreater\ 4) and weak spiral arms ($\frac{L_{\mathrm{arm}}}{L_{\mathrm{inter-arm}}}$ \textless\ 3) and provide more constraints.
The galaxies in our sample do not show a clear correlation between the $D$-values and the stellar mass, global SFR, and the presence of a bar.
A larger sample of spiral galaxies will improve the study of the correlation between azimuthal variations and global galaxy properties.

\subsection{Impacts of radial streams and radial migration}\label{sec:radial_migra}
Previous simulations have discussed the impacts of radial streams and radial migration on azimuthal variations of gas and stars \citep[e.g., ][]{Sellwood_2002, Grand_2015, Grand_2016, Orr_2022}. 
Radial stellar migration can move metal-rich star particles outward along the trailing edge, while bringing metal-poor star particles inward along the leading edge \citep[e.g.,][]{Grand_2016Auriga}.
Radial gas streams can lead to azimuthal variations of gas-phase metallicity distribution, with metal-rich gas concentrated in the trailing side of the stellar spiral arm \citep[e.g.,][]{Khoperskov_2023}. 
\cite{Sanchez-Menguiano_2016} found azimuthal variations of gas metallicity in NGC~6754, where large-scale gas radial migration is also detected.
Radial migration is expected to be stronger in a bar-spiral coupled system \citep{Minchev_2010}.
Given the varying bar strength in our sample, ranging from strong to weak, it is important to discuss the impacts of radial migration in our study of azimuthal variation.

Measuring and modelling the velocity field to investigate the radial gas migration is out of the scope of this paper.
Instead, we adopt the radial metallicity gradient as an indicator for radial gas migration.
We measure the radial metallicity gradient with a single linear function, listed in Tab~\ref{tab:zgradient}, each showing a negative metallicity gradient.
The negative metallicity gradients are consistent with an inside-out galaxy formation.
Seven of our galaxies (except for NGC~2442 and NGC~5236) have comparable radial metallicity gradients with those reported in previous studies, such as the CALIFA survey \citep[$-0.1$ dex/$R_e$;][]{SanchezS_2015}, and the MaNGA survey \citep[$-0.14$ dex/$R_e$;][]{Francesco_2017}.
The shallow metallicity gradients in NGC~2442 and NGC~5236 could be indicative of large-scale radial gas flows dispersing and flattening the metal distribution. 
A large sample of spiral galaxies is needed to establish a stronger relation between metallicity gradients and azimuthal variations.

The final panel in Fig~\ref{fig:global} compares the radial metallicity gradients of our sample and their $D$-values from metallicity CDF in the leading and trailing edge.
We cannot draw a conclusive correlation between the metallicity radial gradient and $D$-values.
In the shallow radial gradient regime ($\lesssim$ $-0.05$dex/\re), where strong and large-scale radial gas flows are expected to exist, we find both galaxies with high (NGC~2442) and low (NGC~5236) $D$-values.
This result suggests that large-scale radial gas flow does not necessarily lead to strong azimuthal offset in metallicity.

A truncated metallicity radial gradient, especially a flattening gradient in the outer parts, is possibly a phenomenon driven by radial gas mixing \citep[][Kewley et al. In Preparation]{Minchev_2011, Sanchez-Menguiano_2018, Garcia_2023}.
We summarise the best fit of a piecewise linear function on gas-phase metallicity (Sec~\ref{sec:ana_z}) in Table~\ref{tab:zgradient}.
We find a shallow-steep metallicity radial profile in NGC~1566, NGC~2997 and NGC~6744.
In NGC 1365, NGC~2442, NGC 2835, NGC 4536, NGC 5236 and NGC~5643, we observe a flattening of the metallicity radial gradient, i.e., the outer gradient is less than half as steep as the inner gradient. 
This phenomenon could be indicative of radial gas migration outside the break radius \citep{Minchev_2011} and/or satellite accretion \citep{Qu_2011}.
Among the six spiral galaxies with a flattening metallicity radial gradient, an azimuthal offset is observed in NGC~1365 and NGC~2442.
In NGC~1365, we cannot distinguish the dominant mechanisms driving the azimuthal variation, with the potential involvement of both density wave theory and radial gas motion.
The merging event in NGC~2442 can be attributed to the strong radial gas mixing and meanwhile azimuthal variations in metallicity.

\begin{table*}
\scriptsize
\newcommand{\tabincell}[2]{\begin{tabular}{@{}#1@{}}#2\end{tabular}}
\centering
    \begin{tabular}{cccccc}
    \hline
    Galaxy & NGC~1365 & NGC~1566 & NGC~2442 & NGC~2835 & NGC~2997 \\
    \hline
    Single linear (dex/\re) & $-0.136$ & $-0.144$ & $-0.014$ & $-0.119$ & $-0.077$ \\
    Piecewise fits (dex/\re) & \textbf{$-$0.664, $-$0.093} & $-$0.118, $-$0.156 & \textbf{$-$0.816, $-$0.00} & \textbf{$-$0.152, $-$0.078} & $-$0.037, $-$0.097  \\
    Break radius (kpc) & 4.82 & 8.00 & 1.93 & 6.00 & 6.73 \\
    \hline
    Galaxy & NGC~4536 & NGC~5236 & NGC~5643 & NGC~6744\\
    \hline
    Single linear (dex/\re) & $-0.086$ & $-0.038$ & $-0.066$ & $-0.132$\\
    Piecewise fits (dex/\re) & \textbf{$-$0.281, $-$0.068} & \textbf{$-$0.075, 0.038} & \textbf{$-$0.262, $-$0.061} & 0.030, $-$0.160\\
    Break radius (kpc) & 2.77 & 5.37 & 1.85 & 7.65\\
    \hline
    \end{tabular}
    \caption{This table shows the metallicity gradient fitted by a single linear function and piecewise linear function, with both the inner gradient and outer gradient listed sequentially.
    The break radius of the piecewise fits is listed in the last row. All of the observed spiral galaxies show a negative metallicity gradient, consistent with an inside-out galaxy formation. The galaxies showing a flattening metallicity gradient truncated outside 2~kpc are in bold.}
    \label{tab:zgradient}
\end{table*}

\subsection{Inside and outside the co-rotation radius}\label{sec:CR}
The asymmetric distributions of \SigSFR\ and gas-phase metallicity in NGC~1365 and NGC~1566 (Sec~\ref{sec:results_sfr} and Sec~\ref{sec:results_z}) suggest that the density wave theory explains the origin of their spiral features.
According to the density wave theory, the material surpasses the spiral density wave inside the CR while lagging behind the density wave outside the CR.
A simulation by \citet{Spitoni_2019} shows that the density perturbation of a disc model can result in stronger oxygen abundance fluctuations in the outer region compared to the inner regions of a galaxy. 
When an analytic spiral arm\footnote{Analytic spiral arms show regular gravitational perturbation with a fixed pattern speed. The surface arm density can be described by the radial distance and the azimuth.} is included in the simulation, and the fluctuations near the co-rotation resonance are enhanced.
To further compare the observations and simulations, it is essential to assess the behaviour of ISM inside/outside the CR.

Although it is challenging to measure the CR, astronomers have devoted numerous efforts to measuring the CR of nearby spiral galaxies with various methods.
Three of our observed spiral galaxies (NGC~1365, NGC~1566, and NGC~5236) have their CR reported in previous works and collected in Tab~\ref{tab:cr}. 

\begin{table}
    \centering
    \footnotesize
    \begin{tabular}{ccc}
    \toprule[1pt]
        Galaxy & CR (kpc) & References\\
    \hline
        NGC~1365 & 13.8 & \citet{Lindblad_1996, Elmegreen_2009} \\
        NGC~1566 & 8.8 & \citet{Scarano_2013,Abdeen_2020}\\
        NGC~5236 & 8.5 & \citet{Scarano_2013, Abdeen_2020}\\
    \toprule[1pt]
    \end{tabular}
    \caption{The CR of NGC~1365, NGC~1566 and NGC~5236 reported in previous publications. We take the mean values as the CR in the analysis in this work.}
    \label{tab:cr}
\end{table}

\begin{figure*}
    \centering
    \includegraphics[width=0.4\textwidth]{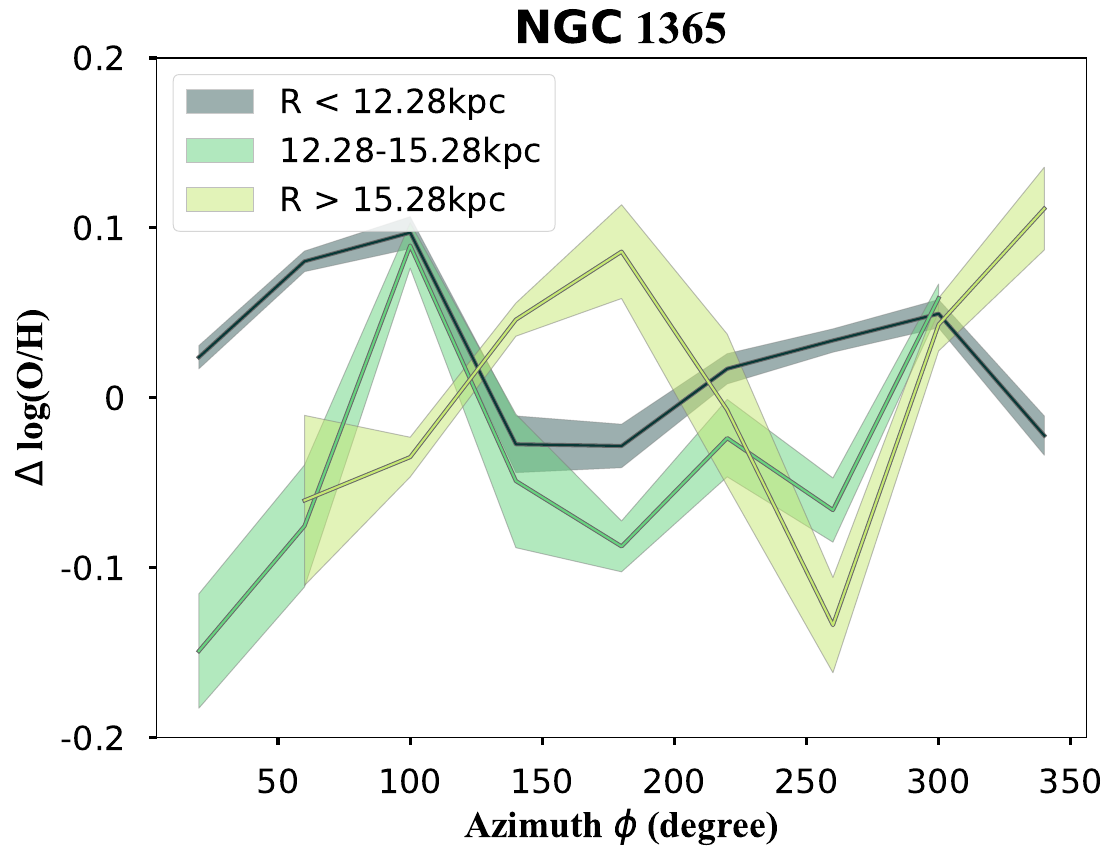}
    \includegraphics[width=0.4\textwidth]{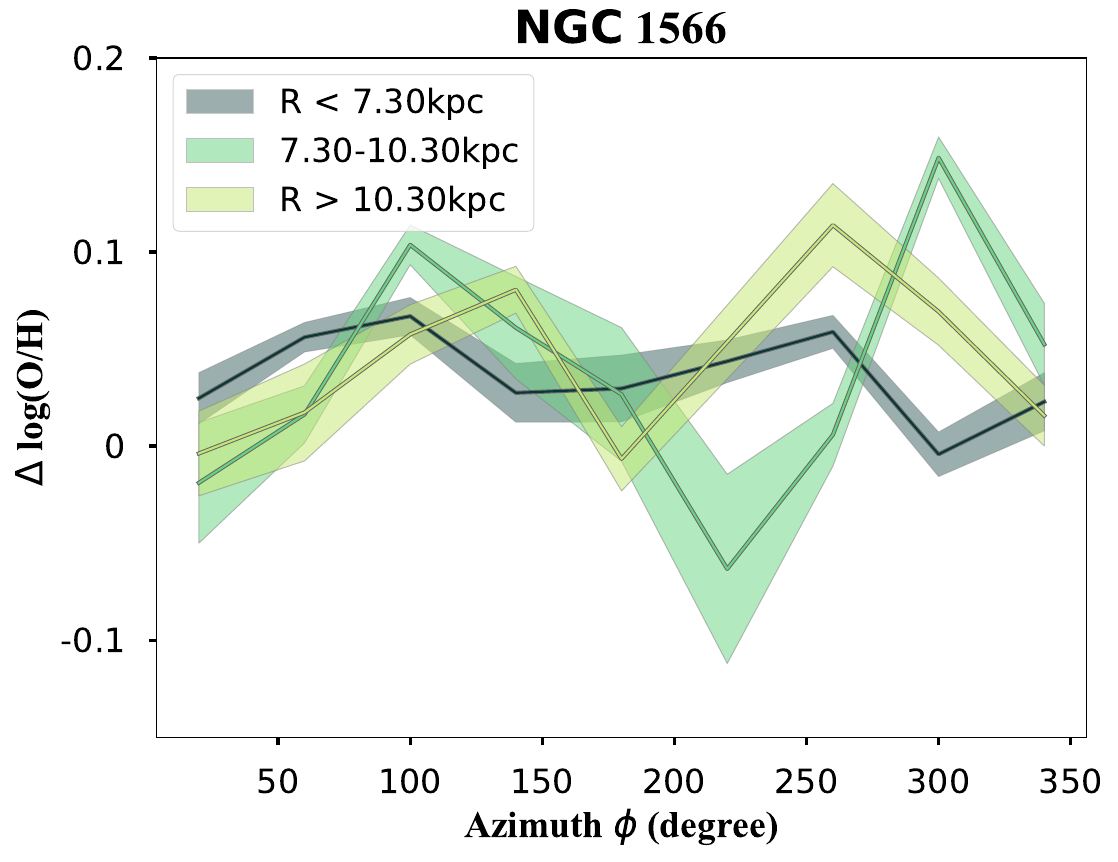}
    \caption{Residual of gas-phase metallicity as a function of azimuth inside the CR (grey), near the CR (green), and beyond the CR (lime).
    The azimuth starts from the position angle (Table~\ref{tab:info}) and increases counter-clockwise.
    We apply bootstrapping for 500 iterations and show the medians with 1$\sigma$ uncertainty in the figure.
    The shadow of the lines indicates 1$\sigma$ of the medians.}
    \label{fig:1365_CR}
\end{figure*}

With the location of CR, we divide the spiral galaxies into three sections: 
inside the CR, near the CR, and outside the CR.
We exclude NGC~5236 in the following discussion since the CR of NGC~5236 is more than 3~kpc beyond the observed optical disc in the TYPHOON survey.
Fig~\ref{fig:1365_CR} compares the behaviour of metallicities at different azimuths and radii, indicated by colours. 
The azimuth starts from the position angle in Table~\ref{tab:info} and increases counter-clockwise to 360$^{\circ}$.

We observe the smallest azimuthal fluctuation (0.13~dex for NGC~1365; 0.07~dex for NGC~1566) of metallicity in the inner region (grey line in Fig~\ref{fig:1365_CR}), in agreement with the observations of NGC~6754 \citep[Fig~7 in][]{Sanchez_2015}.
In NGC~1365, we find that the fluctuations in metallicity with azimuth are comparable when near (0.24~dex) and outside the CR (0.25~dex), which is significantly larger than those within the CR.
As an analytic spiral arm \citep{Spitoni_2019} predicts the largest metallicity fluctuation near the CR, our observation indicates the presence of a density perturbation in NGC~1365, resulting in a greater metallicity fluctuation beyond the CR.
Additionally, we notice that the \dOH\ offset between the leading and trailing edge (Fig~\ref{fig:dphi_dz}) is slightly larger in the outskirts (R \textgreater 10.98~kpc) than in the inner region (R \textless 10.98~kpc).
The absent opposing behaviour inside versus outside the CR (13.8~kpc) does not align with the prediction of density wave theory.
However, there is a large uncertainty in the detection of CR. 
In \citet{Ho_2017}, they adopt a larger CR where all observed spaxels in the TYPHOON field are within the CR.
In this case, the consistent \dOH-\dphi\ trend at different radii (dashed and dotted lines in Fig~\ref{fig:dphi_dz}) aligns with the prediction of density wave theory within the CR.

In NGC~1566, the metallicity shows the highest fluctuation near the CR (0.21~dex) when compared with the metallicity fluctuation outside the CR (0.12~dex) and within the CR (0.07~dex).
This is in line with the predictions from \citet{Spitoni_2019}, indicating that the spiral arms in NGC~1566 are also analytic.

Inside the CR, the metallicity fluctuation can be decreased or wiped out by the rotation of material, due to the decreased travel distance between spiral arms.
We calculate the timescale required to travel between spiral arms, in order to assess the reliability of the small-scale metallicity fluctuations we measured.
We calculate the angular distance that stars will travel in 10~Myr, a typical life span of O-type stars, using equation~\ref{eq:V_ang}.
The $V_{\mathrm{circ}}$ of NGC~1365  is $\sim$ 300~km s$^{-1}$ at $\sim$~100\arcsec ($\sim$~8~kpc; \citeauthor{Jorsater_1995} \citeyear{Jorsater_1995}), while NGC~1566 has a $V_{\mathrm{circ}}$ of $\sim$ 180~km s$^{-1}$ at 6~kpc \citep{Elagali_2019}.
Following equation~\ref{eq:v}, we find the rotation angle within 10~Myr is $\lesssim$ 20$^{\circ}$ inside the CR for both NGC~1365 and NGC~1566, which is smaller than the azimuthal distance between two spiral arms.
Considering the flattening of the rotation curve, we find that the rotation angle is \textgreater 20$^{\circ}$ outside the CR of NGC~1365 and NGC~1566.
Hence, the fluctuation of metallicity cannot be wiped out by the rotational motion of the material, even inside the CR.
Therefore, the observed smaller metallicity fluctuation within the CR can be considered reliable and meaningful.

\subsection{Dominant mechanisms driving spiral features of each galaxy}\label{sec:dominant}

In this section, we discuss the underlying mechanisms driving the formation of spiral arms in our galaxies, taking into account the distributions of \SigSFR, 12 + log(O/H) from the TYPHOON survey, as well as the environmental factors reported in previous works.

NGC~1365 is a grand-design spiral galaxy in the Fornax cluster. 
The small fragments near the two prominent spiral arms infer that NGC~1365 has undergone tidal interaction.
In this work, we find slightly higher \SigSFR\ (Sec~\ref{sec:results_sfr}) and generally higher metallicity (Sec~\ref{sec:results_z}) in the trailing edge (\dphi\ > 0) of the spiral arms.
These scenarios are supportive evidence for density wave spiral arms.
However, the radial gas motion, indicated by the flattening metallicity gradient (Sec~\ref{sec:radial_migra}), can also result in the azimuthal variation in \SigSFR\ and \dOH.
We notice a comparable fluctuation in metallicity near (0.24~dex) and beyond (0.25~dex) the CR, which is twice as pronounced as the fluctuation within the CR (13.8~kpc).
This observation supports the influence of tidal interactions, by comparison with the simulated model in \citet{Spitoni_2019}.

According to the asymmetric H{\sc{i}} distribution \citep{Elagali_2019}, NGC~1566 is possibly experiencing ram-pressure interaction with the intergalactic medium (IGM) in the Dorado cluster. 
\citet{Slater_2019} report the strong outflows observed in ionised and molecular gas in the central kpc along the bar and the spiral arms.
Our observations show generally higher \SigSFR\ and higher metallicity in the trailing edge (\dphi\ > 0) of the spiral arms.
This evidence supports that the spiral arms in NGC~1566 follow the density wave theory.
The strongest metallicity fluctuations near the CR (Fig~\ref{fig:1365_CR}) are consistent with the simulated galaxy with an analytic spiral arm.

NGC~2442 is a system undergoing a merger \citep{Mihos_1997, Pancoast_2010} and the south and north spiral arms are distorted fragments from the same galaxy (NGC~2442 and NGC~2443 respectively).
Our observations find significantly lower \SigSFR\ and lower metallicity in the trailing edge (\dphi\ $> 0$) of the spiral arms, which is unexpected by either density wave theory or dynamic spiral theory.
The spiral features in NGC~2442 are dominated by the gravity perturbation from the ongoing tidal interactions and the induced strong gas radial migration (Sec~\ref{sec:radial_migra}).

NGC 2835 is a multi-armed spiral galaxy in a small galaxy group \citep{Anand_2021}, with ESO 497-035 and ESO 565-001.
The flattened metallicity radial profile, happening at $\sim$ 7~kpc, indicates the gas accretion from the circumgalactic medium to the outskirts of the galaxy \citep{Chen_2023, Garcia_2023}.
We observe negligible offset in \SigSFR\ and 12 + log(O/H), given the scatters of the spaxels.
The $D$-value is not significant while the KS test and AD test suggest the metallicity on both sides of the spiral arms are drawn from different distributions.
This result suggests the spiral arms in NGC~2835 follow the dynamical spiral theory, potentially under the density perturbation from the accreted gas, flattening the metallicity gradient in the outskirts (Table~\ref{tab:zgradient}).

NGC 2997 belongs to a loose galaxy group and has undergone tidal interaction according to the anomalous H{\sc i} distribution \citep{Hess_2009}.
We observe generally no offset between both sides of the spiral arms, either in \SigSFR\ or 12 + log(O/H).
This suggests that tidal interaction does not necessarily lead to azimuthal variation in \SigSFR\ and 12 + log(O/H).
The tidal-induced spiral arms in NGC~2997 behave similarly to dynamic spiral arms, instead of density-wave-like structures.

NGC 4536 is located in the Virgo cluster without evident hints of tidal interaction. 
The only kinematic perturbation is the bar-induced inflows observed in the H{\sc i} map \citep{Davies_1997}.
We observe slightly lower \SigSFR\ and lower metallicity in the trailing edge (\dphi\ $> 0$).
However, the CDF diagram shows no offset between the downstream and upstream.
The $p$-value (3.78 $\times\ 10^{-1}$) from the KS test suggests that the metallicity on both sides of the spiral arms is drawn from the same parent distribution.
Therefore, we conclude that the current TYPHOON data suggest NGC~4536 follows the dynamic spiral theory, with insufficient spaxels to discern the azimuthal variations.

NGC 5236 (M83) is the largest member in its galaxy group.
There is a large optically detected tidal stream to the north of M83 \citep[][]{Malin_1997, Pohlen_2004, Jarrett_2013}, tracing the disruption of a dwarf galaxy in the strong gravitational field of M83.
The penetrating gas stream may be attributed to the flattening metallicity gradient beyond 5.37~kpc (Table~\ref{tab:zgradient}).
Unlike studies on star clusters \citep[][]{Silva-Villa_2012, Bialopetravicius_2020, Abdeen_2022} and full-spectral fitting (Sextl et al. In Preparation), we observe no significant offset in either \SigSFR\ or 12 + log(O/H) between the two sides of the spiral arms. Our observations do not support the density wave theory but rather show a preference for the dynamic spiral theory.

As a type-{\sc ii} Seyfert galaxy \citep{Simpson_1997}, NGC~5643 is in a small galaxy group with NGC~5530 and has a dwarf satellite ESO 273-014.
After excluding the hard component contaminated spaxels, we find slightly higher \SigSFR\ in part of the trailing edge (\dphi\ > 70) and lower metallicity in the trailing edge (\dphi $> 0$).
However, the CDF diagram shows no offset and the KS test agrees that the metallicity in the downstream and upstream is drawn from the same distribution.
This finding implies that no statistical offset is found in NGC~5643 based on the TYPHOON data, due to the limited spaxel in the inter-arm regions (especially \dphi $< 0$).
We agree that NGC~5643 follows the dynamic spiral theory, instead of the density wave theory.

NGC 6744 is a spiral galaxy in the Virgo supercluster.
The H{\sc i} in NGC~6744 is possibly connected to a companion galaxy, ESO~104$-$g44 \citep{Ryder_1999}.
We observe a slightly increasing trend in \SigSFR\ and \dOH\ when crossing the spiral arms from the trailing edge (\dphi\ > 0) to the leading edge (\dphi\ < 0).
However, the difference in \dOH\ is not statistically evident in the CDF and the KS test.
Our findings suggest that under environmental influences, NGC~6744 shows a preference for dynamic spiral theory with absent azimuthal variation.

\section{Summary}
We map the 2D ISM properties, \SigSFR\ and gas-phase metallicity, of nine spiral galaxies in the TYPHOON survey.
The 3D dataset and wide FoV, covering most of the star-forming disc of each galaxy, allows us to constrain spaxel-by-spaxel fluctuations in ISM properties as a function of azimuthal distance from the spiral arms in each galaxy.
These azimuthal distributions constrain the dominant mechanism driving the spiral arms, which can assess the density wave theory.

Considering the azimuthal ISM distribution observed in TYPHOON and the environment reported in previous works (Sec~\ref{sec:dominant}), 
we discuss the dominant theory/theories that drive the spiral features in our samples.
We find higher \SigSFR\ (Fig~\ref{fig:dphi_dsfr}) and higher gas-phase metallicity (Fig~\ref{fig:dphi_dz}) in the trailing edge of NGC~1365 and NGC~1566, which is in line with expectations from density wave theory driving the observed spiral features in these two galaxies.
Additionally, the higher \SigSFR\ in the trailing edge indicates that star formation occurs when gas clouds approach the density wave \citep[right spiral arms in Fig 1 of][]{Pour-Imani_2016} in NGC~1365 and NGC~1566.
The interacting galaxy, NGC~2442, presents significantly lower metallicity in part of the trailing edge (\dphi $< 50$), opposite to the expectation of density wave theory, which can be attributed to the ongoing merging event.
We do not observe statistically significant offset in the inter-arm regions of the remaining six galaxies, in line with the prediction of dynamic spiral theory.

We investigate the global properties of galaxies and the $D$-values from metallicity CDFs, indicating the significance of azimuthal variations.
We find that more prominent spiral arms, more open-armed galaxies and more extended galaxies tend to show stronger azimuthal variations in metallicity.

We collect the co-rotation radius (CR) from the earlier works and compare the azimuthal variation in metallicity within and beyond the CR (Sec~\ref{sec:CR}).
In NGC~1365, we find the smallest metallicity fluctuation inside the CR and comparable fluctuation near and outside the CR.
This is consistent with a simulated density wave spiral galaxy under gravity perturbation in the outer region \citep{Spitoni_2019}.
In NGC~1566, we observe the greatest metallicity fluctuation near the CR, aligning with an analytic spiral arm where gravity perturbation is regular.

Our work highlights the importance of azimuthal variations in ISM, which constrain the dominant mechanism driving spiral features.
Different theories, including the density wave theory, dynamic spiral theory and tidal interactions, can explain the formation of spiral arms in various galaxies in the local Universe.
Despite our handful of galaxy samples, we observe a positive relation between azimuthal metallicity variation and T-type, galaxy size, and arm strength.
A larger sample of observations and a comparison between simulations and observations are essential for more constraints on the dominant mechanism driving spiral arms.

\section*{Acknowledgements}
We gratefully acknowledge the anonymous referee for their highly constructive and valuable comments, which have significantly improved the scientific quality of this paper.
This paper includes data obtained with the du Pont Telescope at the Las Campanas Observatory, Chile, as part of the TYPHOON programme, which has been obtaining optical data cubes for the largest angular-sized galaxies in the Southern Hemisphere. We thank past and present Directors of The Observatories and the numerous time assignment committees for their generous and unfailing support of this long-term programme.

This research has made use of NASA's Astrophysics Data System Bibliographic Services (ADS). 
This research made use of Astropy\footnote{\href{http://www.astropy.org}{http://www.astropy.org}}, a community-developed core Python package for Astronomy \citep{astropy13, astropy18}. 
This research has made use of the NASA/IPAC Extragalactic Database (NED) which is operated by the Jet Propulsion Laboratory, California Institute of Technology, under contract with NASA.

KG is supported by the Australian Research Council through the Discovery Early Career Researcher Award (DECRA) Fellowship DE220100766 funded by the Australian Government. 
KG is supported by the Australian Research Council Centre of Excellence for All Sky Astrophysics in 3 Dimensions (ASTRO~3D), through project number CE170100013. 

ES acknowledges support by the Munich Excellence Cluster Origins funded by the Deutsche Forschungsgemeinschaft (DFG, German Research Foundation) under Germany’s Excellence Strategy EXC-2094 390783311.

\section*{Data Availability}
The TYPHOON team is planning for a public data release on Data Central\footnote{\href{https://datacentral.org.au/}{https://datacentral.org.au/}} once the observation is complete.
The TYPHOON data and the code in this article will be shared on reasonable request to the corresponding author.


\bibliographystyle{mnras}
\bibliography{example} 

\appendix
\section{\texorpdfstring{\SigSFR}{Sigma SFR}\ maps and \texorpdfstring{\dSFR}{Delta Sigma SFR}\ maps}\label{appendix_sfr}
The maps of \SigSFR\ and \dSFR\ are shown in Fig~\ref{appendix_sfr}.
\begin{figure*}
    \centering
    \includegraphics[width=0.23\textwidth, height=1.8in]{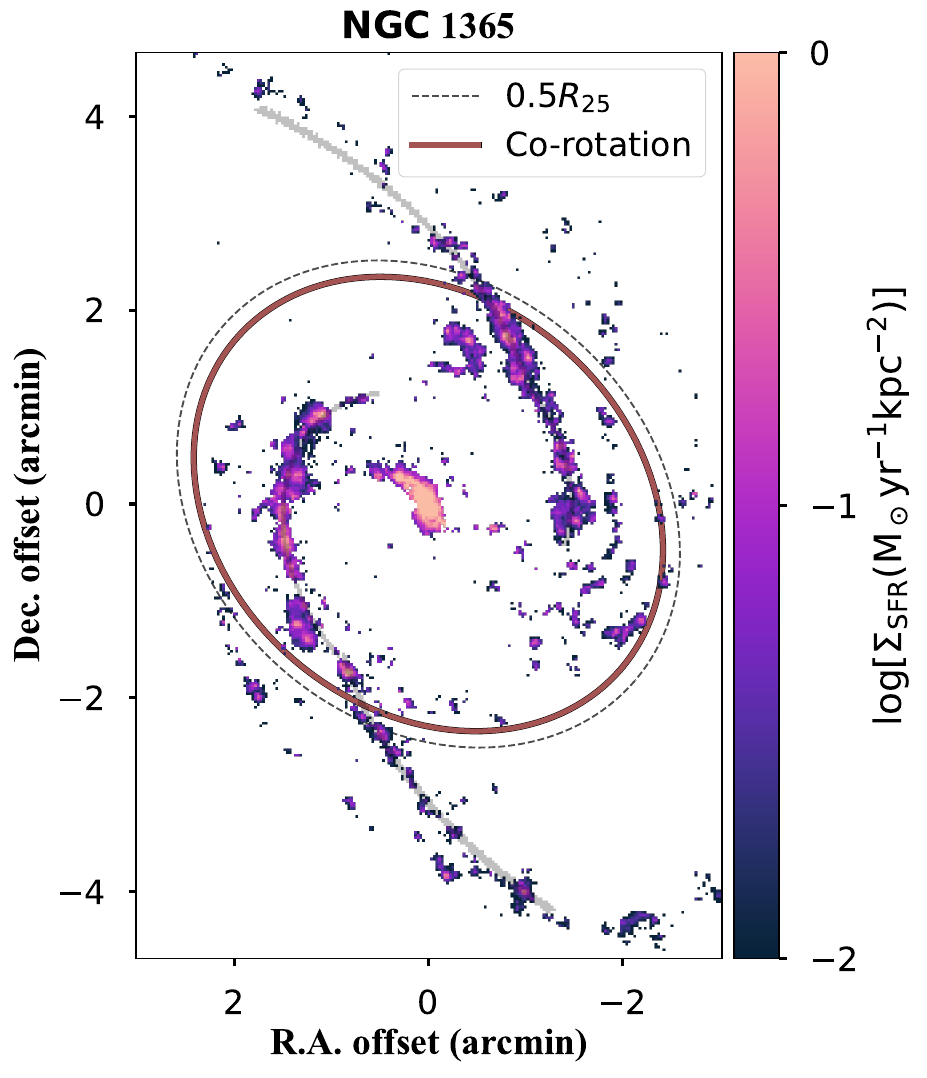}
    \includegraphics[width=0.23\textwidth, height=1.8in]{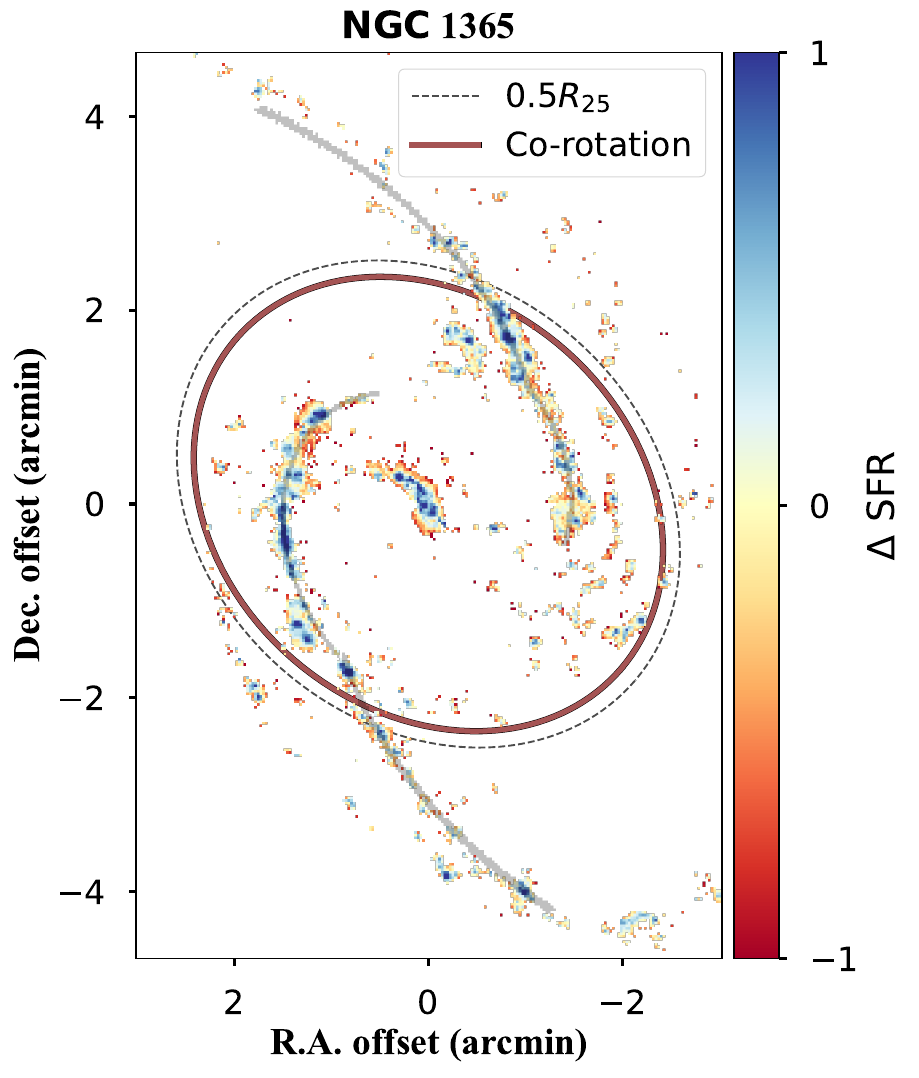}
    \includegraphics[width=0.23\textwidth, height=1.8in]{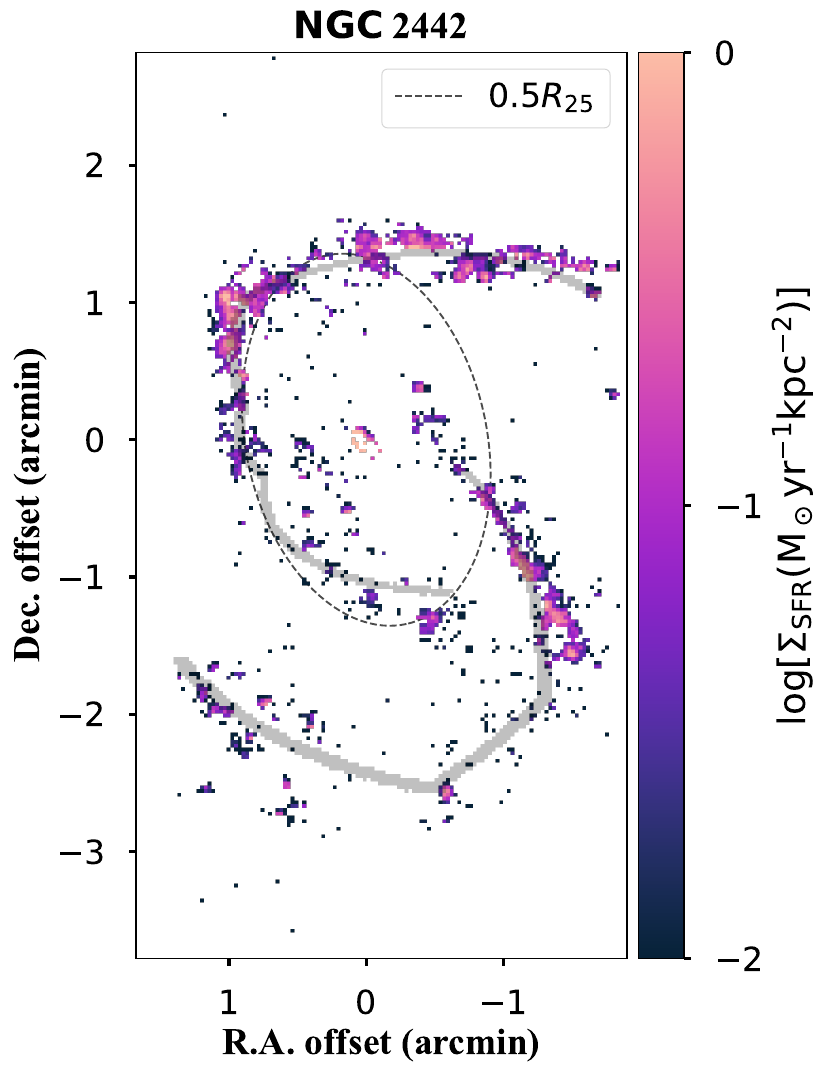}
    \includegraphics[width=0.23\textwidth, height=1.8in]{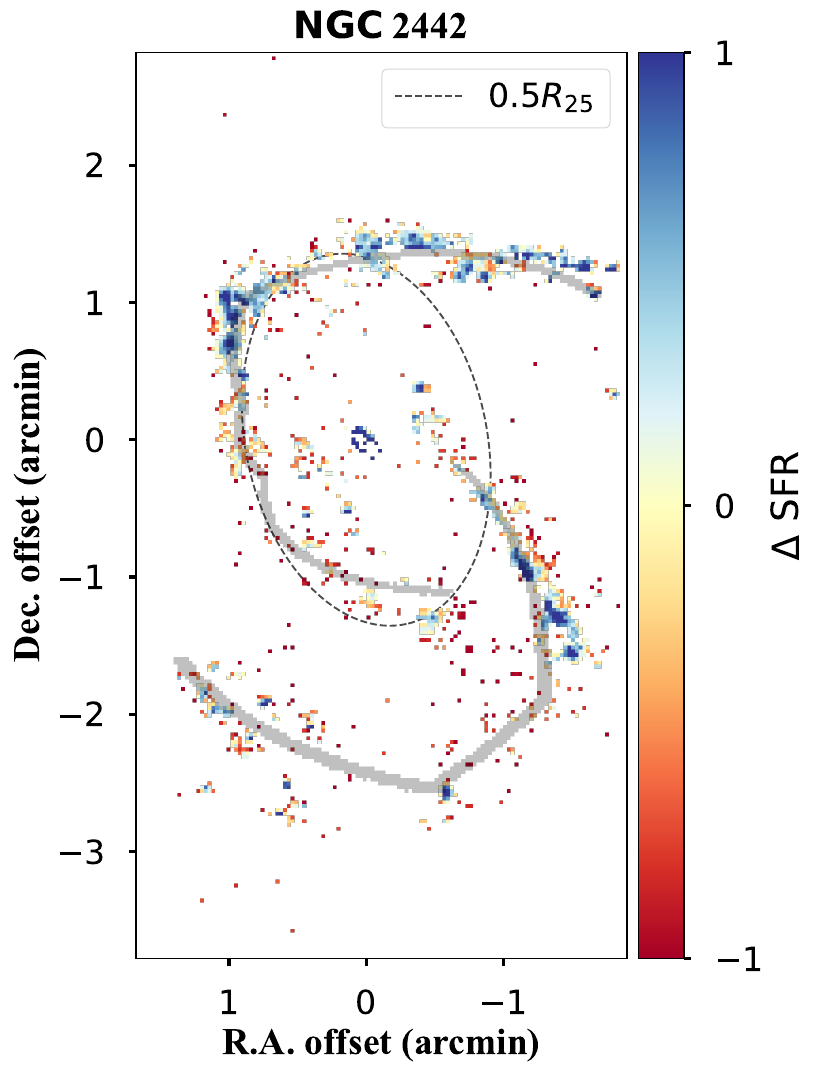}

    \includegraphics[width=0.23\textwidth, height=1.8in]{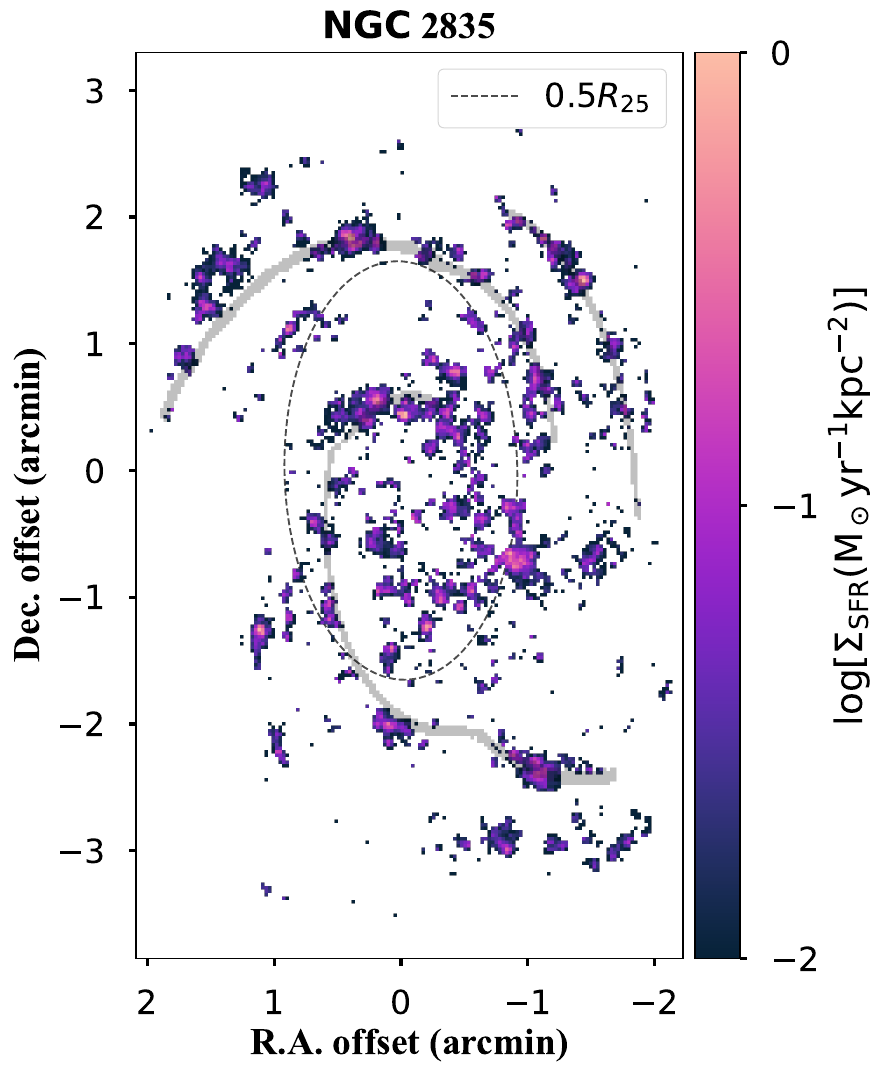}
    \includegraphics[width=0.23\textwidth, height=1.8in]{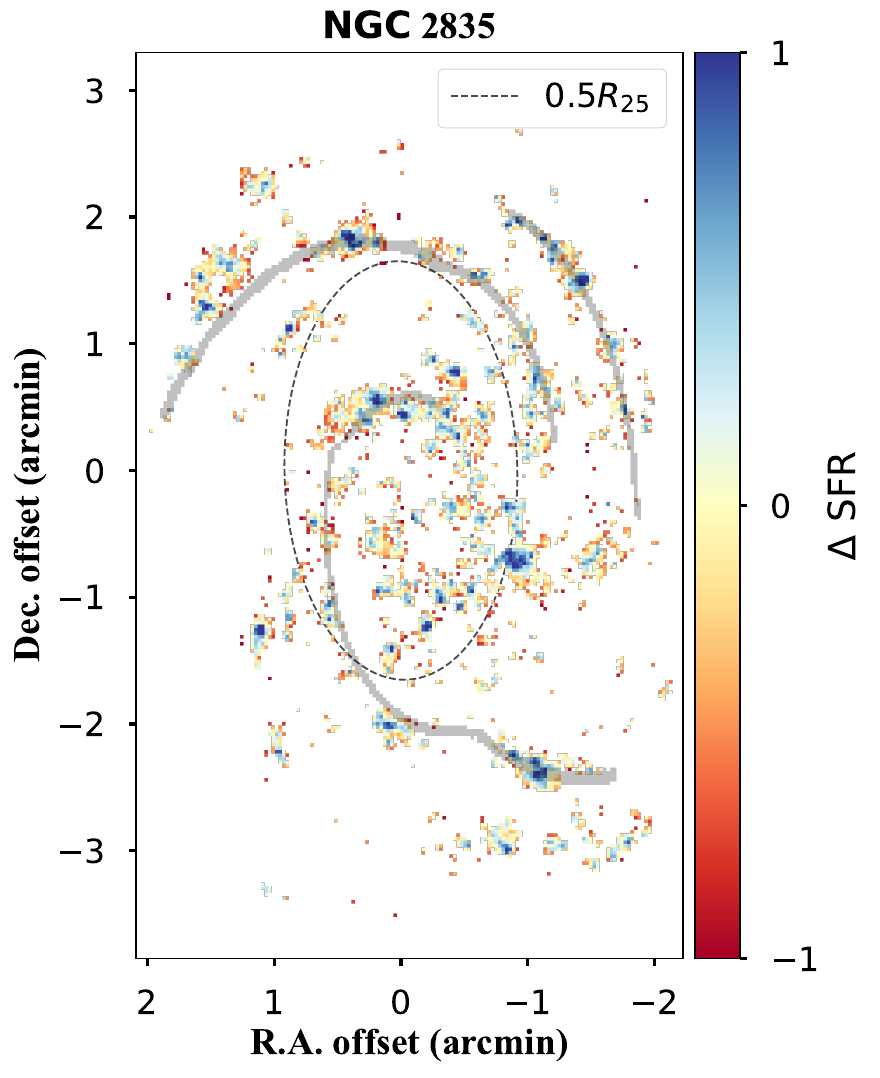}
    \includegraphics[width=0.24\textwidth, height=1.8in]{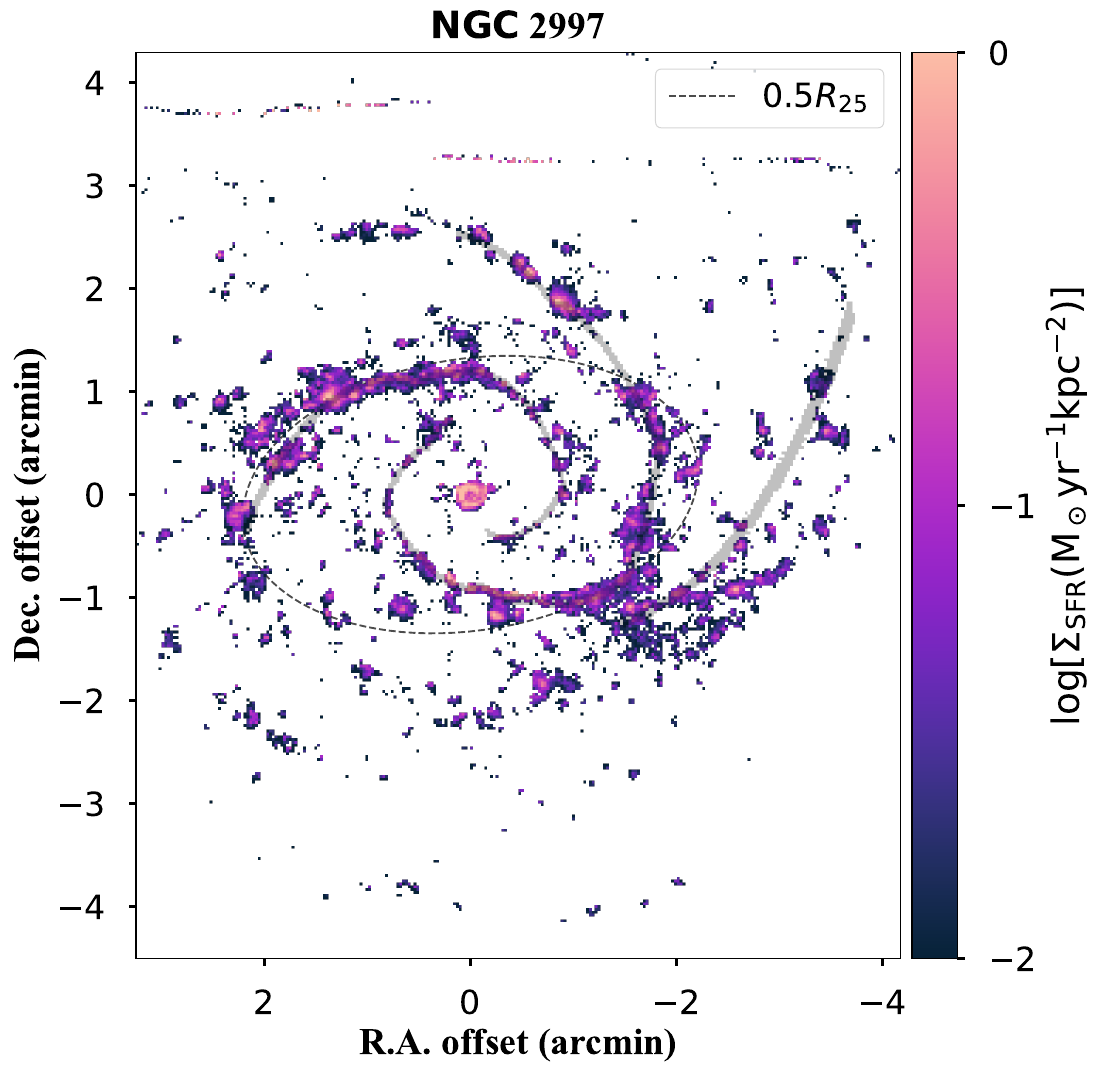}
    \includegraphics[width=0.24\textwidth, height=1.8in]{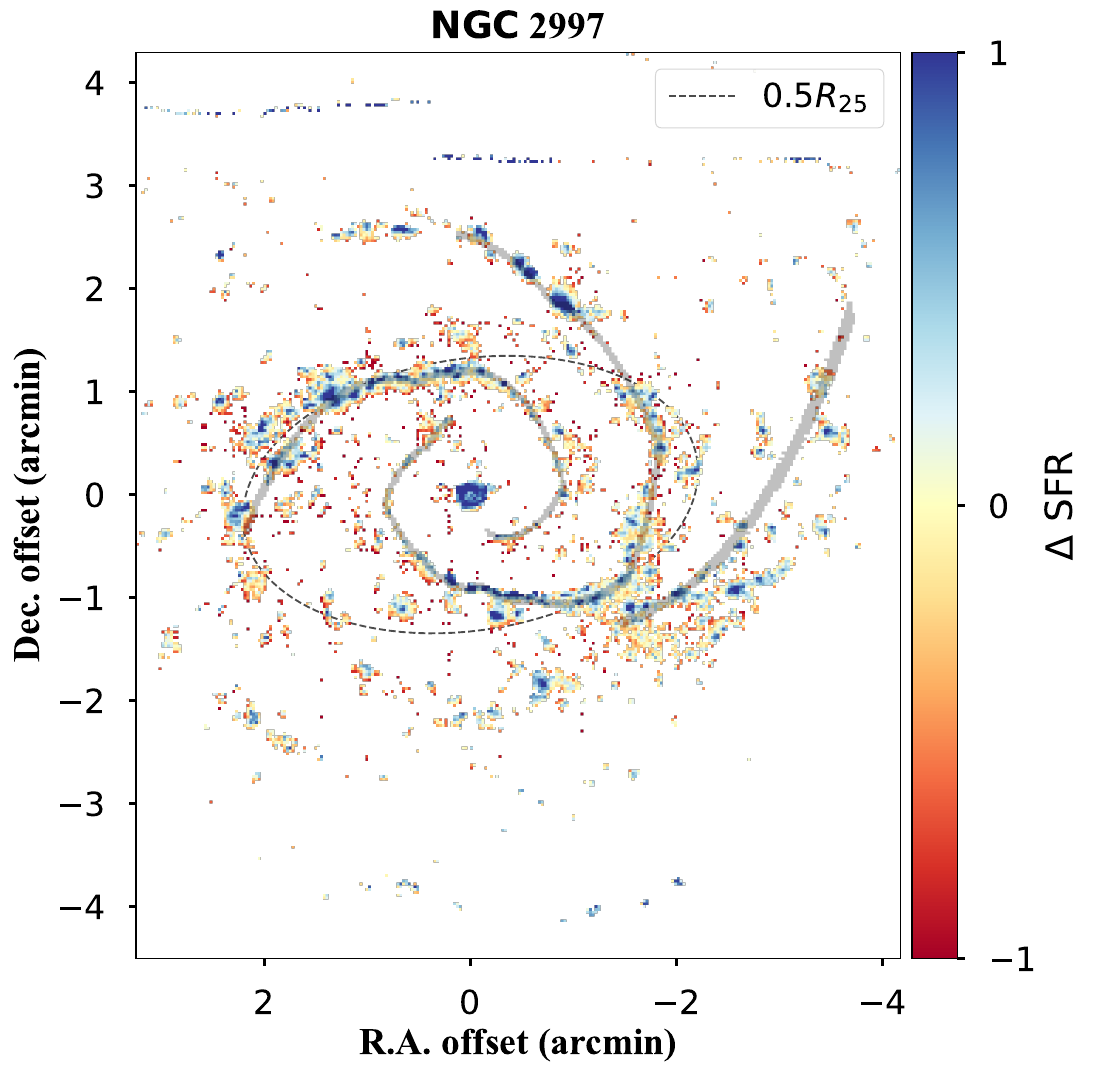}
    
    \includegraphics[width=0.20\textwidth, height=1.8in]{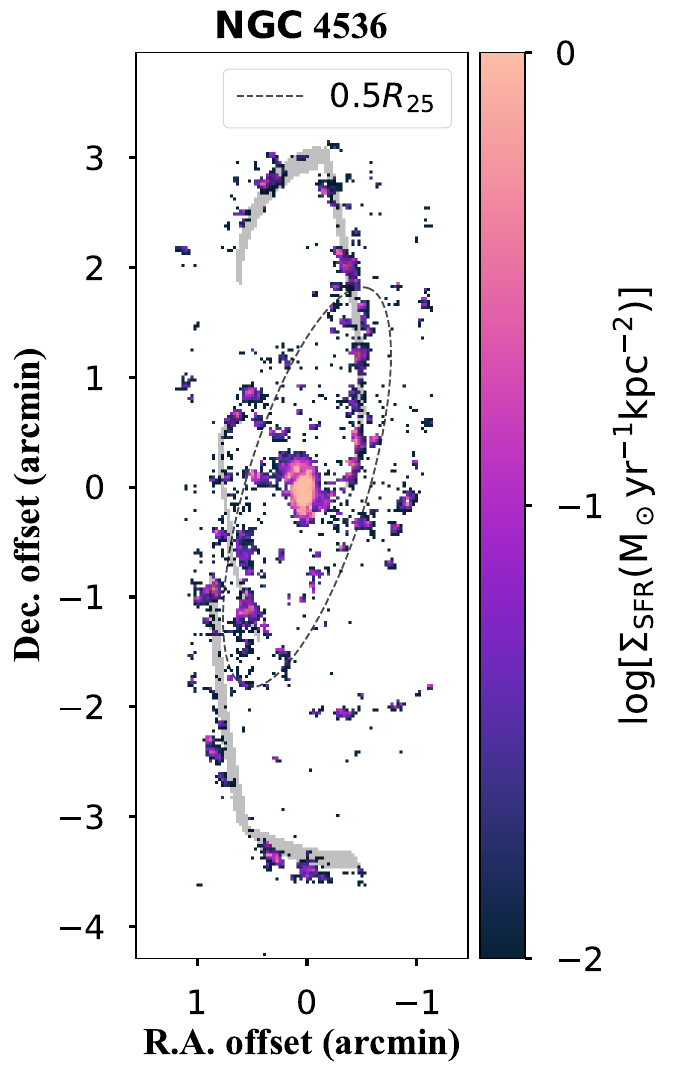}
    \includegraphics[width=0.20\textwidth, height=1.8in]{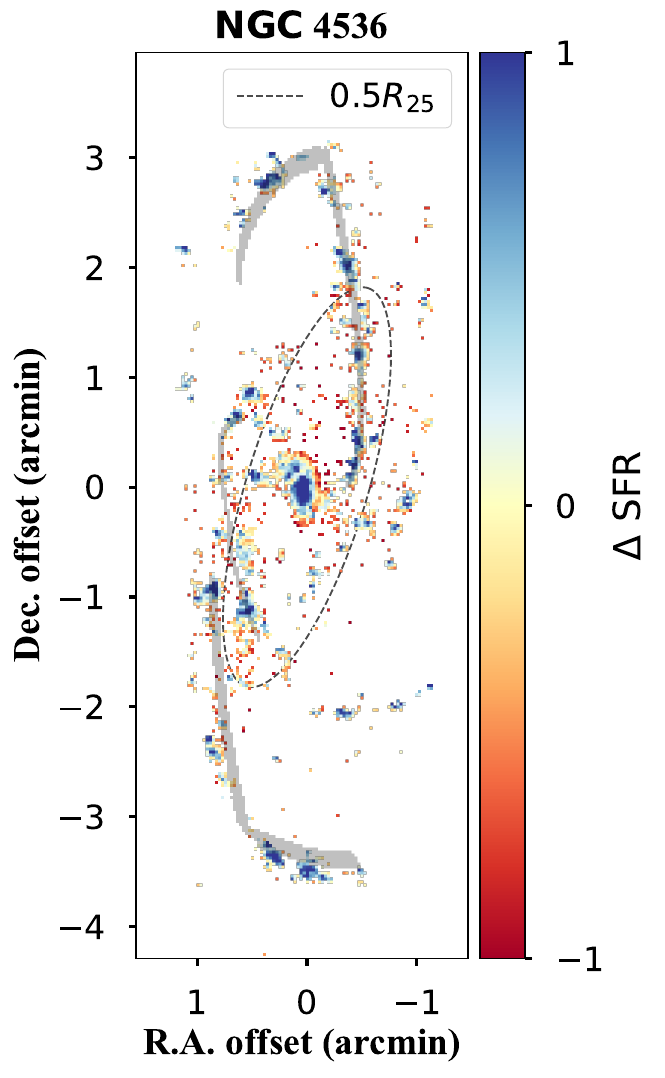}
    \includegraphics[width=0.23\textwidth, height=1.8in]{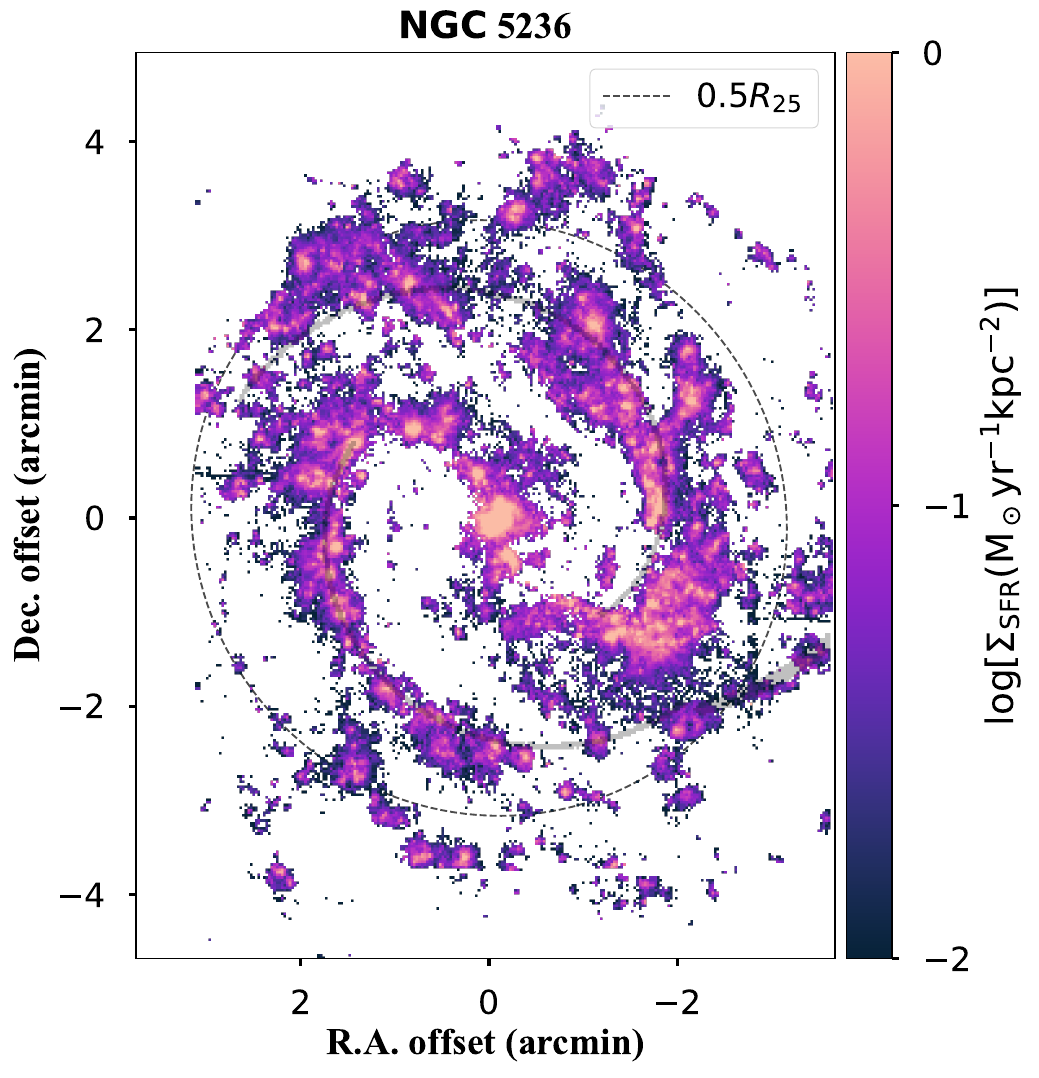}
    \includegraphics[width=0.23\textwidth, height=1.8in]{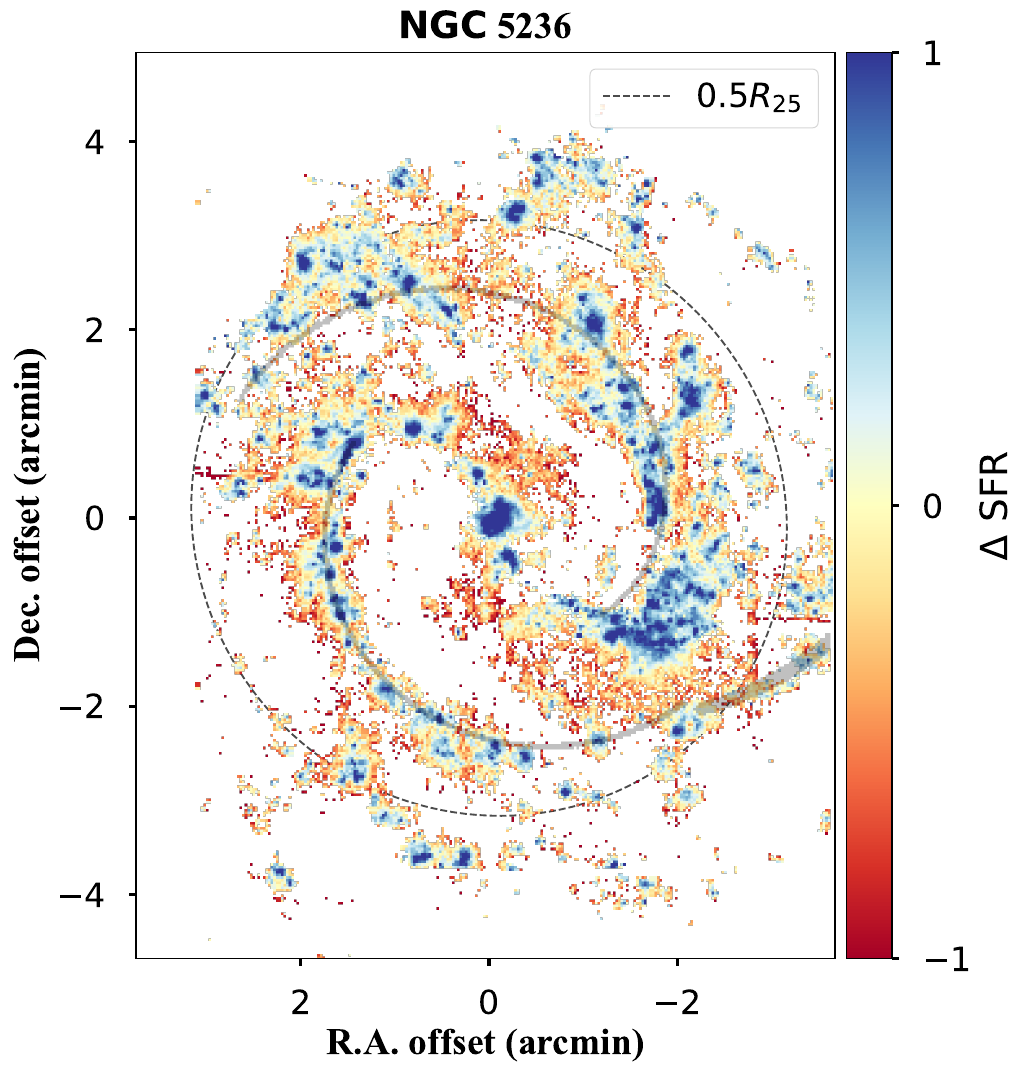}

    \includegraphics[width=0.20\textwidth, height=1.8in]{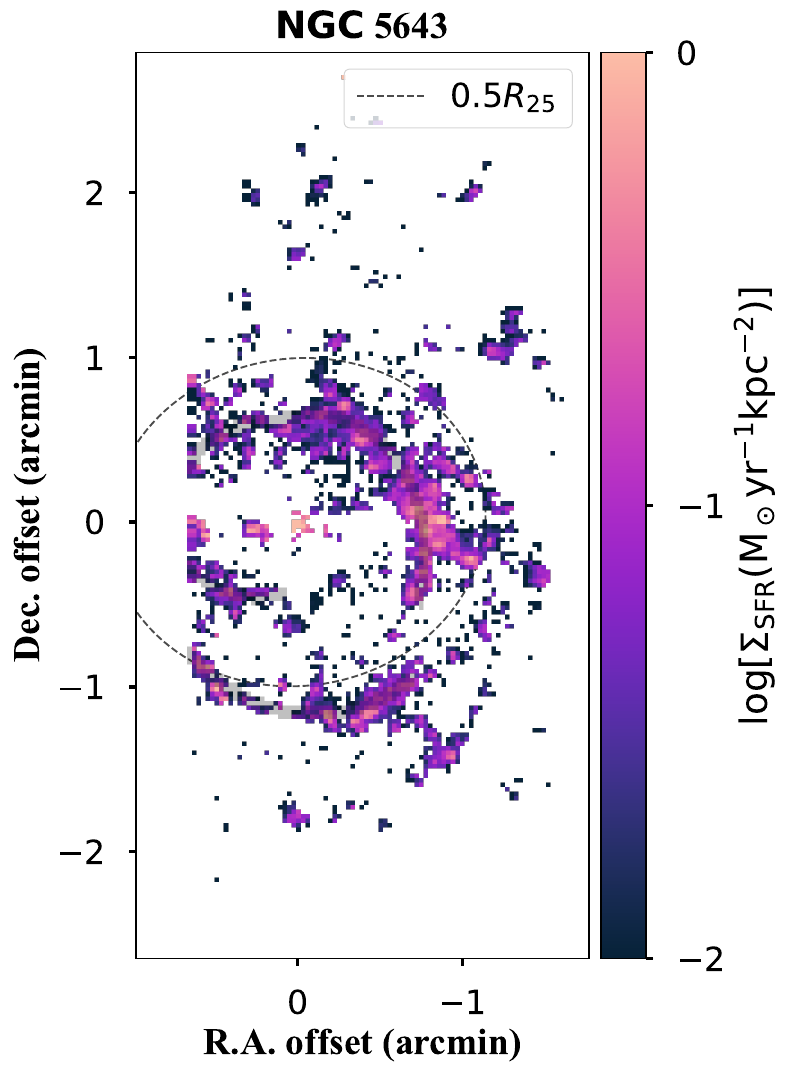}
    \includegraphics[width=0.20\textwidth, height=1.8in]{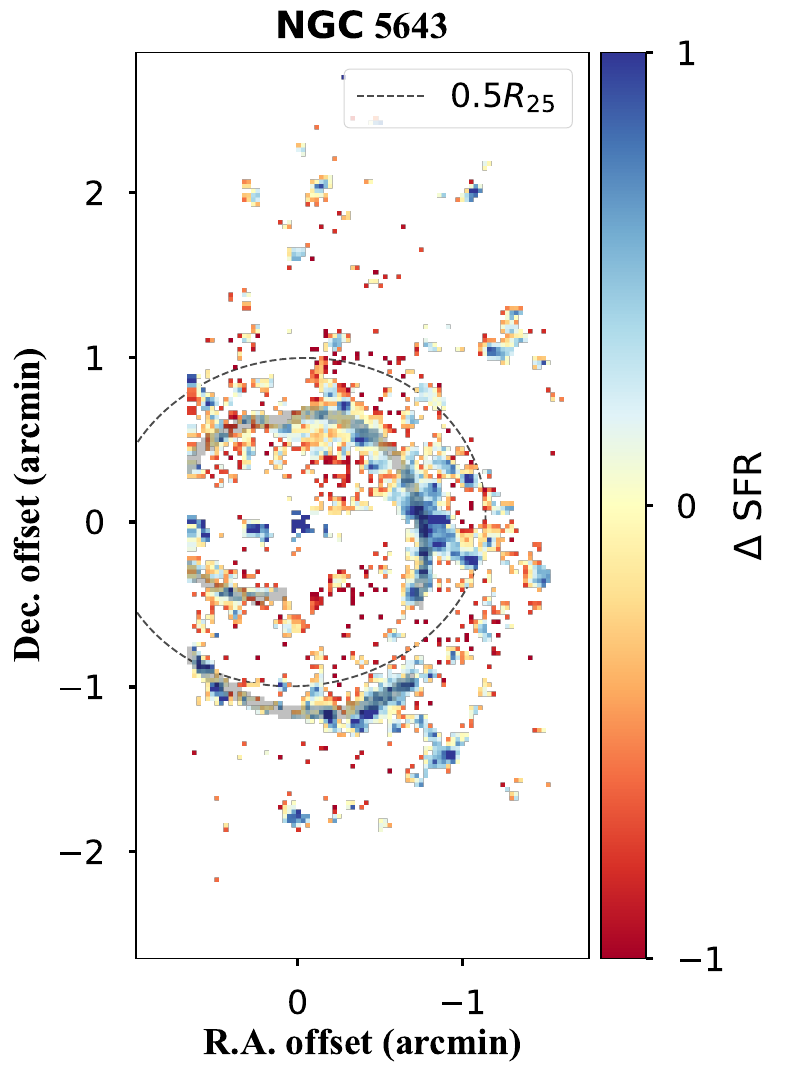}
    \includegraphics[width=0.2\textwidth, height=1.8in]{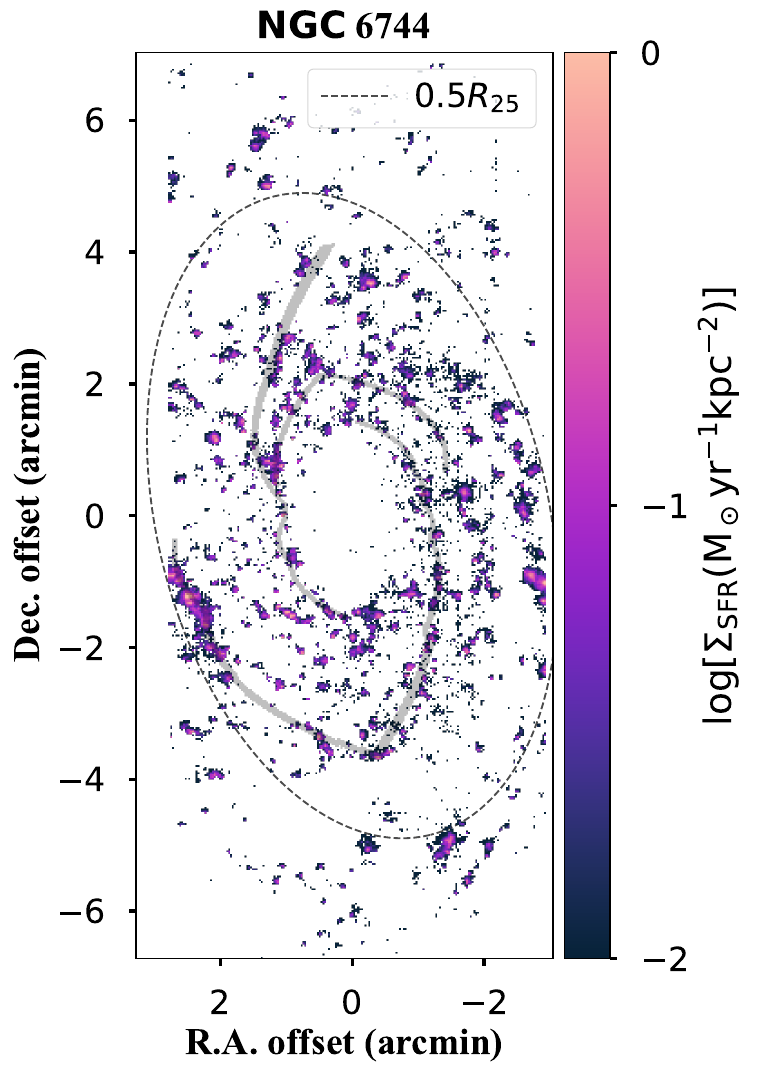}
    \includegraphics[width=0.2\textwidth, height=1.8in]{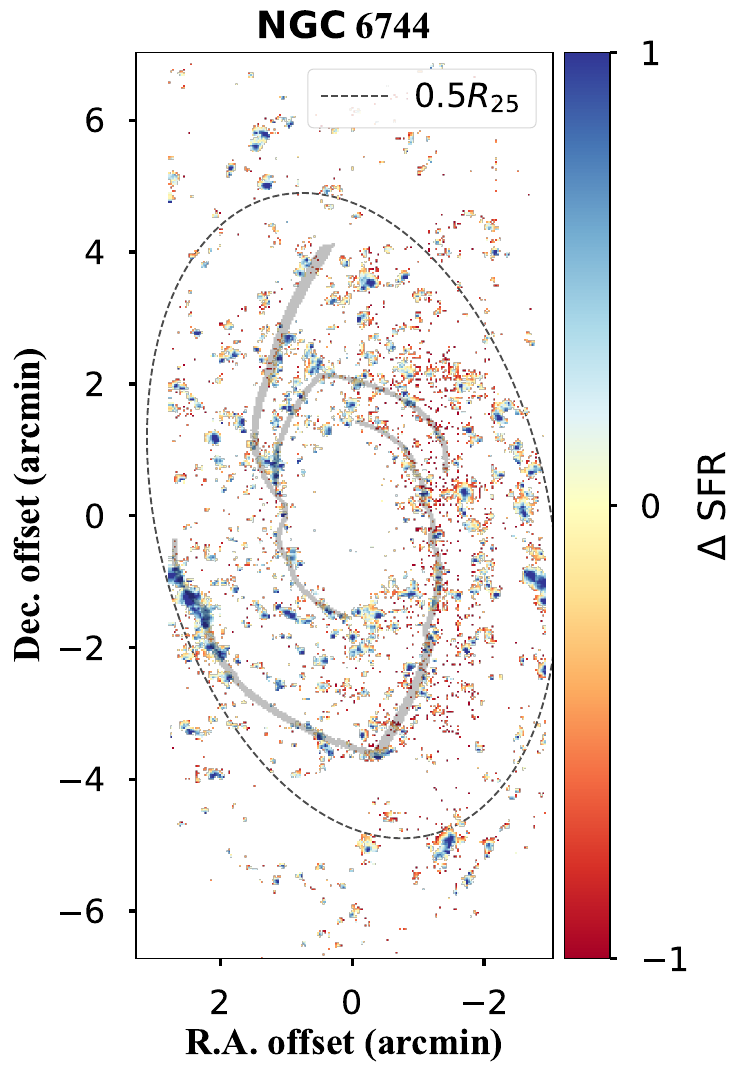}
    
    \caption{{\bf Column 1 \& 3}: 2D \SigSFR\ maps of our galaxies, except for NGC~1566 (Fig~\ref{fig:sfr_map}).
    {\bf Column 2 \& 4}: Residual of \SigSFR\ by subtracting the radial gradients.
    More detail on this analysis is in Sec~\ref{sec:ISM}.
    The dashed ellipse marks the location of half $R_{25}$ and the red ellipse denotes the CR, if applicable.}
\end{figure*}

\section{12 + log(O/H) maps and \texorpdfstring{\dOH}{Delta log O/H}\ maps}\label{appendix_Z}
The maps of 12 + log(O/H) and \dOH\ are shown in Fig~\ref{appendix_Z}.
\begin{figure*}
    \centering
    \includegraphics[width=0.23\textwidth, height=1.8in]{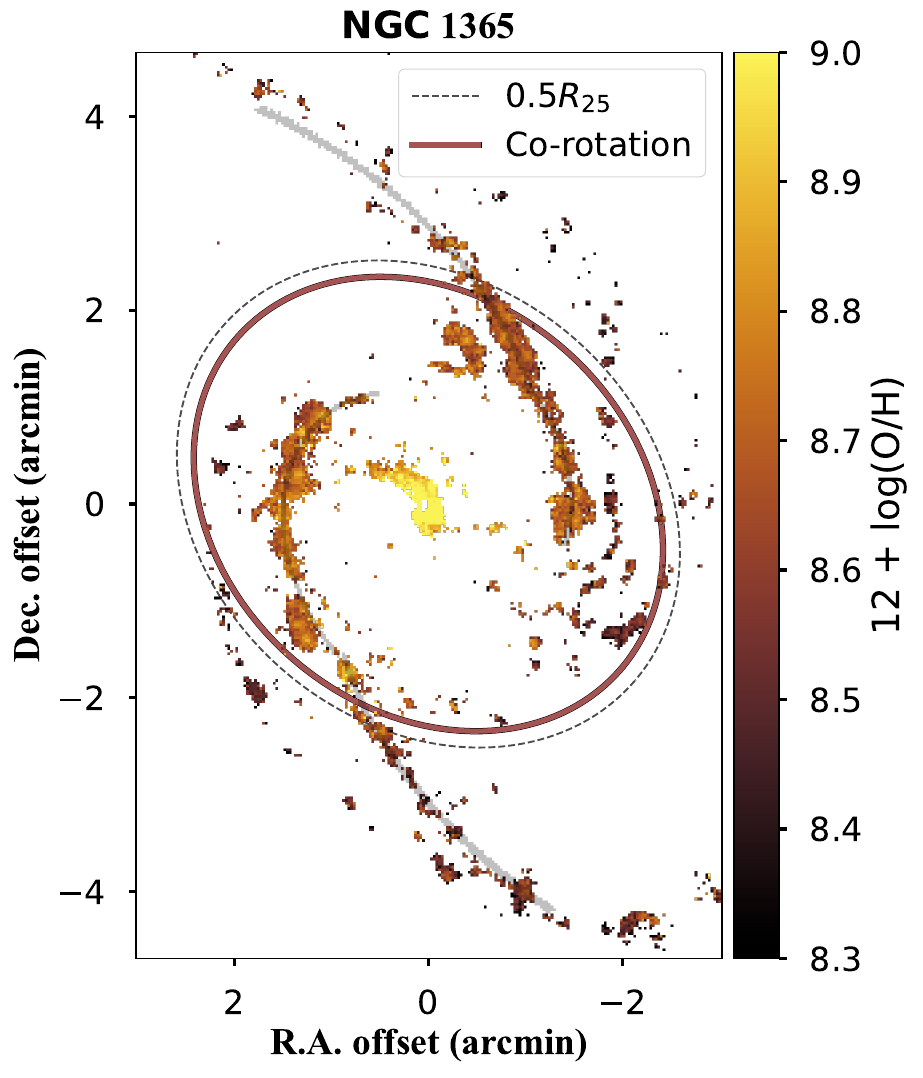}
    \includegraphics[width=0.23\textwidth, height=1.8in]{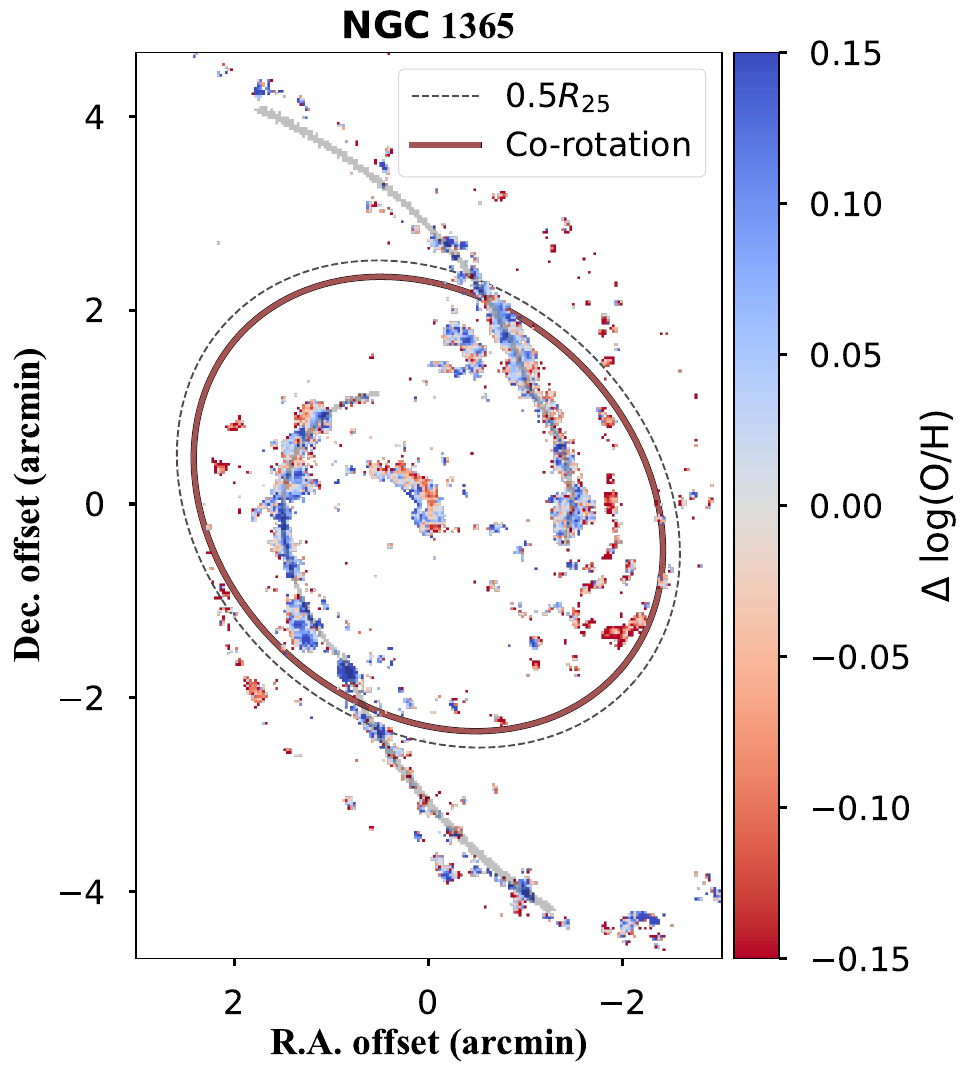}
    \includegraphics[width=0.23\textwidth, height=1.8in]{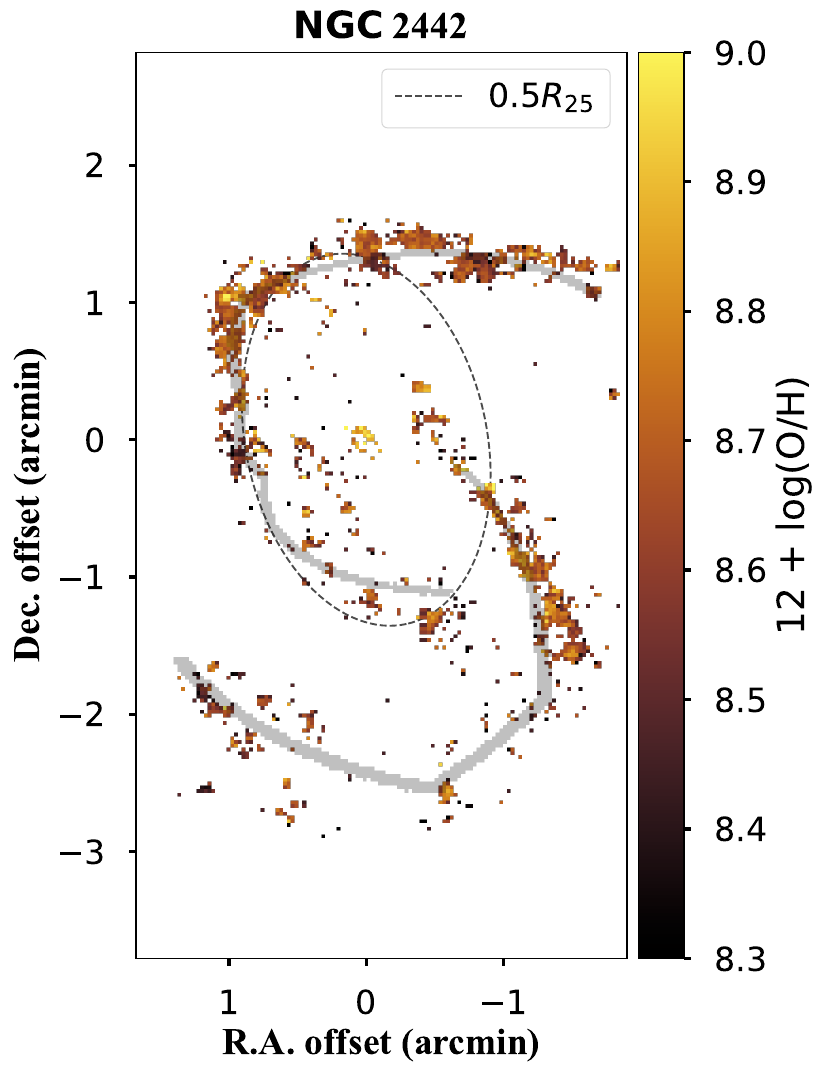}
    \includegraphics[width=0.23\textwidth, height=1.8in]{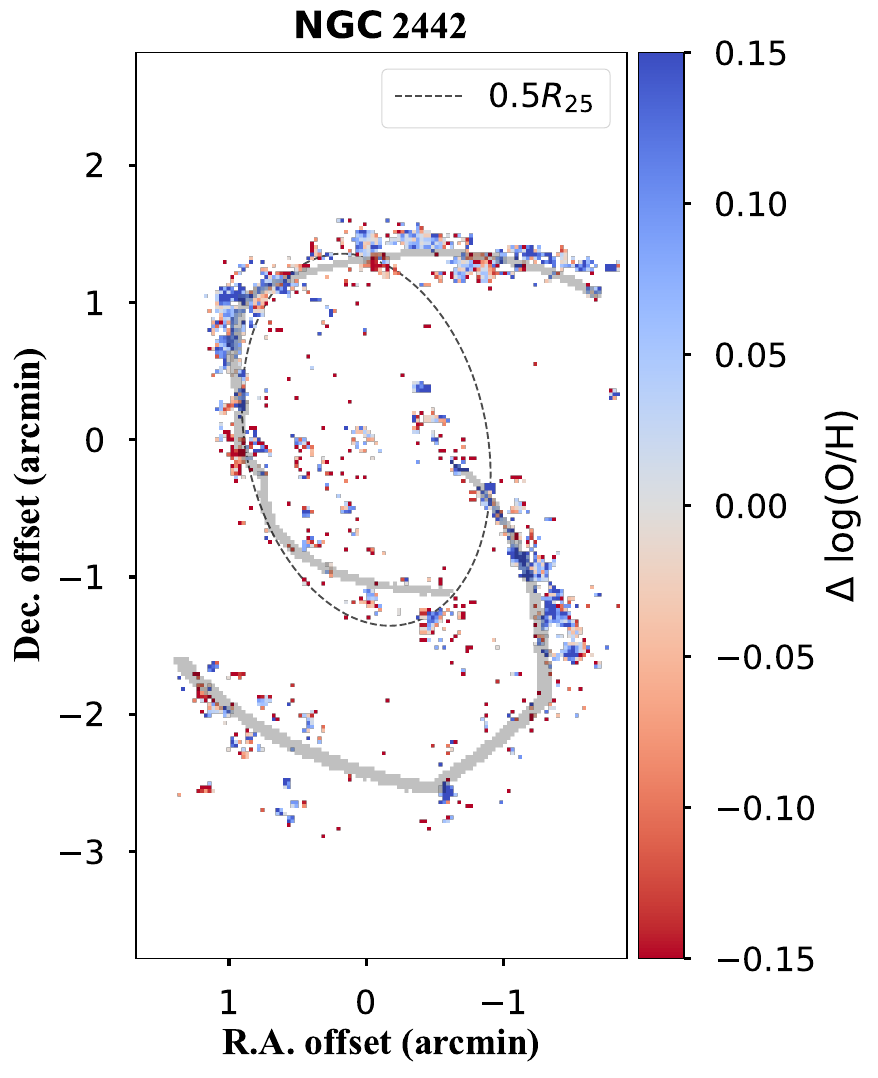}

    \includegraphics[width=0.23\textwidth, height=1.8in]{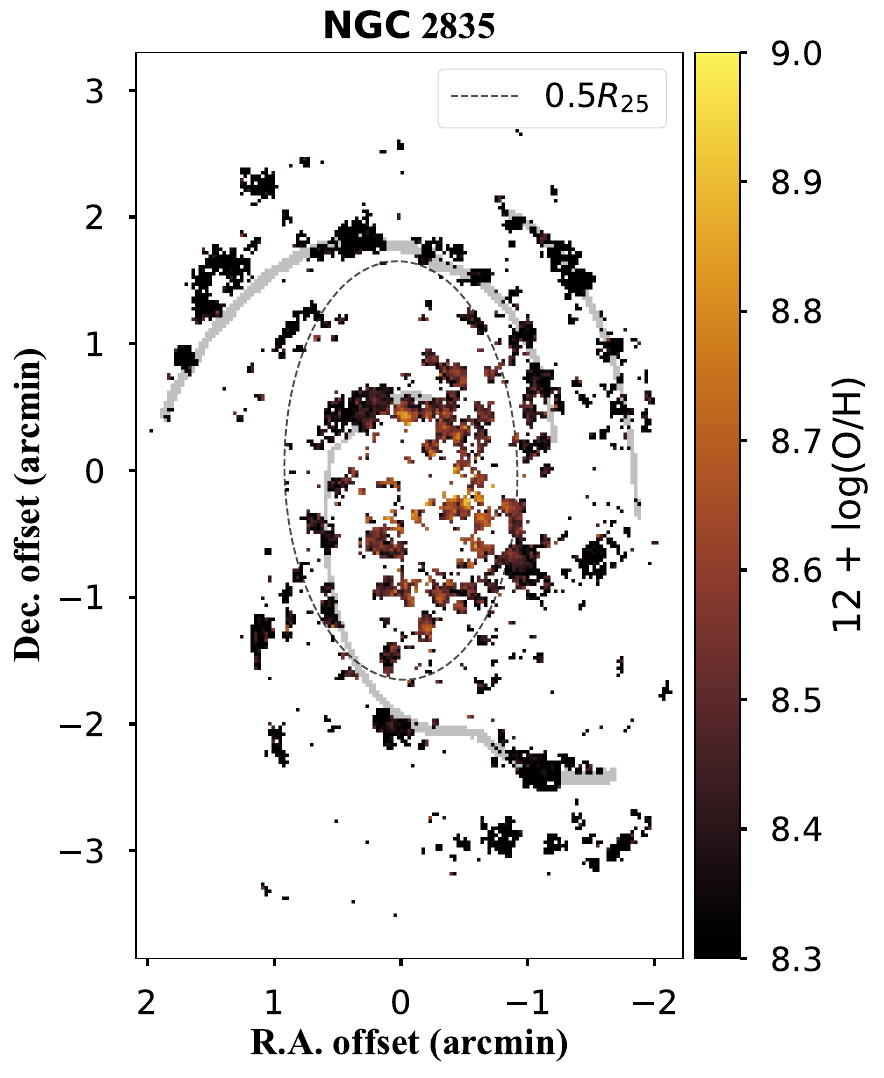}
    \includegraphics[width=0.23\textwidth, height=1.8in]{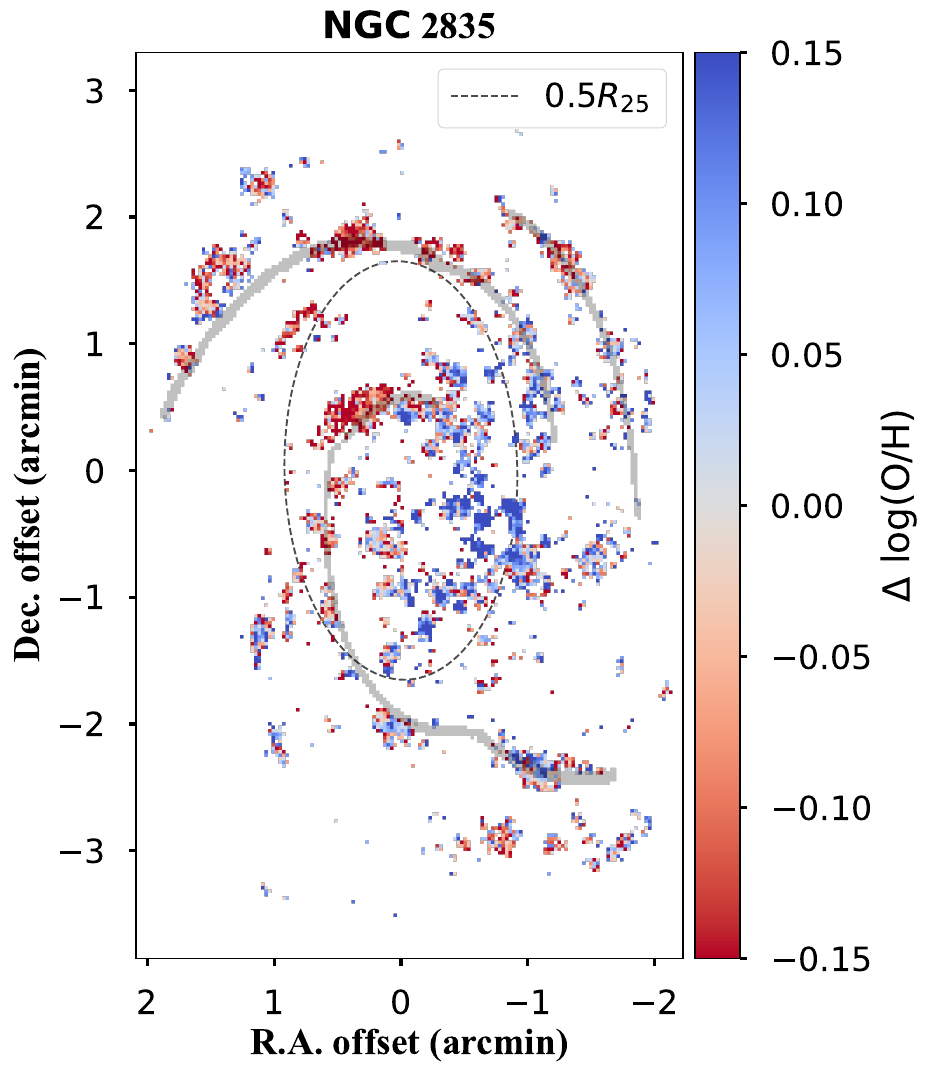}
    \includegraphics[width=0.24\textwidth, height=1.8in]{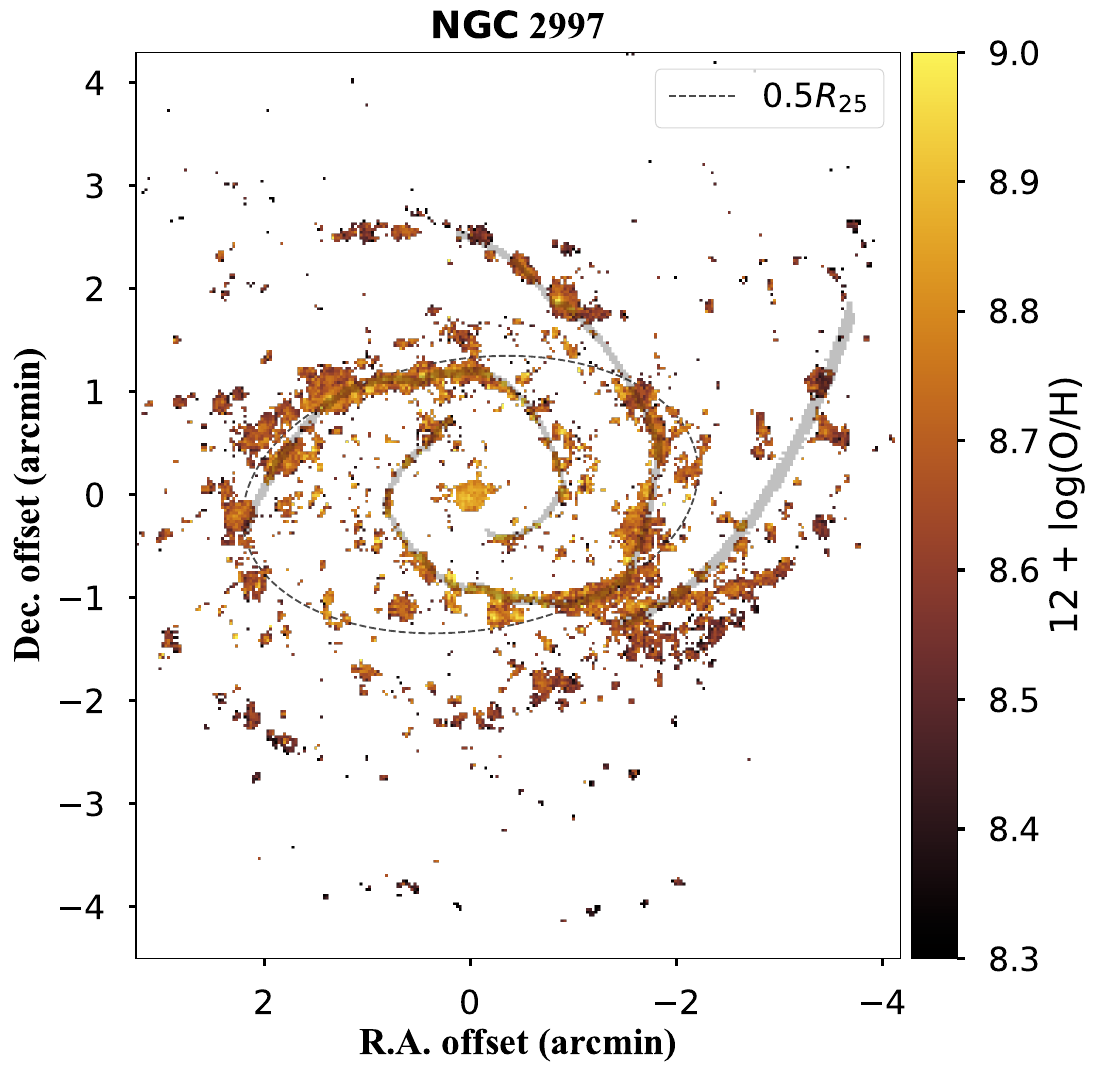}
    \includegraphics[width=0.24\textwidth, height=1.8in]{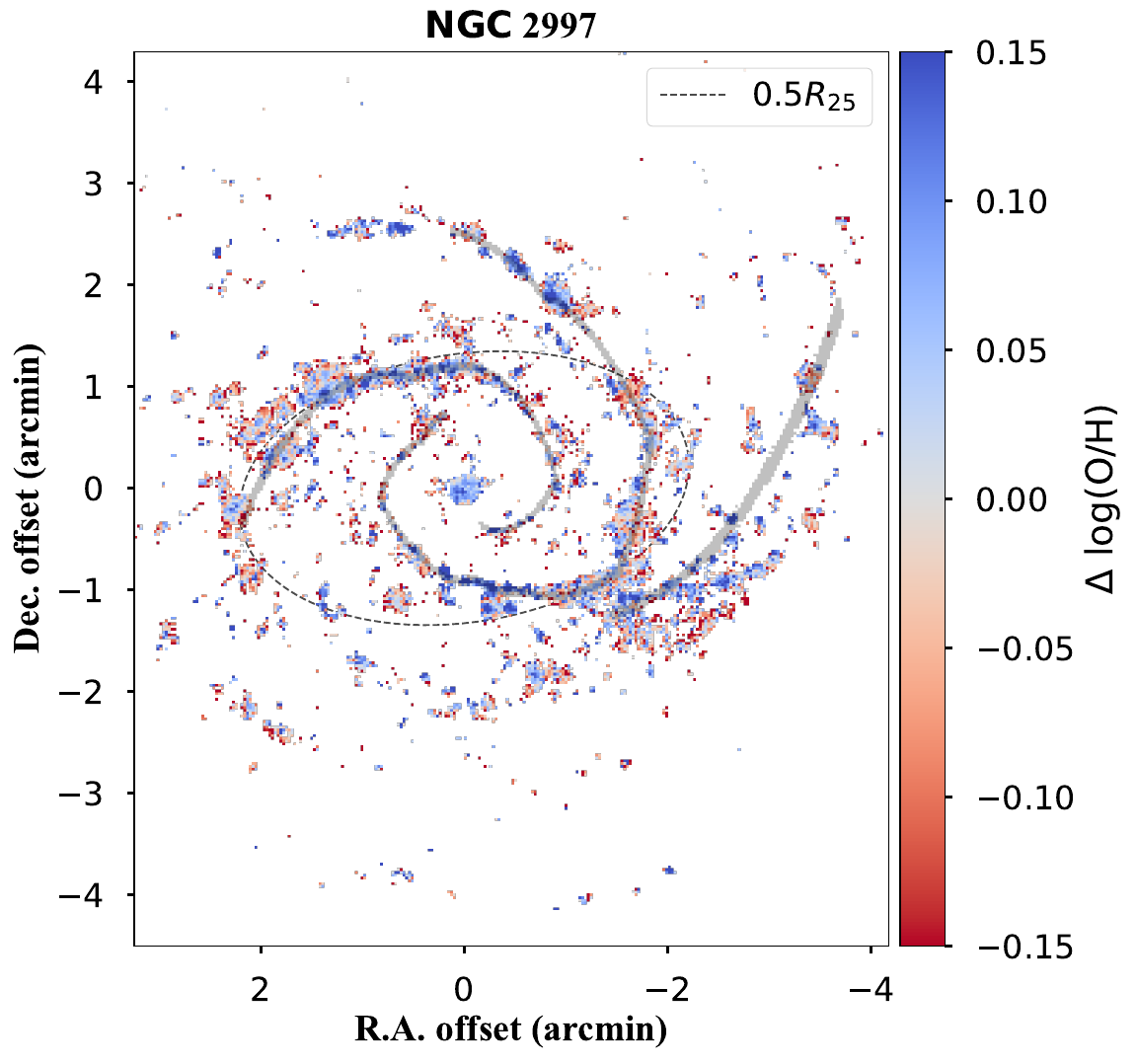}

    \includegraphics[width=0.2\textwidth, height=1.8in]{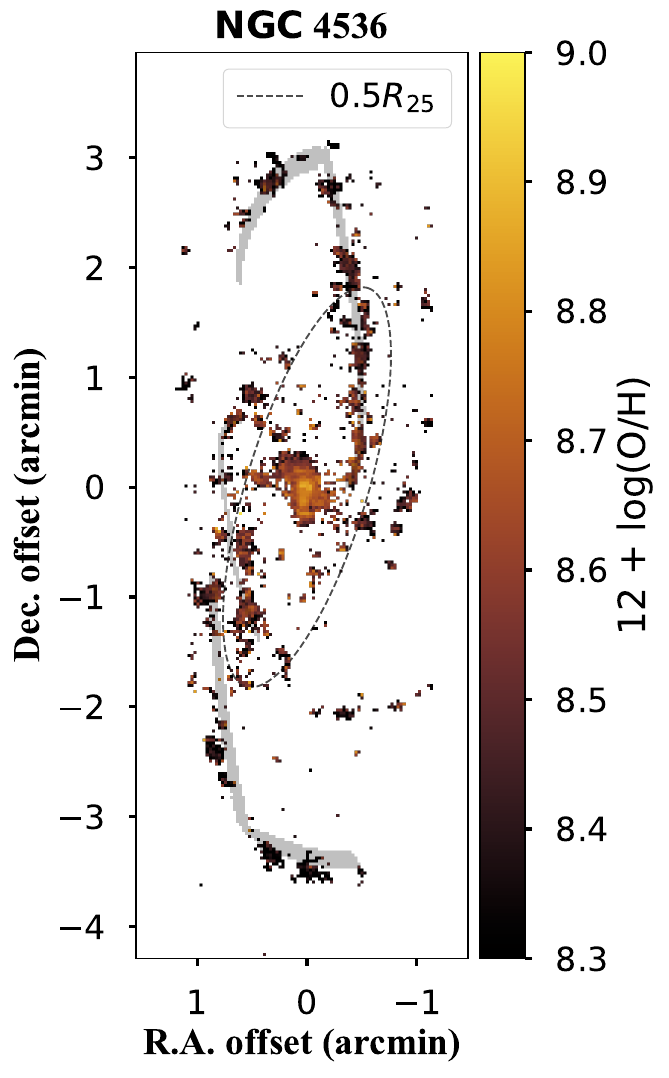}
    \includegraphics[width=0.2\textwidth, height=1.8in]{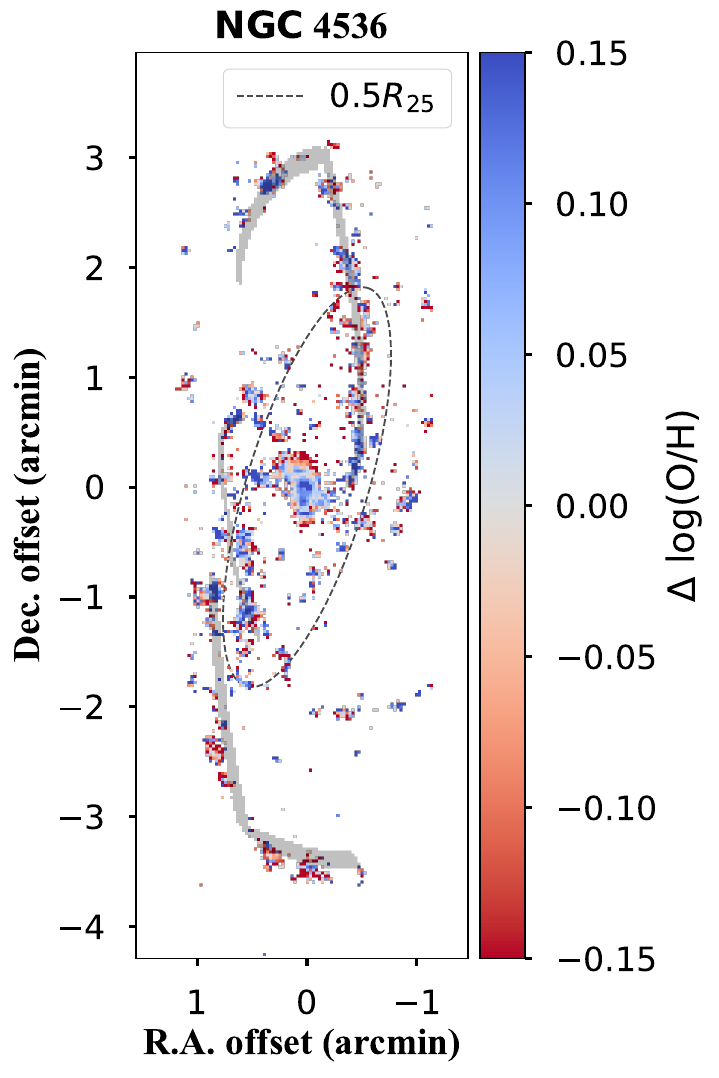}
    \includegraphics[width=0.23\textwidth, height=1.8in]{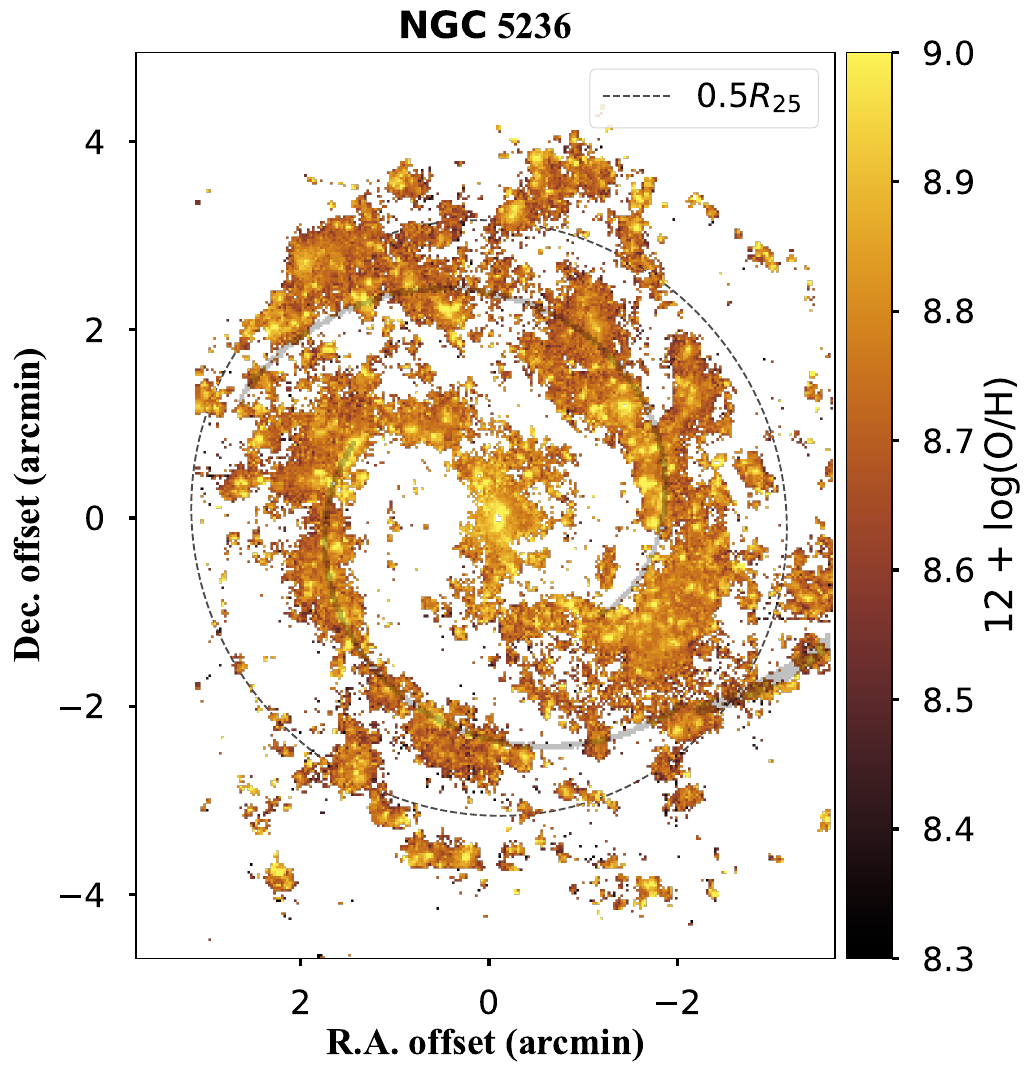}
    \includegraphics[width=0.23\textwidth, height=1.8in]{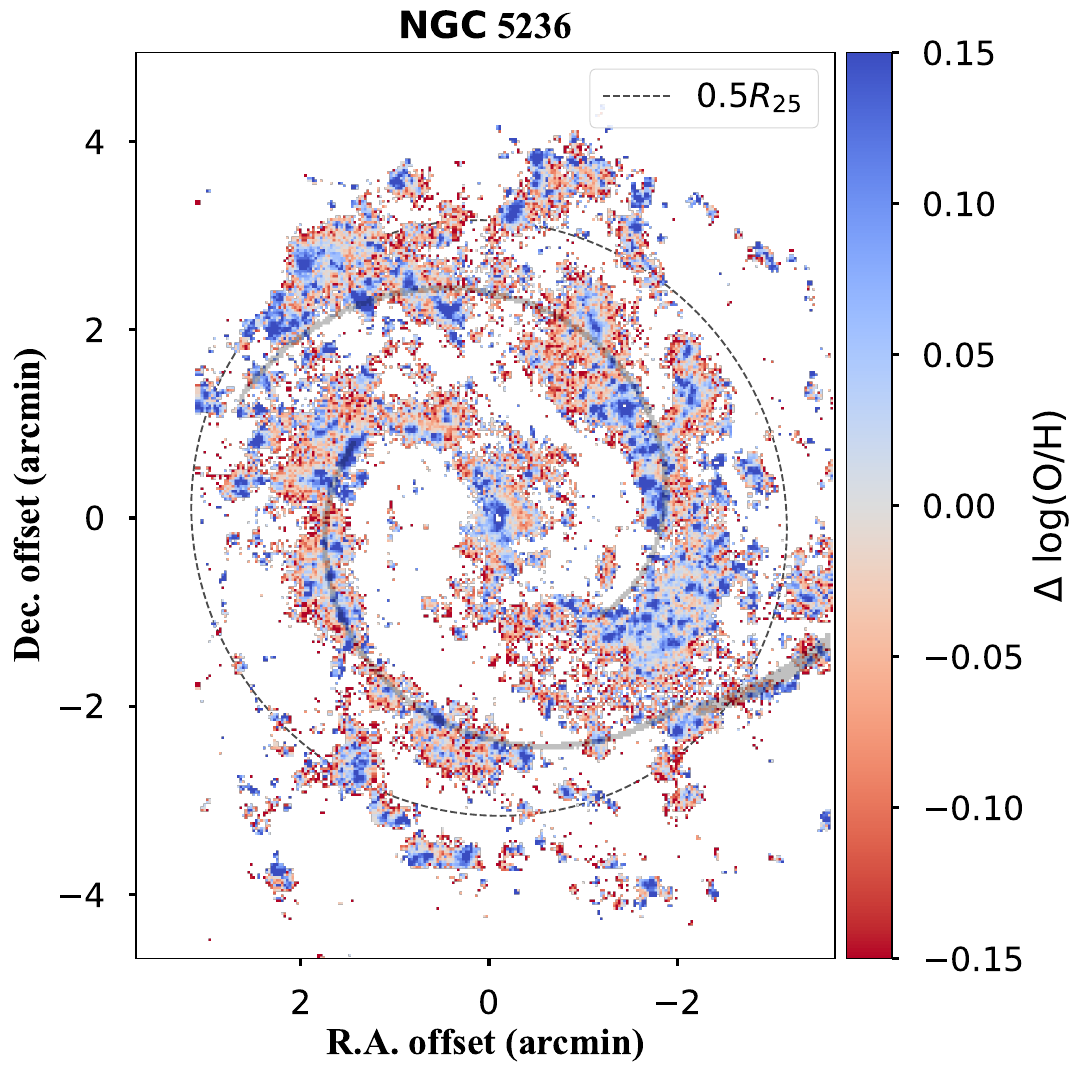}

    \includegraphics[width=0.2\textwidth, height=1.8in]{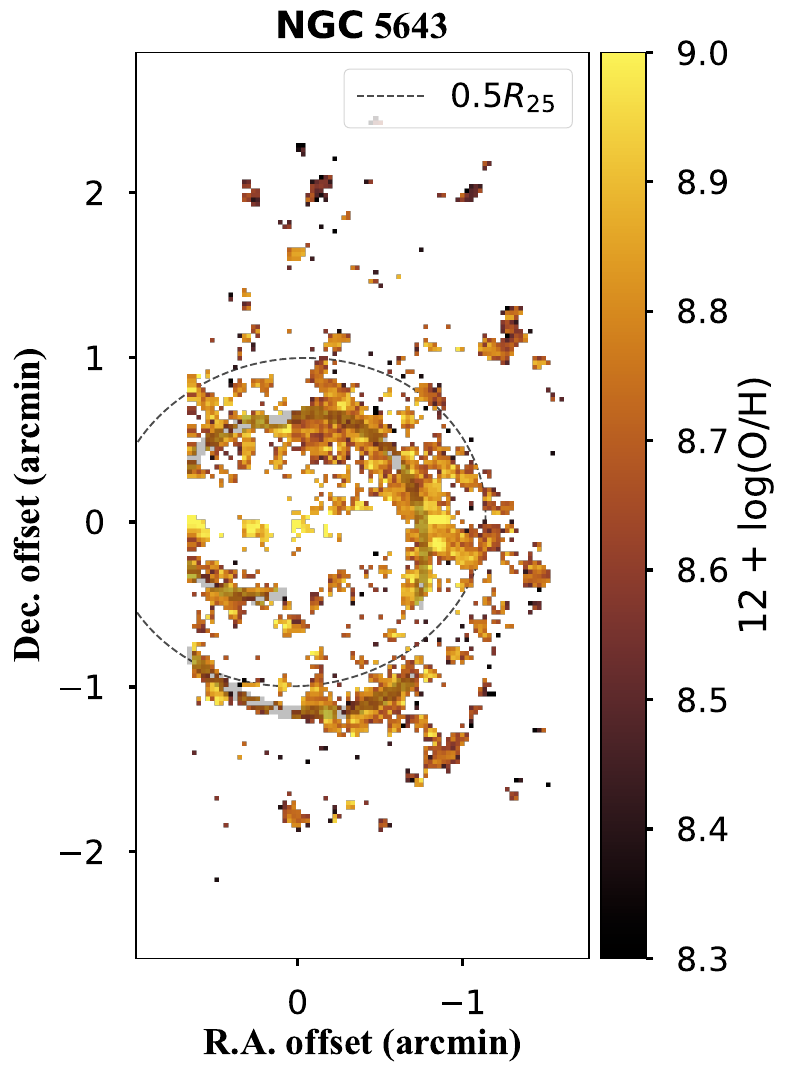}
    \includegraphics[width=0.2\textwidth, height=1.8in]{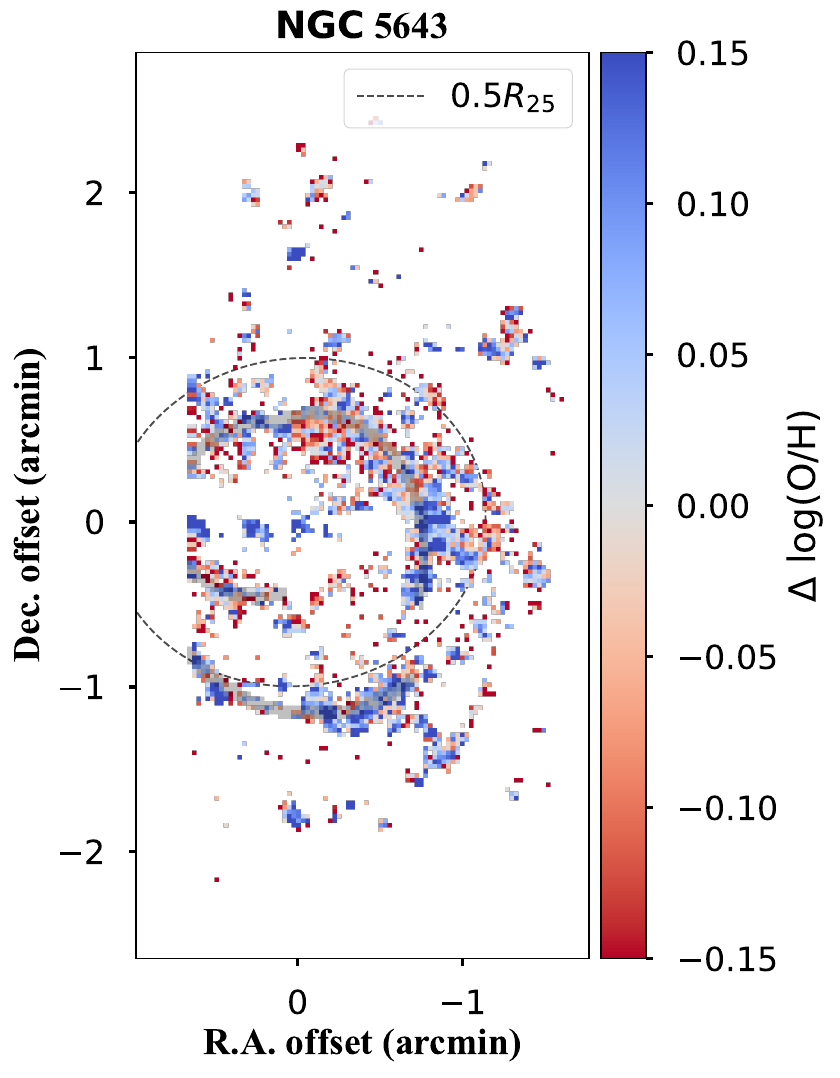}
    \includegraphics[width=0.2\textwidth, height=1.8in]{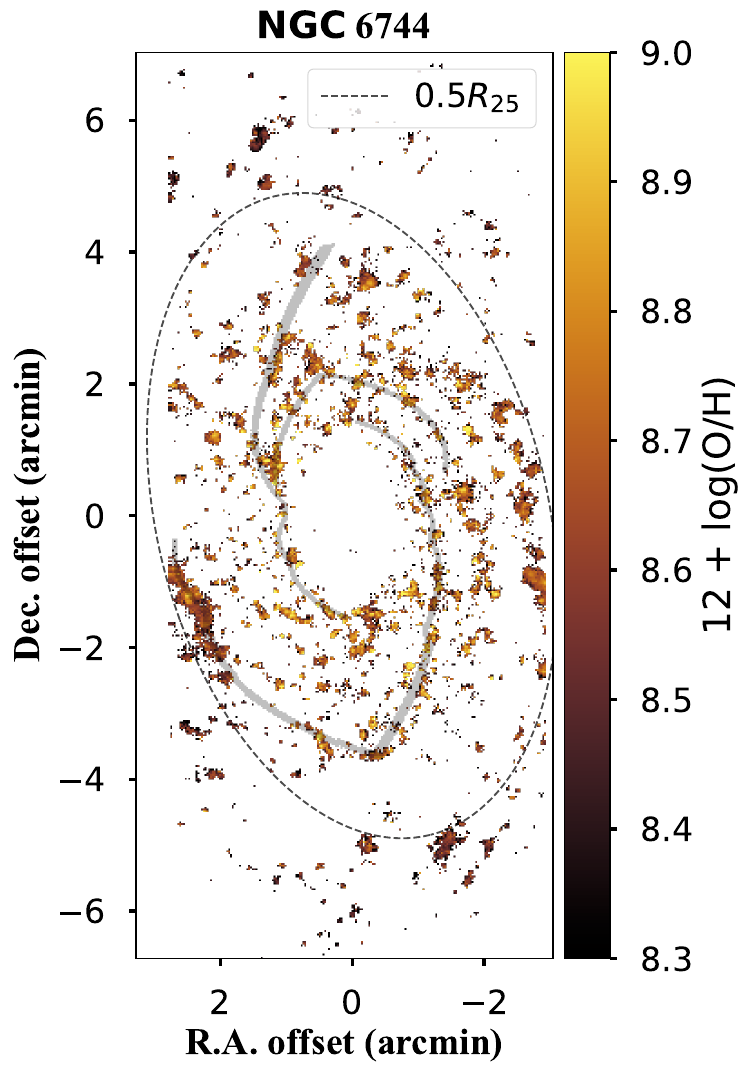}
    \includegraphics[width=0.2\textwidth, height=1.8in]{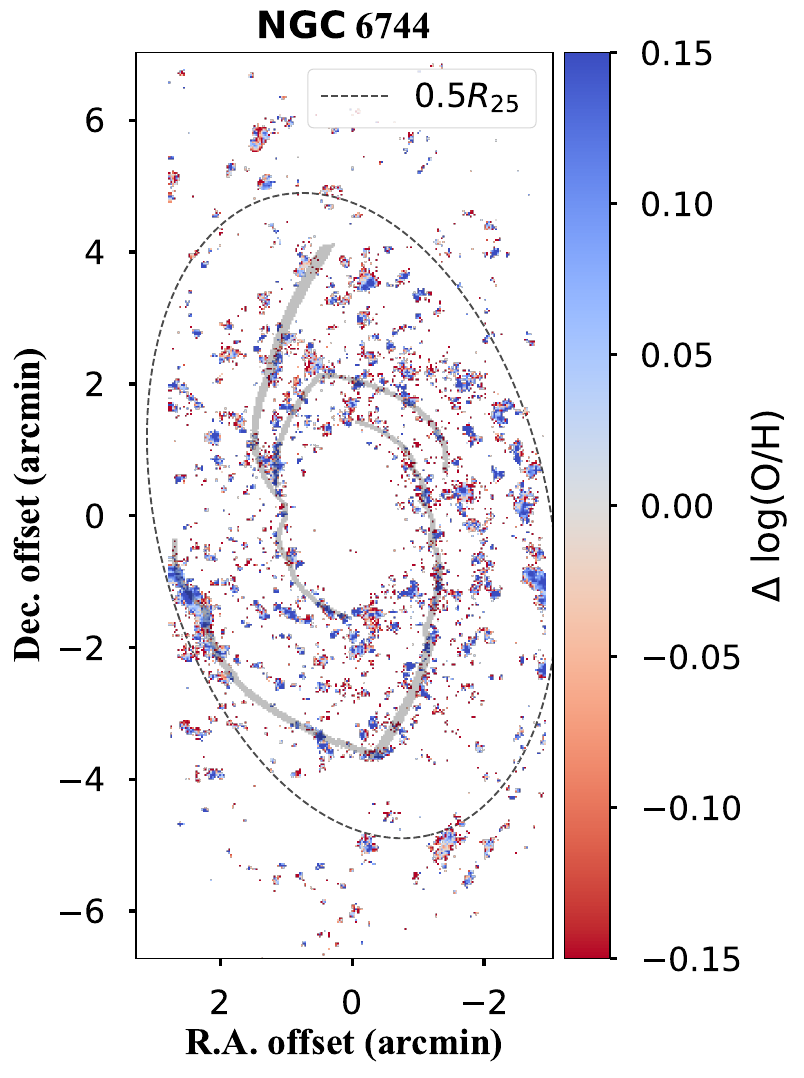}    
    \caption{{\bf Column 1 \& 3}: 2D 12 + log(O/H) maps of our galaxies, except for NGC~1566 (Fig~\ref{fig:z_map}).
    {\bf Column 2 \& 4}: Residual of metallicity (\dOH) by subtracting the radial gradients.
    The dashed ellipse marks the location of half $R_{25}$ and the red ellipse denotes the CR, if applicable.
    More detail on this analysis is in Sec~\ref{sec:ISM}.}
    \label{fig:z_map_other}
\end{figure*}

\section{Gas-phase metallicity with S-cal calibration}\label{appendix_scal}
In the main text, we use N2S2-N2\ha\ diagnostic from D16 to determine the gas-phase metallicity of our spiral galaxies.
As the physical resolution of our sample ranges from 145~pc to 39~pc, comparable to the typical size of H~{\sc ii} region, it is a caveat that the metallicities of some spaxels are contaminated by DIG. 
\citet{Poetrodjojo_2019} find that all diagnostics are affected by the inclusion of DIG.
As the current knowledge about modelling metallicity in DIG is still limited \citep{Kewley_2019}, we cannot separately derive the metallicity in the DIG.

Emission lines in DIG are excited by ionized photons leaked from H~{\sc ii} regions and low-mass evolved stars.
To test the impact of DIG in our results, we adopt S-calibration (Scal) from \citet{Pilyugin_2016} which relies on three standard diagnostic lines: 
\begin{equation}
\begin{split}
    R_2 &= I_{[O III]\lambda4959 + \lambda5007} / I_{H\beta}, \\
    N_2 &= I_{[N II]\lambda\lambda6548, 84} / I_{H\beta}, \\
    S_2 &= I_{[S II]\lambda\lambda6717, 31} / I_{H\beta}. 
\end{split}
\end{equation}
The inclusion of three emission line ratios allows Scal to be corrected for the dependence on ionization parameter \citep{Pilyugin_2016}.

\begin{figure*}
    \centering
    \includegraphics[width=0.45\textwidth]{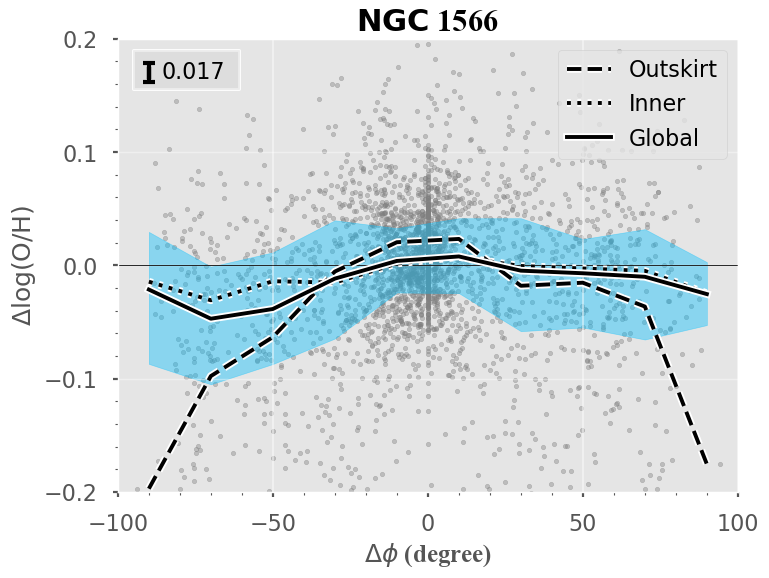}
    \includegraphics[width=0.45\textwidth]{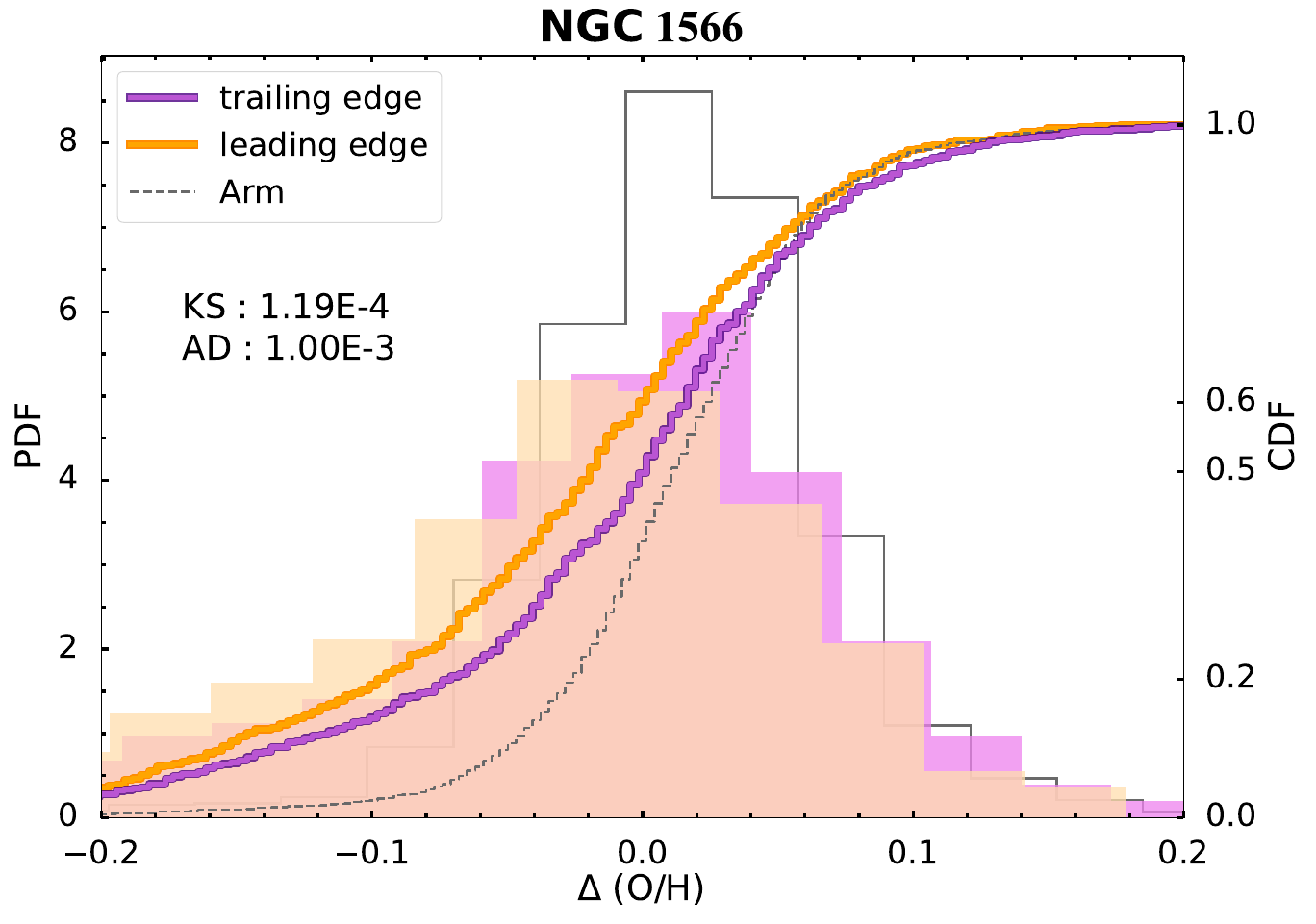}
    \caption{{\bf Left}: density scatter plot of metallicity residual \dOH\ versus azimuthal distance to the spiral arms \dphi.
    \dOH\ is calculated by subtracting the radial gradient from the S-cal metallicity.
    \dphi\ is the same as Fig~\ref{fig:dphi}, described in Sec~\ref{sec:arm_define}.
    We find slightly higher \dOH\ in the trailing edge (\dphi\ $> 0$) than the leading edge (\dphi\ $< 0$), similar with the main results using N2S2-N2\ha\ diagnostic (Fig~\ref{fig:dphi_dz}).
    {\bf right}: comparing the CDFs of \dOH\ from Scal among the trailing edge, leading edge and spiral arms.
    The $p$-value from the KS test and AD test are shown in the upper left.
    We find higher \dOH\ in the trailing edge, with a $p$-value of 1.19$\times 10^{-4}$, suggesting that the metallicity on both sides of the spiral arms are drawn from different distributions.
    }
    \label{fig:scal}
\end{figure*}

We take NGC~1566 for an explanation in this section.
The left panel of Fig~\ref{fig:scal} shows the fluctuation of metallicity residual \dOH\ when crossing the spiral arms, with positive \dphi\ indicating the trailing edge.
We use the same definition of \dphi\ in the main test, shown in Fig~\ref{fig:dphi} and described in Sec~\ref{sec:arm_define}.
The moving medians of each 20$^\circ$ \dphi\ are shown as a solid black line, with 25\% and 75\% quantiles as blue shadows.
We repeat the calculation of moving medians to the inner region and outer region, with each region containing half of the spaxels.
Similar to the result from N2S2-N2\ha\ (Fig~\ref{fig:dphi_dz}), we find slightly lower metallicity in the leading edge (\dphi\ $< 0$) in NGC~1566 using Scal (Fig~\ref{fig:scal}).
The magnitude of the azimuthal variation and the metallicity scattering from Scal are smaller than those from N2S2-N2\ha, which is expected \citep{Pilyugin_2016, Kreckel_2019}.

We further compare the metallicity distributions on both sides of the spiral arms by applying the KS-test and AD-test to their CDFs.
Both tests reject that the metallicity distribution in the leading and trailing edges are drawn from the same parental distribution.
Although Scal brings in a smaller intrinsic scatter, we obtain a $D$-value of 0.127 from the Scal CDFs, comparable to the $D$-value from N2S2-N2\ha.
We observe azimuthal variation in the metallicity of NGC~1566 in both Scal and N2S2-N2\ha\ diagnostics.
The highly similar trend of \dOH\ along \dphi\ suggests the limited impacts of DIG on our TYPHOON data.

\end{CJK} 
\label{lastpage}
\end{document}